\documentclass[12pt]{article}
\usepackage{cite}
\usepackage{graphicx}
\usepackage{multicol}
\usepackage{amsfonts}
\usepackage{amssymb}
\usepackage{amsmath}
\usepackage{bbm}
\usepackage{heck}
\usepackage{color}
\usepackage{hyperref}
\hypersetup{colorlinks=true}
\hypersetup{linkcolor=black}
\hypersetup{citecolor=black}
\hypersetup{urlcolor=black}
\usepackage{setspace}
\usepackage[all]{xy}
\usepackage{verbatim}
\usepackage{color}

\numberwithin{equation}{section}

\newcommand{\Tr}{\, {\rm Tr}}
\newcommand{\ident}{{\rm 1\hspace*{-0.4ex}%
\rule{0.1ex}{1.52ex}\hspace*{0.2ex}}}

\begin{document}

\date{October 2010}


\date{October, 2010}

\institution{SISSA}{\centerline{${}^{1}$Scuola Internazionale Superiore di Studi Avanzati, Via Bonomea 265 34100 Trieste, ITALY}}

\institution{HarvardU}{\centerline{${}^{2}$Jefferson Physical Laboratory, Harvard University, Cambridge, MA 02138, USA}}

\institution{IAS}{\centerline{${}^{3}$School of Natural Sciences, Institute for Advanced Study, Princeton, NJ 08540, USA}}

\institution{MIT}{\centerline{${}^{4}$Center for Theoretical Physics, Massachusetts Institute of Technology, Cambridge, MA 02139, USA}}

\title{T-Branes and Monodromy}

\authors{Sergio Cecotti\worksat{\SISSA}\footnote{e-mail: {\tt cecotti@sissa.it}}, Clay C\'{o}rdova\worksat{\HarvardU}\footnote{e-mail: {\tt clay.cordova@gmail.com}}, \\[4mm]Jonathan J. Heckman\worksat{\IAS}\footnote{e-mail: {\tt jheckman@ias.edu}} and Cumrun Vafa\worksat{\HarvardU,\MIT}\footnote{e-mail: {\tt vafa@physics.harvard.edu}}}

\abstract{We introduce T-branes, or ``triangular branes,'' which are novel non-abelian bound states of branes characterized by the condition that on some loci, their matrix of normal deformations, or Higgs field, is upper triangular.  These configurations refine the notion of monodromic branes which have recently played a key role in F-theory phenomenology.  We show how localized matter living on complex codimension one subspaces emerge, and explain how to compute their Yukawa couplings, which are localized in complex codimension two. Not only do T-branes clarify what is meant by brane monodromy, they also open up a vast array of new possibilities both for phenomenological constructions and for purely theoretical applications. We show that for a general T-brane, the eigenvalues of the Higgs field can fail to capture the spectrum of localized modes. In particular, this provides a method for evading some constraints on F-theory GUTs which have assumed that the spectral equation for the Higgs field completely determines a local model.}

\maketitle

\enlargethispage{\baselineskip}

\setcounter{tocdepth}{2}
\begin{spacing}{1}
\tableofcontents
\end{spacing}

\section{Introduction}

One of the most intriguing facts to emerge in the post-duality era is that gauge fields and matter can be trapped on branes and their intersections.  This idea of localization opened up the possibility that perhaps the observed particles in our universe may have their origin in a small region of an internal space, leading to a potentially dramatic simplification in the search for our corner of the vast string landscape.   Combined with a few key features, in particular the assumption of a supersymmetric grand unification of forces, this has naturally led to F-theory as a promising corner of the string landscape \cite{DWI,BHVI,BHVII,DWII}.\footnote{See \cite{Heckman:2010bq,Weigand:2010wm} for recent reviews.} In these models seven-branes trap the gauge fields and their intersections lead to matter localized on complex curves in the internal space.  At points where these matter curves meet, one finds a Yukawa coupling among the localized modes.

The physics of these systems is captured by a topologically twisted eight-dimensional gauge theory which describes a stack of space-filling seven-branes wrapping a compact four-cycle \(S\).  This leads to a 4\(D\) \(\mathcal{N}=1\) supersymmetric gauge theory whose low energy dynamics are governed by a generalization of Hitchin's equations \cite{VafaWitten,BHVI}.  The explicit field theory description enables many properties of intricate configurations of intersecting seven-branes to be computed with relative ease.  The key fact is that the eight-dimensional gauge theory supports an adjoint Higgs field \(\Phi\) whose expectation value parameterizes normal motion of the seven-brane stack.  Configurations of supersymmetric intersecting seven-branes are then obtained by studying solutions to the equations of motion where \(\Phi\) has a holomorphically varying vacuum expectation value.  Matter fields are described by fluctuations around a background \(\langle \Phi \rangle\) and Yukawa couplings measure the obstruction to extending these solutions beyond linear order.  A simple class of backgrounds which exhibit this general structure is to take \(\langle \Phi \rangle\) to reside in the Cartan subalgebra.  Along codimension one loci the background \(\langle\Phi\rangle\) degenerates and the local effective gauge group is partially unHiggsed.  This enhancement of the symmetry group leads to matter curves.  On codimension two loci where the local symmetry group enhances further one finds trilinear Yukawa couplings.

Localization of the particle physics degrees of freedom, especially the interaction terms, appears to be a promising framework for string based phenomenology. The real world exhibits Yukawa couplings which display striking patterns and hierarchies.  In our current understanding of nature these couplings are mysterious parameters and one task of beyond the standard model physics is to explain them.  In the seven-brane gauge theory the Yukawa term is a superpotential interaction and as such is a holomorphic object, invariant under the complexified group of gauge transformations and insensitive to the metric on the seven-brane worldvolume.\footnote{Of course the physical Yukawa coupling does in addition depend on the metric coming from the D-term normalizations.}  When three seven-branes intersect, the fields living at the three matter curves are paired together to form a coupling.  From the geometry it is clear that this interaction is concentrated at the triple intersection of branes and by making full use of the symmetries of the superpotential one can make this exact: the contribution to the superpotential from a triple intersection of branes is localized to an arbitrarily tiny neighborhood of the triple intersection.  This coupling is therefore a universal object.  It depends only on the dynamics of the theory in a small patch containing the triple intersection and hence is independent of the four-cycle \(S\).  Yukawa couplings in seven-brane gauge theories are thus incredibly robust physical quantities and this gives us hope that perhaps seven-branes provide an avenue for string theory to make contact with the theory of flavor \cite{HVCKM,BHSV,HVCP,FGUTSNC}.

\begin{figure}
\begin{center}
\framebox{
\includegraphics[totalheight=0.25\textheight]{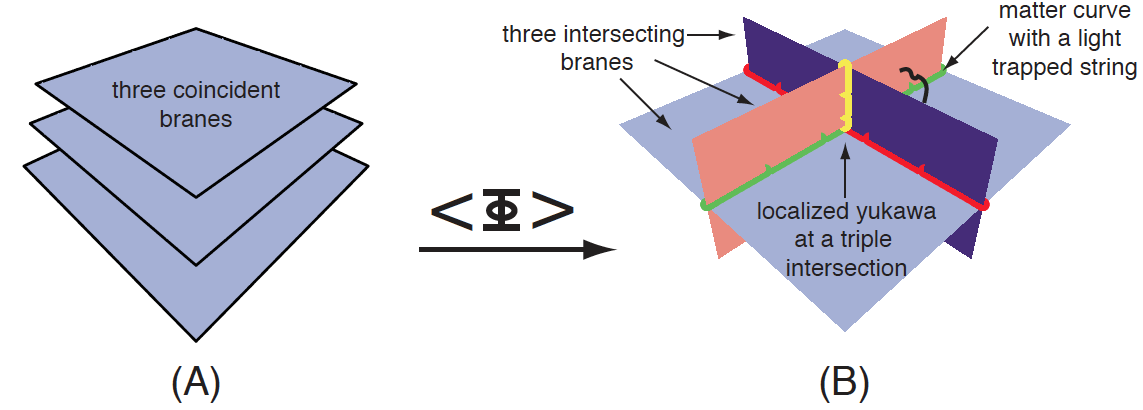}
}
\caption{A configuration of intersecting branes can be studied as a background in a theory of coincident branes.  In (A) we have a stack of three branes supporting a \(U(3)\) gauge group.  In (B), the Higgs field \(\Phi\) develops a vev and describes three intersecting branes with gauge group \(U(1)^{3}\).  At the intersection of branes are trapped charged fields.  At triple intersections a Yukawa coupling is generated.}
\label{fig:intbranes}
\end{center}
\end{figure}

An important observation by \cite{Hayashi:2009ge} was that to achieve exactly one heavy generation of up type quarks in these models, the phenomenon of seven-brane monodromy is required.  Subsequent works \cite{BHSV, EPOINT, Marsano:2009gv} showed that seven-brane monodromy is a helpful ingredient for other aspects of F-theory models as well. The notion of a monodromic brane was originally interpreted as field configurations \(\langle \Phi \rangle\) which are valued in the Cartan but have branch cuts and undergo monodromy by elements of the Weyl group.  This characterization in terms of just the eigenvalues of $\Phi$ turns out to be physically inadequate in many situations. For example it was found in \cite{Funparticles, FCFT} that a three-brane probing a configuration of ``monodromic branes'' is sensitive to far more than just the eigenvalues of $\Phi$. In this paper we show that the correct picture of monodromy is that the background Higgs field can be described globally without branch cuts, but that there are loci where \(\langle\Phi \rangle\) cannot be gauge rotated to lie in the Cartan. Away from such loci one can view the Higgs field as lying in the Cartan, but then in principle it is only single valued up to the action of the Weyl group.  Thus a well defined, single-valued Higgs field can lead to a monodromic brane. For example in the case of a $U(N)$ gauge group, there can be loci where \(\langle \Phi \rangle\) is upper triangular, and thus \emph{non-diagonalizable}.  We call such configurations of seven-branes ``T-branes.''

T-branes can be viewed as certain non-abelian bound states of branes, whose description is not completely captured by the position-dependent eigenvalues of the Higgs field.  The most dramatic possibility is to contrast a nilpotent non-zero \(\langle \Phi \rangle\) with the zero matrix.  Both of these have vanishing eigenvalues but lead to strikingly different physics in the worldvolume gauge theory. The map between T-branes and monodromic branes is many to one, and for a given choice of brane monodromy group there are many choices of T-branes with the same monodromy action.  In this way we can view T-brane configurations as a refinement and clarification of the idea of a monodromic brane. This fills a conceptual gap in the current literature, for although many papers have explored (in a different language) some examples of monodromic branes, a systematic and general analysis of these systems has been lacking.

A general method for studying such Higgs fields involves the technology of spectral covers, reviewed for example in \cite{DonagiSpec}. Here it is important to draw a distinction between a general spectral cover, which is specified by a choice of a matrix $\langle \Phi \rangle$, and the spectral equation, which is specified by the characteristic polynomial for $\langle \Phi \rangle$. It was found in \cite{Hayashi:2009ge,DWIII} that the spectral equation for $\Phi$ can be interpreted as defining some aspects of a local elliptically fibered Calabi-Yau fourfold, and thus in F-theory language, as capturing some aspects of a configuration of seven-branes.  Unfortunately, in the physics literature it has been often assumed that the spectral equation carries complete information about a local F-theory compactification. Our aim in this paper will be to determine when such assumptions are warranted, when they are not, and in all cases, how to analyze the corresponding T-brane configurations using $\langle \Phi \rangle$.

In section \(\ref{ReviewB}\) we begin with a review of the case where the Higgs field is valued in the Cartan.  For concreteness we specialize to the case of $U(N)$ gauge theory and consider position dependent eigenvalues of \(\langle\Phi \rangle \).  This is the gauge theory description of intersecting branes.  The unbroken gauge group is a product of $U(k_i)$'s and the system is governed by a diagonal background Higgs field \(\langle \Phi \rangle\).  Here we also review the residue calculus which enables one to compute \emph{exactly} and \emph{explicitly} the localized contributions to the superpotential.  Much of this material is known from \cite{BHVI,FGUTSNC} and we review it here as we will need it when we generalize the discussion to T-branes.

Following this preliminary analysis original results begin in earnest.  We generalize section \ref{ReviewB} by turning to the more interesting case with fundamentally non-abelian T-brane solutions where \(\langle \Phi \rangle\) is non-diagonalizable. Our first task in section \ref{Mono} is to describe in detail examples of such triangular backgrounds and the associated spectrum of massless fluctuations.  In contrast to the abelian intersecting brane solutions of section \ref{ReviewB} the equations of motion are now non-linear and obtaining exact solutions is already a non-trivial task.  As evidence of this fact we find that in the course of studying the simplest possible examples the famous differential equations of Liouville and Painlev\'{e} make a surprise appearance.  Next, we develop the study of the spectrum of massless matter in these backgrounds from a number of points of view.  Of particular importance is our description of the general notion of a matter curve where a fluctuation is trapped.  In contrast to the abelian case there is now no simple geometric picture like figure \ref{fig:intbranes} which makes the existence of localized matter obvious. Nevertheless, we find that non-diagonalizable Higgs backgrounds frequently support trapped charged matter, and that matter curves are again characterized by loci where the complexified gauge group is partially unHiggsed.

With some examples under our belt, in section \ref{formalism} we turn to an abstract description of the localized spectrum and their superpotential couplings.  The goal we accomplish there is to develop a holomorphic formalism where the full symmetries of the superpotential are manifest and where exact answers are available for the universal localized Yukawa couplings which occur when matter curves in T-brane backgrounds intersect.  This section forms the technical core of the paper.  The most significant conceptual point that we address is to understand the precise match between the 8\(D\) fields of the seven-brane gauge theory which describe a localized mode and the actual \(6D\) field which resides on a matter curve.  Once this correspondence is developed, it is straightforward to generalize the residue calculus of intersecting branes in section \ref{ReviewB} to a wide class of background Higgs fields.

The remaining sections of the paper apply the holomorphic formalism of section \ref{formalism}.  Section \ref{Higgsing} explains in some detail how some T-brane backgrounds have a simple interpretation in terms of brane recombination.  We match the massless spectra in both the original and recombined frame and determine the conditions under which such an alternative picture is applicable.  One particularly useful result of this analysis is the determination of exactly when a background Higgs field can be reconstructed from its eigenvalues.

Section \ref{weylorbits} is devoted to a more detailed analysis of the monodromy group.  We explain when the notion of a T-brane collapses to that of a monodromic brane and explain how, under appropriate assumptions, the monodromy group provides a useful picture of the spectrum of localized charged matter and various selection rules in the superpotential.

In section \ref{examples} we compute a number of simple examples of superpotentials including the phenomenologically interesting Yukawas generated at \(E_{6}, E_{7}\), and \(E_{8}\) points conjectured to be responsible for the mass of the top quark. In this section we also discuss a wide variety of novel physical properties which cannot be seen from the eigenvalues of \(\langle \Phi \rangle\). Of significant practical importance for F-theory phenomenology, we present examples which show that the spectral equation for the Higgs field does \textit{not} in general determine the spectrum of massless charged matter. These counterexamples provide a general way to bypass various constraints on the matter content of F-theory GUTs found in \cite{Marsano:2009gv ,Dudas:2009hu}, which assumed that the spectral equation provides complete information on the localized matter content.
\textit{Let us repeat: to specify the physical theory, one must in general indicate an explicit choice of} $\langle \Phi \rangle$.

Finally, section \ref{CONCLUDE} contains our conclusions and possible directions for further investigation.  Some additional technical material is collected in the Appendices.

\section{The Field Theory of Intersecting Seven-Branes}
\label{ReviewB}
In this section we review how intersecting branes are described by background field configurations in a fixed field theory. Much of this material can be found throughout the literature, and we shall follow in particular the discussion in \cite{KatzVafa, BHVI, FGUTSNC}.  We consider seven-branes which fill a four-dimensional Minkowski spacetime and wrap a compact four-manifold \(S\) inside our compactification.  To preserve supersymmetry \(S\) should be a complex and K\"{a}hler manifold, with K\"{a}hler form \(\omega\).  For applications to type IIB string theory it is natural to consider a unitary gauge group \(U(n)\) with \(n\) the number of seven-branes.  However our considerations have a wider application to the non-perturbative case of F-theory.  There one may consider a seven-brane which supports an arbitrary compact Lie group $G$ as the gauge group, and for now we will frame the discussion in this more general setting. 

On a flat brane worldvolume, the field content and Lagrangian of this gauge theory is simply that of minimal 8\(D\) \(\mathcal{N}=1\) super-Yang-Mills.  The bosonic fields are then a gauge field \(A\) and a complex adjoint scalar \(\Phi\).  As usual with brane field theories, the adjoint scalar describes normal fluctuations of the brane worldvolume in the ambient space of the compactification.  When we wrap the branes on the curved background \(\mathbb{R}^{3,1}\times S\), supersymmetry demands that the field theory be topologically twisted in such a way that \(\Phi\) is now a (2,0) form on \(S\), and when this is so the resulting effective 4\(D\) theory in Minkowski space has an unbroken \(\mathcal{N}=1\) supersymmetry \cite{BSV,BershadskyFOURD,BHVI}.  Our interest is in studying field configurations in this theory which preserve the \(SO(3,1)\) Lorentz group, thus we take the expectation values of the connection \(A\) in the Minkowski directions to vanish.  Requiring \(\mathcal{N}=1\) supersymmetry in four-dimensions then enforces the following BPS equations of motion on the fields \cite{VafaWitten, BHVI}\footnote{A word on conventions.  We define a matrix \(X\) to be in the Lie algebra if \(e^{X}\) is in the gauge group.  Thus for example, if our gauge group is \(SU(n)\) then the Lie algebra consists of traceless \emph{anti}hermitian matrices.  Also one should be aware that in comparison with the notation used elsewhere in the literature \(\overline{\Phi}_{\rm{there}}\equiv-\Phi^{\dagger}_{\rm{here}}\).}
\begin{eqnarray}
F^{0,2}_{A} & = & 0, \label{integrable} \\
\bar{\partial}_{A}\Phi & = & 0 ,\label{phieom}\\
\omega \wedge F_{A}+\frac{i}{2}[ \Phi^{\dagger}, \Phi] & =  & 0. \label{dterm11}
\end{eqnarray}
The first two of these equations are determined by F-flatness conditions.  The relation \((\ref{integrable})\) is an integrability requirement which tells us that the gauge field \(A\) is a connection on a \emph{holomorphic} bundle.  Equation \((\ref{phieom})\) then states that, with respect to the holomorphic structure defined by \(A\), the Higgs field \(\Phi\) is holomorphic. One can obtain these two equations by minimizing the superpotential \cite{BHVI,DWI}:
\begin{equation}
W_{8D}=\int_{S}\mathrm{Tr}\left(F_{A}^{0,2}\wedge \Phi \right). \label{firstw}
\end{equation}
An important property of this superpotential is that it is insensitive to any K\"{a}hler data.  By virtue of the topological twist, the superpotential density is naturally a (2,2) form and can be integrated over \(S\) without reference to the metric.  As a consequence of this the two F-term equations \((\ref{integrable})\) and \((\ref{phieom})\) are invariant under the complexified group of gauge transformations, and throughout this work we will make heavy use of this fact.  Finally, the third constraint \((\ref{dterm11})\) is the D-flatness condition for this gauge theory.  It is explicitly sensitive to the K\"{a}hler form \(\omega\) and as in similar gauge systems it plays the role of a stability condition.  These three equations are well-known \cite{VafaWitten} and constitute the basic tool which we shall use to study general configurations of seven-branes. They define a rich moduli space of field configurations on \(S\) which one should think of as a generalization to two complex dimensions of the celebrated Hitchin system \cite{HitchinSelf}.

Our basic paradigm in this paper will be to study seven-brane gauge theories in the presence of a BPS background \((\langle \Phi\rangle, \langle A \rangle)\).  The massless matter content of such a configuration can then be deduced by studying small fluctuations around the given solution.  We define first order quantities \(\varphi\) and \(a\) by
\begin{eqnarray}
 \Phi & = & \langle \Phi \rangle +\varphi , \\
 A^{0,1} & = &\langle A^{0,1}\rangle+ a.
\end{eqnarray}
From now on, the total quantum fields \(\Phi\) and \(A\) will not appear and for simplicity of notation we will denote the background value \(\langle \Phi \rangle\) by \(\Phi\) and \(\langle A\rangle\) by \(A\).   We linearize the BPS equations to find the equations satisfied by the fluctuation fields \((\varphi, a)\)
\begin{eqnarray}
 \bar{\partial}_{A}  a & = & 0, \label{eomda} \\
\bar{\partial}_{A} \varphi + [a,  \Phi ] & = & 0, \label{aeom} \\
 \omega \wedge \left(\partial_{A} a-\bar{\partial}_{A}a^{\dagger}\right)+\frac{i}{2}\left([ \Phi ^{\dagger},  \varphi]+[\varphi^{\dagger}, \Phi ]\right) & = & 0. \label{oeom}
\end{eqnarray}
To get a correct count of the physically distinct modes, we must quotient the space of solutions to the above by the action of the linearized group of gauge transformations.  As one readily checks, the effect of a gauge transformation with small parameter \(\chi\) is to change the fields as
\begin{eqnarray}
a & \rightarrow & a+\bar{\partial}_{A}\chi, \label{agauge}\\
\varphi & \rightarrow & \varphi+[ \Phi  , \chi]  \label{phigauge}.
\end{eqnarray}
To deduce the spectrum of the theory we must then determine the space of solutions to the fluctuation equations \((\ref{eomda})-(\ref{oeom})\) modulo the linearized gauge transformations \((\ref{agauge})-(\ref{phigauge})\).

Once we have determined the matter spectrum we can move on to study their F-term interactions.  These are dictated by the \(8D\) superpotential (\ref{firstw}).  Since the fluctuation fields solve the linearized BPS equations, at leading order the superpotential gives a cubic coupling.  We expand the fields about their vacuum expectation values and use the equations of motion satisfied by the background to find
\begin{equation}
W_{Y}=\int_{S}\mathrm{Tr}\left(a\wedge a\wedge\varphi \right). \label{wydef}
\end{equation}
This is a trilinear Yukawa coupling, and one primary aim in the remainder of this work is to elucidate its structure both abstractly and explicitly in a variety of examples.  One can see directly that this coupling is gauge invariant under the linearized gauge transformations (\ref{agauge})-(\ref{phigauge}).  If \(\chi\) is the gauge parameter, then the F-term equations of motion for the fluctuations imply that the first order change in \(W_{Y}\) is given by
\begin{equation}
\delta W_{Y}=\int_{S}\bar{\partial}_{A} \ \mathrm{Tr}\left(a\wedge [\varphi,\chi]\right) \label{cmpctwvan}
\end{equation}
Since the change in the superpotential is \(\bar{\partial}_{A}\) exact, by virtue of the fact that the surface \(S\) is compact we conclude that \(\delta W_{Y}\) vanishes.

The problem of extracting the low energy behavior of a given seven-brane configuration is now reduced to the study of the solutions to the fluctuation equations and their renormalizable superpotential couplings as computed by \((\ref{wydef})\).  In general the physics depends in an intricate way on the background fields, and it useful to organize the study of solutions by the complexity of the Higgs field \(\Phi\).  The simplest class of widely studied  backgrounds are the \emph{intersecting brane solutions}.  We define these by the condition that
\begin{equation}
[\Phi,\Phi^{\dagger}]=0. \label{intbranedef}
\end{equation}
In such a situation the Higgs field \(\Phi\) can be brought to a gauge where it is valued in the Cartan subalgebra.  In the simplest case of a \(U(n)\) gauge group this means that \(\Phi\) is diagonal. For simplicity, it is also common to assume that no background gauge field flux has been switched on. In this case, the physics is totally dictated by the behavior of the eigenvalues of \(\Phi\) with each eigenvalue controlling the position of one of the branes. For the remainder of this section we present a detailed review of these intersecting brane solutions.  Following this, in section \ref{Mono} we undertake the study of seven-branes where the simplifying assumption \eqref{intbranedef} is dropped.

\subsection{Matter}
\label{ReviewMat}
\subsubsection{Unitary Gauge}
\label{MatU}
Although realistic applications of seven-brane gauge theory require that the complex surface \(S\) should be compact, much of the intuition for solutions can be seen in the non-compact limit where we study the equations on a small flat patch \(\mathbb{C}^{2}\subset S\) with complex coordinates \((x,y)\).   We work in the simplest case of an intersecting brane solution with a unitary gauge group \(U(n)\).  As usual, the overall \(U(1)\) center of mass decouples and allows us to restrict our attention to \(SU(n)\) field theory. Here we further simplify the discussion by taking the background gauge field \(A\) to vanish.  The only non-trivial constraint on the background is then
\begin{equation}
\bar{\partial}\Phi=0.
\end{equation}
The simplest class of solutions consists of a diagonal Higgs field \(\Phi\) with constant eigenvalues \(\lambda_{i}\).  Since the Higgs field parameterizes deformations of the brane stack, its eigenvalues control the relative positions of each of the branes and this vacuum has the familiar interpretation of moving the branes off of each other.  If all the eigenvalues are distinct then the gauge group is Higgsed from \(SU(n)\) to \(U(1)^{n-1}\).

Still focusing on a small patch \(\mathbb{C}^{2}\), we can obtain a more interesting solution by taking the eigenvalues of \(\Phi\) to be holomorphic functions, \(\lambda_{i}\rightarrow \lambda_{i}(x,y)\).  A basic fact is that this background now describes a configuration of supersymmetric intersecting seven-branes.  Along the loci where pairs of eigenvalues coincide the relative separation of a pair of branes shrinks to zero and the seven-branes collide.  It is exactly in this situation that one expects to find massless charged matter at the intersection given by open strings stretched between the two branes, and one of the virtues of the gauge theory description is that these states are easily visible.

Following \cite{KatzVafa,BHVI} and especially \cite{FGUTSNC}, let us now study this phenomenon concretely in the simplest possible example of an \(SU(2)\) gauge theory with background Higgs field
\begin{equation}
 \Phi =\left(
\begin{array}{c c}\frac{x}{2} & 0 \\
0 & -\frac{x}{2}
\end{array}
\right)dx \wedge dy. \label{su2ex}
\end{equation}
This background breaks the gauge symmetry from \(SU(2)\) to \(U(1)\).  Away from the complex line \(x=0\) the eigenvalues of \( \Phi \) are distinct and this field describes a pair of separated branes.  Along \(x=0\) these branes intersect.

To find the spectrum of massless matter, we study the spectrum of small fluctuations around the background (\ref{su2ex}).  Since \(A=0\), all covariant derivatives become ordinary derivatives, and the F-term equations read
\begin{eqnarray}
\bar{\partial}a & = & 0 \label{naeoma}\\
\bar{\partial}\varphi  & = & [\Phi,a]. \label{nphieoma}
\end{eqnarray}
We can solve these equations by noting that since we are working locally on a patch \(\mathbb{C}^{2}\subset S\), the \(\bar{\partial}\) operator is exact.  Thus any differential form which, like \(a\),  is \(\bar{\partial}\) closed is also \(\bar{\partial}\) exact.  So we can solve \((\ref{naeoma})\) by introducing an \(sl(2,\mathbb{C})\) matrix \(\xi\) with
\begin{equation}
\bar{\partial} \xi = a.
\end{equation}
Since \(a\) transforms as a \((0,1)\) form on \(\mathbb{C}^{2}\), \(\xi\) transforms as a scalar.  The next step in solving the linearized equations is to integrate \((\ref{nphieoma})\)
\begin{equation}
 \varphi = [ \Phi , \xi]+h, \label{phiint}
\end{equation}
with \(h\) is an arbitrary holomorphic adjoint \((2,0)\) form.  Given the data \((\xi, h)\) the final D-term equation gives us a second order differential equation relating them.
\begin{equation}
\omega \wedge (\partial a -\bar{\partial}a^{\dagger})+\frac{i}{2}\left([\Phi^{\dagger},\varphi]+[\varphi^{\dagger},\Phi]\right)=0
\end{equation}
To study this equation it is natural to decompose \(\xi\) into eigenvectors under commutation with the background Higgs field \( \Phi  \).  The possible \(sl(2,\mathbb{C})\) eigenmatrices and their eigenvalues are listed below.
\begin{itemize}
\item A diagonal \(\xi \) which commutes with \( \Phi \):
\begin{equation}
\xi=\left(
\begin{array}{cc}
\xi_{0} & 0 \\
0 & -\xi_{0}
\end{array}
\right).
\end{equation}
\item An off-diagonal \(\xi\) with eigenvalues \(\pm x\):
\begin{equation}
\xi=\left(
\begin{array}{cc}
0 & \xi_{+} \\
0 & 0
\end{array}
\right)  \ \ \mathrm{eigenvalue} \  x, \ \ \ \ \ \ \ \xi=\left(
\begin{array}{cc}
0 & 0 \\
\xi_{-} & 0
\end{array}
\right)  \ \ \mathrm{eigenvalue} \  -x.
\end{equation}
\end{itemize}
Focusing now on the first possibility of a diagonal \(\xi\), a short calculation shows that we can reach a unique gauge where the holomorphic matrix \(h\) is simultaneously diagonal and the non-vanishing matrix entries \(\pm \xi_{0}\) are real.  The linearized D-term equation then reduces to the fact that \(\xi_{0}\) is harmonic.  Thus the solution is specified by
\begin{equation}
a=\left(
\begin{array}{cc}
\bar{\partial}\xi_{0} & 0 \\
0 & -\bar{\partial}\xi_{0}
\end{array}
\right), \hspace{.4in}
\varphi=\left(
\begin{array}{cc}
h_{0} & 0 \\
0 & -h_{0}
\end{array}
\right)dx \wedge dy, \hspace{.5in}  \Delta \xi_{0}=\bar{\partial}h_{0}=0
\end{equation}
where $\Delta$ denotes the usual Laplacian. These modes are nothing but the gauge multiplet of the unbroken \(U(1)\) gauge group that remains in the presence of our \(SU(2)\) background \((\ref{su2ex})\).  Geometrically, the perturbation \(a\) encodes the freedom to turn on a flat connection, while the \(\varphi\) mode further deforms the brane configuration.

More interesting solutions are found by choosing \(\xi\) to be off-diagonal corresponding to the generators of the gauge group which are broken by the background Higgs field.  As a representative example take
\begin{equation}
\xi=\left(
\begin{array}{cc}
0 & \xi_{+} \\
0 & 0
\end{array}
\right), \hspace{.4in}
h=\left(
\begin{array}{cc}
0 & h_{+} \\
0 & 0
\end{array}
\right)dx \wedge dy.
\end{equation}
And let us further equip our brane worldvolume with a flat K\"{a}hler form which is, up to a scale \(\ell\), just the flat metric on \(\mathbb{C}^{2}\).
\begin{equation}
\omega=\frac{i\ell^{2}}{2}\left(dx \wedge d\bar{x}+dy\wedge d\bar{y}\right).
\end{equation}
The linearized D-flatness condition (\ref{oeom}) then amounts to
\begin{equation}
\left(\Delta-\frac{|x|^{2}}{\ell^{2}}\right)\xi_{+}= \frac{\bar{x}}{\ell^{2}}h_{+}.
\end{equation}
This equation admits solutions where the holomorphic function \(h_{+}\) depends only on \(y\), the complex coordinate on the brane intersection.  Explicitly we find that
\begin{equation}
a=\left(
\begin{array}{cc}
0 & -\frac{h_{+}(y)}{\ell}e^{-|x|^{2}/\ell}\\
0 & 0
\end{array}
\right)d\bar{x} , \hspace{.4in}
\varphi=\left(
\begin{array}{cc}
0 & h_{+}(y)e^{-|x|^{2}/\ell} \\
0 & 0
\end{array}
\right)dx \wedge dy. \label{su2hypersol}
\end{equation}
These modes are the massless strings stretched between intersecting branes.  From the solution we can see that these modes are sharply concentrated at \(x=0\), the matter curve, where the branes intersect and the classical length of the string shrinks to zero.  The fact that the solution depends on an arbitrary holomorphic function \(h_{+}(y)\) is a signal that in the effective \(U(1)\) theory in the presence of the background \((\ref{su2ex})\) these light strings comprise the degrees of freedom of a 6\(D\) quantum field which lives at the intersection of the two seven-branes.  The solution (\ref{su2hypersol}) of charge \(+1\), together with the linearly independent transposed solution of charge \(-1\) comprise the bosonic fields of a hypermultiplet.

Based on this example one can easily deduce the spectrum of charged trapped matter for an arbitrary intersecting brane background.  To avoid writing many matrices, it is useful to introduce some notation for the \(sl(n,\mathbb{C})\) Lie algebra.  We set conventions such that \(H_{i}\) denotes an \(n\times n\) diagonal matrix with a one in the \(i\)-th slot and zeros elsewhere.  The Cartan subalgebra, \(\mathfrak{h}\), is given by matrices of the form
\begin{equation}
\mathfrak{h}=\left\{c_{1}H_{1}+c_{2}H_{2}+\cdots c_{n}H_{n}|\sum_{i}c_{i}=0\right \}.
\end{equation}
The roots vectors of the algebra will be denoted \(R_{ij}\).  They are elementary matrices with a one in the \(i\)-th row and \(j\)-th column and zeros elsewhere, and have definite eigenvalues under commutation with a Cartan element
\begin{equation}
[\sum_{k}c_{k}H_{k}, R_{ij}]=(c_{i}-c_{j})R_{ij}.
\end{equation}
The roots of the algebra are the associated eigenvalues; for \(R_{ij}\) the root is \(c_{i}-c_{j}\).  

Now in the case of a general diagonal background we have
\begin{equation}
\Phi=\lambda_{1}H_{1}+\cdots +\lambda_{n}H_{n} \hspace{1in} \sum_{k}\lambda_{k}=0. \label{gendiaghiggs}
\end{equation}
If all the holomorphic eigenvalues \(\lambda_{i}\) are distinct then the gauge group is Higgsed to \(U(1)^{n-1}\).  Along complex curves in the brane worldvolume pairs of eigenvalues become equal and the branes intersect.  These are the matter curves.  We see that in general they are given by roots of the algebra.  We consider a matter fields \((\varphi,a)\) which are the generalization of the off-diagonal modes of our example \eqref{su2ex}, whose form in matrix space is given by a root \(R_{ij}\).  The F-term equations are then integrated exactly as above
\begin{equation}
a=\bar{\partial}\xi, \hspace{1in} \varphi=(\lambda_{i}-\lambda_{j})\xi+h.
\end{equation}
Solving the D-term equations as before we then obtain a solution where \(h\) depends only on the coordinate along the matter curve, and the modes vanish exponentially fast away from \(\lambda_{i}=\lambda_{j}\).
\subsubsection{Holomorphic Gauge \label{MatH}}

Section \ref{MatU} provides a complete calculation of the spectrum an \(SU(n)\) gauge theory in an intersecting brane background.  We have linearized the equations of motion and obtained the solutions to the F- and D-flatness conditions in a physical unitary gauge.  As is frequently the case in supersymmetric theories, the analysis can be simplified by working with the complexified group of gauge transformations.  The F-term equations are invariant under this larger group, and by a standard argument the full system of F- and D-flatness conditions modulo unitary gauge transformations has the same space of solutions as the F-flatness conditions modulo complexified gauge transformations.\footnote{Stability conditions, which add important subtleties to this statement, will not play a role in our discussion.}  In the example of the spectrum on a small patch \(\mathbb{C}^{2}\subset S\) this is a major simplification.

Working with the complexified gauge group, we can obtain a clear picture of the theory by passing to what is known as \emph{holomorphic gauge}.  This gauge is characterized by the fact that the \((0,1)\) part of the connection vanishes so that in a holomorphic gauge
\begin{equation}
\bar{\partial}_{A}=\bar{\partial}+A^{0,1}=\bar{\partial}.
\end{equation}
The integrability condition for this equation is simply \(F_{A}^{0,2}=0\) and thus our ability to reach a holomorphic gauge is guaranteed by one of the basic BPS equations \((\ref{integrable})\). In a physical unitary gauge, one must also keep track of the $(1,1)$ component of the field strength as well as the K\"{a}hler form, which may in general be non-trivial.   However when we work with the complexified gauge group we get to neglect this data, and indeed the entire D-term equation \eqref{dterm11}.  The price we pay for this simplification is that in holomorphic gauge, we will not be able to obtain the detailed profile of the perturbations.  Nevertheless for intrinsically holomorphic questions, such as superpotential calculations, the wave function profile as determined by the D-term data is irrelevant.

Now, to study the spectrum of small fluctuations about a given background in holomorphic gauge we note that it is still true that we can take \(A^{0,1}\), and hence the perturbation \(a\) to vanish.  This fact can be easily seen directly.  In a holomorphic gauge for the background, the general F-term fluctuation equations are
\begin{eqnarray}
\bar{\partial}a & = & 0 \\
\bar{\partial}\varphi & = & [\Phi,a],
\end{eqnarray}
and our analysis in the previous section implies that these have as a general solution
\begin{equation}
a =\bar{\partial}\xi \hspace{1in} \varphi = [\Phi,\xi]+h
\end{equation}
To obtain a correct count of the degrees of freedom we must now quotient this space of solutions to the F-term equations by the complexified group of gauge transformations.  According to \((\ref{agauge})\) an infinitesimal gauge transformation with parameter \(\chi\) has the effect of shifting \(\xi\)
\begin{equation}
\xi\longrightarrow \xi +\chi.
\end{equation}
\(\xi\) is valued in the complexified Lie algebra $\frak{g}_{\mathbb{C}}$ and therefore if we work with the unitary form of the gauge group, the gauge parameter \(\chi\) has only half as many degrees of freedom as \(\xi\).  However, if we work with complexified gauge transformations then \(\xi\) and \(\chi\) are valued in the same space and there is no loss in generality in setting \(\chi=-\xi\) and thereby gauging \(a\) to zero.

Once we go to holomorphic gauge, the F-term equation for the Higgs field perturbation \(\varphi\) implies that \(\varphi=h\) is simply a holomorphic (2,0) form.   Keeping \(a\) gauged to zero, we still have the freedom to make complexified gauge transformations with a \emph{holomorphic} infinitesimal parameter \(\chi\).  Under such a transformation the field \(\varphi\) shifts by the commutator of \(\chi\) with the background Higgs field
\begin{equation}
\varphi\longrightarrow \varphi+[\Phi,\chi].
\end{equation}
Thus in a holomorphic gauge the calculation of the spectrum is reduced to a completely algebraic problem.  The space of gauge inequivalent modes is given by all possible holomorphic matrices modulo those matrices which are commutators with the background \(\Phi\).

Let us see what this means in the context of the simple example \((\ref{su2ex})\).  In a holomorphic gauge we have
\begin{equation}
\varphi=\left(
\begin{array}{cc}
h_{0}(x,y) & h_{+}(x,y) \\
h_{-}(x,y) & -h_{0}(x,y)
\end{array}
\right)dx \wedge dy.
\hspace{.5in} \bar{\partial}h_{\alpha}=0.
\end{equation}
Now we quotient by the remaining holomorphic gauge transformations.  Using this freedom we can reach a gauge where the off-diagonal elements of \(\varphi\) depend only on \(y\), the complex coordinate along the brane intersection
\begin{equation}
\varphi=\left(
\begin{array}{cc}
h_{0}(x,y) & h_{+}(y) \\
h_{-}(y) & -h_{0}(x,y)
\end{array}
\right)dx \wedge dy.
\hspace{.5in} \bar{\partial}h_{\alpha}=0.
\end{equation}
In this form of the solution the fact that the off-diagonal elements of \(\varphi\) depend only on \(y\) is the holomorphic description of the fact that the light strings which they represent are confined to the matter curve \(x=0\).  Meanwhile the diagonal mode \(h_{0}(x,y)\) depends on both coordinates and is a bulk field.

Now that we understand this simple example, the general case of an \(SU(n)\) gauge theory broken to \(U(1)^{n-1}\) by a diagonal Higgs field \((\ref{gendiaghiggs})\) has more indices but is no more complicated.  In a holomorphic gauge a mode \(\varphi\) given by a root \(R_{ij}\)
shifts under a gauge transformation as
\begin{equation}
\varphi \longrightarrow \varphi + (\lambda_{i}-\lambda_{j})\alpha \label{alphashift}
\end{equation}
with \(\alpha\) an arbitrary holomorphic function.  The space of gauge inequivalent perturbations can then be described abstractly by introducing \(\mathcal{O}\) the ring of holomorphic functions in two complex variables \((x,y)\).  In a holomorphic gauge \(\varphi \in \mathcal{O}\) and according to equation (\ref{alphashift}) this description is redundant up to an arbitrary multiple of the root \((\lambda_{i}-\lambda_{j})\).  If we denote by \(\mathcal{I}_{ij}\) the ideal generated by the root, then we see that the space of gauge inequivalent perturbations in the matrix direction \(R_{ij}\) is exactly the quotient space
\begin{equation}
\mathcal{O}/\mathcal{I}_{ij}.
\end{equation}
The above has an intuitive meaning.  The matter curve is defined by setting all functions in the ideal \(\mathcal{I}_{ij}\) to zero and all gauge invariant data in the perturbation \(\varphi\) is contained in its behavior on this curve.  One should contrast this with the corresponding statement for a bulk mode.  If we consider a diagonal perturbation \(\varphi\), then since \(\varphi\) and \(\Phi\) commute one cannot change \(\varphi\) by a gauge transformation and the behavior of the mode over the entire worldvolume carries physical information.

As explained in \cite{FGUTSNC}, a second important use of complexified gauge transformations is that they allow us to make precise the notion that a mode is confined to curve.  We consider matter fields \((a, \varphi)\) whose form in matrix space is given by a root vector \(R_{ij}\).  Then the solution \(\varphi\) is a smooth \((2,0)\) form which satisfies the linearized BPS equation:
\begin{equation}
\bar{\partial} \varphi=(\lambda_{i}-\lambda_{j})a. \label{newfterm}
\end{equation}
An absolutely key fact is that we can find a representative for \(\varphi\) which is in the same complexified gauge group orbit and which has \emph{arbitrarily narrow support near the matter curve \(\lambda_{i}=\lambda_{j}\).}  The proof of this statement is purely formal.  We simply let \(T_{\epsilon}\) be a small tube of radius \(\epsilon\) around the matter curve.  Then we can define a new smooth \((2,0)\) form \(\varphi'\) with the properties:
\begin{equation}
\varphi '=\left \{\begin{array}{l l}
\varphi & \mathrm{inside}  \ T_{\epsilon/2}\\
0 & \mathrm{outside}  \ T_{\epsilon}
\end{array}
\right. .
\end{equation}
We can then define a smooth gauge parameter \(\chi\) by:
\begin{equation}
\chi=\frac{\varphi'-\varphi}{\lambda_{i}-\lambda_{j}}.
\end{equation}
By construction a complexified gauge transformation with parameter \(\chi\) takes the mode \(\varphi\) to \(\varphi'\).  Notice that as a consequence of equation \((\ref{newfterm})\) if \(\varphi\) vanishes, then so does the associated gauge field perturbation \(a\).  Thus we have succeeded in constructing a \emph{localized gauge} where the matter modes \((a,\varphi)\) are non-vanishing only inside a parametrically small tube \(T_{\epsilon}\) around the matter curve.

Thus the holomorphic and localized gauges differ in the way that one chooses the completely arbitrary gauge field perturbation \(a\).  For holomorphic gauge we simplify our lives by taking \(a\) to vanish leaving only the holomorphic \(\varphi\).  Meanwhile in the localized gauge we choose a non-zero \(a\) in such a way that the solution vanishes away from the matter curve.  Conceptually, the equivalence between these two perspectives follows from the fact that the only gauge invariant data in the mode is the behavior of \(\varphi\) at the matter curve, and there the localized gauge and the holomorphic gauge agree.  For the purposes of computations of holomorphic quantities like the superpotential, we may freely use whichever gauge is most convenient.
\subsubsection{Matter Curve Actions}
\label{mcact}
The previous two sections give us useful perspectives on the 6\(D\) defect quantum fields localized on the intersection of seven-branes.  The modes we have studied are fluctuation fields which solve the linearized BPS equations.  These are the \emph{on-shell} 6\(D\) fields.  For many questions it is often useful to have a notion of \emph{off-shell} fields and thus an action principle.  Since the fields in question are localized on curves we desire an action which is written as an integral along the matter curve and whose minimization enforces the BPS equations of motion.  As noted for example in \cite{DWI, BHVI}, if we work holomorphically, that is with only the F-terms modulo the complexified gauge group, this action is completely determined by the 8\(D\) superpotential (\ref{firstw})
\begin{equation}
W_{8D}=\int_{S}\mathrm{Tr}\left(F_{A}^{0,2}\wedge \Phi \right). \label{superpotential}
\end{equation}

To obtain the 6\(D\) action \(W_{6D}\) for the modes starting from the above, we simply expand \(W_{8D}\) above to quadratic order in the fluctuation fields and evaluate
\begin{equation}
W_{6D}=\int_{S}\mathrm{Tr}\left(\bar{\partial}a\wedge \varphi+a\wedge a\wedge \Phi\right). \label{wquad}
\end{equation}
Since the fluctuation fields solve the linearized equations, by definition \(W_{6D}=0\) on-shell.  To produce a suitable off-shell coupling we thus put the matter fields only half on-shell.  We envision a situation where the matter is localized on a curve \(\Sigma \subset S\).  We take the F-term fluctuation equations and we separate variables into a coordinate parallel and normal to \(\Sigma\).  We solve the equations in the transverse direction, but we leave the modes off-shell in the parallel direction.  Plugging into \(W_{8D}\) and evaluating the integral then gives the \(6D\) action.\footnote{Though this type of analysis is implicit in \cite{BHVI, DWI, FGUTSNC}, we are not aware of an explicit derivation of this fact in the literature.}

To illustrate this procedure we consider the general case of a diagonal background (\ref{gendiaghiggs}) on \(\mathbb{C}^{2}\) and a matter curve defined by \(x=0\).  Such a situation is described by a 6\(D\) field theory and thus the matter that we find must be in a representation of the \(6D\) superalgebra.  This means that matter must come in the form of 6\(D\) hypermultiplets and hence for each matter field \((a,\varphi)\) there is a conjugate mode \((a^{c},\varphi^{c})\) of opposite charge under the unbroken gauge group.  It is easy to see this explicitly.  If \(x=0\) defines a matter curve for the root \(R_{ij}\) then it also defines a matter curve for the transposed root \(R_{ji}\) which thus supports the conjugate mode.  These fields are localized on the same matter curve and are naturally paired by the quadratic superpotential (\ref{wquad}).

To evaluate \(W_{6D}\) we first separate variables.  In components, the F-term equations are
\begin{eqnarray}
\bar{\partial}_{\bar{x}}\varphi & = & x a_{\bar{x}} \label{x1}\\
\bar{\partial}_{\bar{y}}\varphi & = & x a_{\bar{y}} \label{y1}\\
\bar{\partial}_{\bar{x}}\varphi^{c} & = & -x a_{\bar{x}}^{c} \label{x2}\\
\bar{\partial}_{\bar{y}}\varphi^{c} & = & -x a_{\bar{y}}^{c} \label{y2}
\end{eqnarray}
We now put the modes half on-shell by solving equations \((\ref{x1})\) and \((\ref{x2})\) corresponding to the transverse directions of the matter curve, while we do not enforce the parallel equations \((\ref{y1})\) and \((\ref{y2})\).  Our method of computation is to make use of the localized gauge constructed in the previous section.  Although these modes are not fully on-shell, one can easily see that we can still reach a gauge where \((a_{\bar{x}},\varphi)\) and \((a_{\bar{x}}^{c},\varphi^{c})\) vanish outside a parametrically small tube \(T_{\epsilon}\) around the matter curve.  In contrast to the full solutions of the previous section however, nothing can be said about the localization properties of the components \(a_{\bar{y}}\) and \(a_{\bar{y}}^{c}\).

Let us activate the perturbation:
\begin{equation}
\varphi=\varphi+\varphi^{c} \hspace{1in} a=a+a^{c}
\end{equation}
which is a solution to the transverse BPS equations.  We plug into \((\ref{wquad})\) and obtain
\begin{equation}
W_{6D}=\int_{\mathbb{C}^{2}}\left(\bar{\partial}a\wedge \varphi^{c}+\bar{\partial}a^{c}\wedge\varphi+xa\wedge a^{c}\wedge dx \wedge dy\right). \label{wexpand}
\end{equation}
Our goal is to reduce this quantity to an integral along the matter curve \(x=0\).  The first step is to observe that by localization we can take the \(\varphi\) modes to vanish outside a tube of radius \(\epsilon\) around the matter curve, and hence restrict the domain of integration to \(\mathbb{C}^{2}\cap T_{\epsilon}\).  Since the gauge field perturbations are regular at the matter curve the third term in \((\ref{wexpand})\) vanishes in the localized limit and can be dropped.  Next integrate the remaining terms by parts:
\begin{equation}
W_{6D} =\int_{\mathbb{C}^{2}\cap T_{\epsilon}}\left(a\wedge \bar{\partial}\varphi^{c}+a^{c}\wedge\bar{\partial}\varphi \right).
\end{equation}
Expand in components and make use of the transverse BPS equations to obtain:
\begin{equation}
W_{6D}=\int_{\mathbb{C}_{y}}dy\wedge d\bar{y}\left(\int_{|x|\leq \epsilon}\frac{\bar{\partial}_{\bar{x}}\varphi\bar{\partial}_{\bar{y}}\varphi^{c}-\bar{\partial}_{\bar{x}}\varphi^{c}\bar{\partial}_{\bar{y}}\varphi}{x}d\bar{x}\wedge dx+\cdots\right). \label{newwquad}
\end{equation}
In the above, the remaining contribution ``\(\cdots\)'' involves terms proportional to \(a_{\bar{y}}\) and \(a^{c}_{\bar{y}}\).  Since this is not localized near the matter curve, these pieces vanish as we take the localization parameter \(\epsilon \rightarrow 0\).  Now observe that
\begin{eqnarray}
\bar{\partial}_{\bar{x}}(\varphi\bar{\partial}_{\bar{y}}\varphi^{c}) &= &\bar{\partial}_{\bar{x}}\varphi\bar{\partial}_{\bar{y}}\varphi^{c}+\varphi\bar{\partial}_{\bar{y}}\bar{\partial}_{\bar{x}}\varphi^{c}\\
& = & \bar{\partial}_{\bar{x}}\varphi\bar{\partial}_{\bar{y}}\varphi^{c}-x\varphi\bar{\partial}_{\bar{y}}a_{\bar{x}}^{c}.
\end{eqnarray}
The second term in the last line of the above vanishes in the localized limit so we may freely replace our expression (\ref{newwquad}) by
\begin{equation}
W_{6D}=\int_{\mathbb{C}_{y}}dy\wedge d\bar{y}\left(\int_{|x|\leq \epsilon}\frac{\bar{\partial}_{\bar{x}}\left(\varphi\bar{\partial}_{\bar{y}}\varphi^{c}\right)-\bar{\partial}_{\bar{x}}\left(\varphi^{c}\bar{\partial}_{\bar{y}}\varphi\right)}{x}d\bar{x}\wedge dx\right). \label{nnewwquad}
\end{equation}
Finally we may we note that since \(\varphi\) and \(\varphi^{c}\) vanish outside \(T_{\epsilon}\) we have
\begin{equation}
0=\oint_{|x|= \epsilon}\left(\frac{\varphi\bar{\partial}_{\bar{y}}\varphi^{c}-\varphi^{c}\bar{\partial}_{\bar{y}}\varphi}{x}\right) dx. \label{vancont}
\end{equation}
Add (\ref{vancont}) to (\ref{nnewwquad}) and make use of the Cauchy integral formula.  Up to an overall constant which we ignore the quantity above can be simplified as
\begin{equation}
W=\int_{\mathbb{C}_{y}}\left(\varphi\bar{\partial}_{\bar{y}}\varphi^{c}-\varphi^{c}\bar{\partial}_{\bar{y}}\varphi\right)|_{x=0}  \ dy\wedge d\bar{y}. \label{wquadfinal}
\end{equation}
Equation (\ref{wquadfinal}) is our final expression.  It has the desired form of a pairing between \(\varphi\) and \(\varphi^{c}\) written as an integral over the matter curve which is the complex \(y\) plane, \(\mathbb{C}_{y}\).  As expected this action takes the standard form of a free-chiral Dirac Lagrangian for fermions propagating on the matter curve.  As a consistency check on this result, note that if we minimize the superpotential (\ref{wquadfinal}) we find the equations of motion:
\begin{equation}
\bar{\partial}_{\bar{y}}\varphi|_{x=0}=\bar{\partial}_{\bar{y}}\varphi^{c}|_{x=0}0
\end{equation}
and these are exactly the unenforced F-term BPS equations (\ref{y1}) and (\ref{y2}) restricted to the matter curve.

The structure of the \(6D\) superpotential conceptually clarifies the meaning of off-shell modes.  An off-shell 8\(D\) field which describes a mode on a matter curve is one which satisfies the transverse BPS equations.  The off-shell 6\(D\) fields are given by restricting \(\varphi\) and \(\varphi^{c}\) to the matter curve.  Since these do not minimize (\ref{wquadfinal}) these fields are simply general functions of the matter curve coordinates \((y,\bar{y})\).  Extremizing the 6\(D\) superpotential and putting the fields on-shell then amounts to enforcing holomorphy on the 6\(D\) fields.
\subsection{Interactions}

In section \(\ref{ReviewMat}\) we have given an overview of exactly how matter at a pair of intersecting branes can be seen directly from field theory.  In the gauge theory description these modes are described by the fluctuations \((\varphi, a)\) and couplings between them can be computed by simply evaluating the superpotential integral.  After putting all modes on-shell at leading order the superpotential gives the cubic Yukawa coupling (\ref{wydef}).
\begin{equation}
W_{Y}=\int_{S}\mathrm{Tr}\left( a \wedge a  \wedge \varphi \right). \label{superpotentialf}
\end{equation}
The superpotential is a holomorphic object and is invariant under the complexified gauge group.  It follows that when evaluating the Yukawa coupling we can work either in unitary or holomorphic gauge and in the following we will study \(W_{Y}\) from both perspectives.
\subsubsection{Unitary Gauge}
\label{YukU}
Following \cite{FGUTSNC}, let us study the simplest possible background with a non-zero superpotential.  The presence of the trace in the coupling tells us that to obtain a non-vanishing contribution involving three 6D hypermultiplets we must start with an \(SU(3)\) gauge theory.  We envision a situation, like Figure \(\ref{fig:intbranes}\) of the Introduction, where we have three seven-branes meeting pairwise transversally and having a triple intersection at exactly one point in the compactification.  The first and most intuitive way to evaluate the coupling is to recall from our example in section \((\ref{ReviewMat})\) that the physical unitary solutions to the fluctuation equations are concentrated along the matter curve.   Since the branes meet pairwise transversally, this means that the superpotential density which is integrated in \((\ref{superpotentialf})\) is peaked near the region of triple intersection.  It is thus reasonable to approximate \(W_{Y}\) as an integral over only a small patch \(\mathbb{C}^{2}\subset S\) centered at the triple intersection.  This is quite a useful step since now we can use our local analysis of the fluctuation equations to compute the superpotential.

We can begin as before with a diagonal holomorphic background Higgs field
\begin{eqnarray}
 \Phi  & = &
\frac{1}{3}\left(
\begin{array}{ccc}
-2x+y & 0 & 0 \\
0 & x+y & 0 \\
0 & 0 & x-2y
\end{array}
\right)dx \wedge dy \label{su3yukex} \\
& = & \left(\phantom{\int}\hspace{-.175in}(-2x+y)H_{1}+(x+y)H_{2}+(x-2y)H_{3}\right)\frac{dx \wedge dy}{3}. \nonumber
\end{eqnarray}
This background breaks \(SU(3)\) to \(U(1)\times U(1)\).  The branes intersect at the loci where pairs of eigenvalues become equal, that is the \(x\) and \(y\) axes and the curve \(x=y\).  The triple intersection of the branes where all these curves meet and the coupling is concentrated is the origin \((x,y)=(0,0)\). In the conventions of subsection \ref{MatH} we can write the localized modes as
\begin{eqnarray}
 \varphi_{12}& = & \left(h_{12}(y)e^{-|x|^{2}/\ell}dx\wedge dy\right)R_{12} , \nonumber \\
 \varphi_{23} & = &\left(h_{23}(x)e^{-|y|^{2}/\ell}dx\wedge dy\right)R_{23} , \\
 \varphi_{31}& = & \left(h_{31}(x+y)e^{-|x-y|^{2}/\sqrt{2}\ell}dx\wedge dy\right)R_{31}, \nonumber
\end{eqnarray}
together with the corresponding gauge field perturbation modes
\begin{eqnarray}
a_{12}& = & \left(\frac{h_{12}(y)}{\ell}e^{-|x|^{2}/\ell}d\bar{x}\right)R_{12}, \nonumber \\
a_{23} & = & \left(-\frac{h_{23}(x)}{\ell}e^{-|y|^{2}/\ell}d\bar{y}\right)R_{23}, \\
a_{31} & = & \left(-\frac{h_{31}(x+y)}{\sqrt{2\ell}}e^{-|x-y|^{2}/\sqrt{2}\ell}(d\bar{y}-d\bar{x})\right)R_{31}. \nonumber
\end{eqnarray}
Plugging into the superpotential \((\ref{superpotentialf})\) we find that up to an overall non-zero constant\footnote{In general in the rest of this paper when we write \(W_{Y}\) we will mean up to an overall non-zero pure number. This coefficient can be kept track of but is somewhat uninteresting: when one passes from the holomorphic couplings to the physical unitary ones, the coefficient is rescaled anyway upon canonically normalizing the K\"{a}hler potential.}
\begin{equation}
W_{Y}=\frac{1}{\ell^{2}}\int_{\mathbb{C}^{2}}\left(h_{12}(y)h_{23}(x)h_{31}(x+y)e^{-\frac{1}{\ell}(|x|^{2}+|y|^{2}+|x-y|^{2}/\sqrt{2})}\right) dx\wedge dy \wedge d\bar{x}\wedge d\bar{y}. \label{superdterm}
\end{equation}
The resulting coupling depends on the local holomorphic behavior \(h_{ij}\) of the matter fields along the brane intersections.  These are the wavefunctions in the internal space of the four-dimensional quantum fields.  In any given four-dimensional model the compactness of the cycle \(S\) will imply that there are a finite number of \(h_{ij}\) which are realized, and armed with this data we can plug into the above formula and evaluate the Yukawa.  In our case since we are studying the behavior on a small patch \(\mathbb{C}^{2}\subset S\) we should consider arbitrary behavior of \(h_{ij}\).  In this sense by focusing our attention on a small patch of the gauge theory we also reduce to studying germs of wavefunctions. Possible local behaviors for these germs are given by any possible holomorphic power series depending on the single complex variable along each matter curve.  Thus a convenient basis is given by pure powers
\begin{eqnarray}
h_{12}(y) & \in & \mathrm{Span}\{1, y, y^{2}, \cdots\} \nonumber \\
h_{23}(x) & \in & \mathrm{Span}\{1, x, x^{2}, \cdots\} \\
h_{31}(x+y) & \in & \mathrm{Span}\{1, (x+y), (x+y)^{2}, \cdots\}. \nonumber
\end{eqnarray}
A complete local understanding of the superpotential in this example is then equivalent to computing the value of the integral \((\ref{superdterm})\) for any combination of the monomials above.  Luckily these integrals are trivial.  We find
\begin{equation}
W_{Y} = \frac{1}{\ell^{2}}\int_{\mathbb{C}^{2}}\left(y^{m}x^{n}(x+y)^{k}e^{-\frac{1}{\ell}(|x|^{2}+|y|^{2}+|x-y|^{2}/\sqrt{2})}\right) dx\wedge dy \wedge d\bar{x}\wedge d\bar{y}= \left\{ \begin{array}{ll} 1 & m=n=k=0 \\ 0 & \mathrm{else}\end{array}\right. . \label{unitaryyuk}
\end{equation}
Notice that the K\"{a}hler scale \(\ell\) has dropped out of the final answer as expected.  The result shows that in this simple example, the only non-zero couplings involve those wavefunctions which do not vanish at the point of triple intersection of the branes.  In section \ref{RankT} we will see that this fact has a simple geometric interpretation.
\subsubsection{Holomorphic Gauge: Residue Theory}
\label{YukH}
The fact that the effective \(4D\) \(\mathcal{N}=1\) superpotential is a holomorphic object requires that the Yukawa couplings be insensitive to all \emph{non-holomorphic} data of the background, in particular the K\"{a}hler form \(\omega\) and the gauge flux \(F_{A}\).  The physical unitary gauge employed in the previous section obscures this fact.  We can introduce a set-up manifestly independent of \(\omega\) and \(F_{A}\) by making use of the full power of the invariance of the superpotential under complexified gauge transformations.  As in section \ref{MatH} we consider an \(SU(n)\) gauge bundle which is reducible to a direct sum of line bundles and a background Higgs field \(\Phi\) which is purely diagonal:
\begin{equation}
\Phi = \lambda_{1}H_{1}+\cdots+\lambda_{n}H_{n} \hspace{.5in}\sum_{i}\lambda_{i}=0.
\end{equation}
We assume that the \(\lambda_{i}\) are all distinct so that this describes a breaking of \(SU(n)\) to \(U(1)^{n-1}\).  Triple intersections of branes occur at the points in \(S\) where for some choice of three indices \(i, j, k\)
\begin{equation}
\lambda_{i}=\lambda_{j}=\lambda_{k}.
\end{equation}
At such a point the modes for \(R_{ij}\), \(R_{jk}\), and \(R_{jk}\) can form a coupling \cite{BHVI}.  Without loss of generality let us take \(i=1\), \(j=2\), \(k=3\) and study the associated interaction.  For simplicity we assume that there is one point of triple intersection in \(S\), but the generalization to the case of an arbitrary finite number of such points will be clear.

The superpotential coupling involving the three roots is computed by activating the perturbations
\begin{equation}
\varphi=\varphi_{12}+\varphi_{23}+\varphi_{31}, \hspace{1in} a=a_{12}+a_{23}+a_{31},
\end{equation}
and plugging into the Yukawa integral
\begin{equation}
W_{Y}=\int_{S}\mathrm{Tr}(a_{12}\wedge a_{23}\wedge \varphi_{31}+a_{23}\wedge a_{31}\wedge \varphi_{12}+a_{31}\wedge a_{12}\wedge \varphi_{23}).
\end{equation}
We know that \(W_{Y}\) is invariant under the complexified group of gauge transformations.  Hence to evaluate the superpotential we can go to the localized gauge of section \(\ref{MatH}\) where the perturbations vanish outside a tiny neighborhood of the matter curves. It follows then that integrand in the superpotential simply vanishes outside a small neighborhood of the triple intersection.  Letting \(\mathbb{C}^{2}\subset S\) be a small neighborhood of the triple intersection we thus have
\begin{equation}
W_{Y}=\int_{\mathbb{C}^{2}}\mathrm{Tr}(a_{12}\wedge a_{23}\wedge \varphi_{31}+a_{23}\wedge a_{31}\wedge \varphi_{12}+a_{31}\wedge a_{12}\wedge \varphi_{23}). \label{localizednww}
\end{equation}
It is worthwhile to remark on the meaning of this equation.  In general the computation \((\ref{cmpctwvan})\) shows that gauge invariance of the superpotential requires a compact brane \(S\).  In any given patch of \(S\) we can reach a holomorphic gauge where the gauge field perturbations \(a\) can be set to zero, and hence there is no definite meaning to the superpotential contribution in that patch.  What is significant about \((\ref{localizednww})\) is that it shows that there is one particular class of gauges, the localized gauges, where the superpotential density simply vanishes outside a tiny neighborhood of the Yukawa point.  Thus if we restrict to gauge transformations which preserve the localization condition the value of \(a\) in this neighborhood indeed carries physical information.

The upshot of equation (\ref{localizednww}) is that now that we have reduced the computation to the case where \(S=\mathbb{C}^{2}\) our local analysis of the solutions to the equations of motion again takes over.\footnote{Since we work locally we will freely identify (2,0) forms with scalars by ``dividing'' by \(dx\wedge dy\).}  As in section \ref{MatH} we introduce \(\xi_{ij}\) a \(\bar{\partial}\) antiderivative to \(a_{ij}\) so that the solution for the pair \((a_{ij},\varphi_{ij})\) is given by:
\begin{eqnarray}
a_{ij} & = & \bar{\partial}\xi_{ij}, \label{newlocalsol} \\
 \varphi_{ij}& = & R_{ij}(\Phi)\xi_{ij}+h_{ij}=(\lambda_{i}-\lambda_{j})\xi_{ij}+h_{ij} . \nonumber
\end{eqnarray}
Now we know that when we work modulo complexified gauge transformations the function \(\xi_{ij}\) carries no gauge independent information.  Thus the evaluation of the coupling must yield a result which depends only on the holomorphic wavefunction \(h_{ij}\).  This is indeed the case.  In Appendix \ref{resformgen} we prove a general theorem which implies that the Yukawa coupling is computed by a multidimensional residue \cite{FGUTSNC}
\begin{equation}
W_{Y}=\mathrm{Res}_{(0,0)}\left[ \frac{h_{12}h_{23}h_{31}}{(\lambda_{1}-\lambda_{2})(\lambda_{2}-\lambda_{3})} \right].  \label{finalres}
\end{equation}
Where in the above we have introduced a standard notation for a multidimensional residue integral
\begin{equation}
\mathrm{Res}_{(0,0)}\left[\frac{\alpha}{\beta \gamma}\right] \equiv \frac{1}{(2\pi i)^{2}}\int_{|\beta|=\epsilon_{1}, |\gamma|=\epsilon_{2}}\left(\frac{\alpha}{\beta \gamma}\right)dx \wedge dy.
\end{equation}
This is the final result for the holomorphic calculation of the Yukawa.  One can see that it yields the same result as the local calculation in unitary gauge, by considering the case where \(\lambda_{1}-\lambda_{2}=x\) and \(\lambda_{2}-\lambda_{3}=y\).  The full structure of this formula has a number of features which deserve comment:
\begin{itemize}
\item The final answer depends only on the local holomorphic data in the problem:  the roots \(\lambda_{i}-\lambda_{j}\) and the holomorphic wavefunctions \(h_{ij}\) which specify the profile of the matter fields along matter curves.  In particular no K\"{a}hler data whatsoever is needed to formulate the result.
\item Since the denominator only involves two roots, the coupling appears to privilege one of the matter fields as compared to the other two.  This is an illusion.  The residue depends only on the ideal generated by the holomorphic functions in the denominator, and every pair of two roots generates the same ideal.
\item A standard property of residue integrals is that they vanish whenever the numerator is in the ideal generated by the factors in the denominator.  Thus for example the coupling \((\ref{finalres})\) vanishes if \(h_{12}\) is divisible by the root \(\lambda_{1}-\lambda_{2}\).  In our case this has a natural interpretation in terms of the invariance of the superpotential under complexified gauge transformations.   Indeed recall from section \ref{MatH} that the space of gauge inequivalent perturbations for a root \(R_{ij}\) is the quotient space \(\mathcal{O}/\mathcal{I}_{ij}\).  In particular all modes in the ideal \(\mathcal{I}_{ij}\) are gauge equivalent to zero and consistency demands that couplings involving these modes all vanish.
\end{itemize}
Thus a formal summary of our results is that the Yukawa coupling as computed by the residue \eqref{finalres} yields a trilinear pairing on the space of matter fields
\begin{equation}
\mathcal{O}/\mathcal{I}_{12}\otimes \mathcal{O}/\mathcal{I}_{23} \otimes \mathcal{O}/\mathcal{I}_{31}\rightarrow \mathbb{C}.
\end{equation}
Mathematically this pairing is the local form of the Yoneda pairing of Ext groups studied in \cite{DWIII}.  The advantage of expressing it in this way is that while Ext groups can often be unwieldy, the local residue integral is comparatively easy to compute explicitly.
\subsubsection{Rank Theory and Deformations of Superpotentials}
\label{RankT}
An important feature of the Yukawa is that it varies continuously with parameters specifying the holomorphic background field \(\Phi\).  In particular, any integer valued invariants that we can form out of this pairing can be viewed as a topological property of the background \(\Phi\) which is constant under small perturbations.  In our case there are three such quantities which are the ranks of this pairing.  We view the Yukawa as defining three maps:
\begin{equation}
\mathcal{O}/\mathcal{I}_{ij}\otimes \mathcal{O}/\mathcal{I}_{jk}  \rightarrow \left(\mathcal{O}/\mathcal{I}_{ki}\right)^{\ast},
\end{equation}
and we ask for their ranks.  In the case at hand all of these ranks are the same and the local duality theory of residues \cite{Griffiths} implies they are equal to the topological intersection number of any pair of the matter curves.  This is a natural result: the structure of the superpotential, localized at a triple intersection of seven-branes,  reflects a topological invariant of the intersection.  Notice also that this explains the result of the unitary gauge computation \((\ref{unitaryyuk})\).  In that case we found the the only non-zero couplings involved wavefunctions which were non-vanishing at the Yukawa point enforcing the fact that the pairing is rank one and reflecting the brane geometry of a transverse triple intersection.

\begin{figure}
\begin{center}
\framebox{
\includegraphics[totalheight=0.1\textheight]{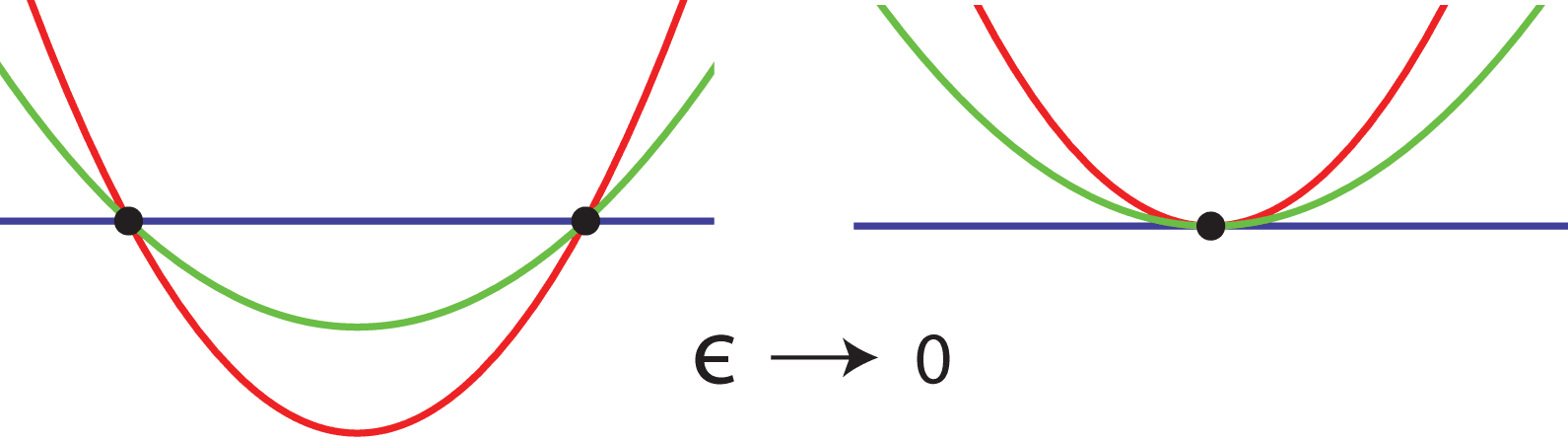}
}
\caption{The Yukawa behaves continuously with respect to deformations of the background parameters.  On the left we have three matter curves, illustrated by the colored lines, meeting transversally at two points leading to a rank two superpotential.  On the right a parameter \(\epsilon \rightarrow 0\) and the Yukawa points collide leading to a rank two contribution from a single intersection.}
\label{fig:deformed}
\end{center}
\end{figure}

It is illuminating to see the invariance of the rank of the superpotential worked out in a specific example.  Consider the one-parameter family of background fields \(\Phi_{\epsilon}\) given by
\begin{equation}
\Phi_{\epsilon}=\left[\left(y-\frac{x(x-\epsilon)}{3}\right)H_{1}-\frac{x(x-\epsilon)}{3}H_{2}+\left(\frac{2x(x-\epsilon)}{3}-y\right)H_{3}\right]dx\wedge dy.
\end{equation}
As in our previous examples, matter is localized on the curves defined by the roots of \(\Phi_{\epsilon}\).  Thus the matter curves are
\begin{equation}
y=0, \hspace{.5in} y=x(x-\epsilon),\hspace{.5in} y=\frac{1}{2}x(x-\epsilon).
\end{equation}
As illustrated in Figure \(\ref{fig:deformed}\), for \(\epsilon \neq 0\) the matter curves have a pair of generic intersections, while for \(\epsilon=0\), these intersections collide yielding a non-transverse intersection with intersection number two.  We let \(W_{Y}(\epsilon)\) denote the superpotential as a function of the the deformation parameter \(\epsilon\).

In this background, the possible holomorphic wavefunctions are given by
\begin{eqnarray}
h_{12} \in \frac{\mathcal{O}}{\langle y \rangle}  \hspace{.5in} h_{23} \in \frac{\mathcal{O}}{\langle y-x(x-\epsilon)\rangle} \hspace{.5in} h_{31} \in \frac{\mathcal{O}}{\langle 2y-x(x-\epsilon)\rangle}.
\end{eqnarray}
Thus a convenient basis of gauge inequivalent states is given by monomials in the \(x\) coordinate
\begin{equation}
h_{ij} \in \mathrm{Span}\{1, x, x^{2}, \cdots\}.
\end{equation}
 According to our residue formula of section \ref{YukH} the Yukawa coupling is rank two and computed by the residue integral
\begin{equation}
W_{Y}(0)=\mathrm{Res}_{(0,0)}\left[\frac{h_{12}h_{23}h_{31}}{(y)(x^{2})}\right].
\end{equation}
For purposes of illustration we fix the wavefunction \(h_{12}\) to be the constant mode \(h_{12}=1\).  We can then view the superpotential as a matrix whose rows and columns label increasing powers of \(x\) for the wavefunctions \(h_{23}\) and \(h_{31}\)
\begin{equation}
W_{Y}(0)(h_{12}=1,h_{23}=x^{k}, h_{31}=x^{j}) = \left( \begin{array}{cccc}
0 & 1 & 0  & \cdots \\
1 & 0 &   \\
0 &  & \ddots  \\
\vdots \\
\end{array} \right) \label{wmatrix}
\end{equation}
As expected this matrix is rank two.

Now consider the situation when \(\epsilon\neq0\).  The non-transverse intersection is then deformed into two transverse intersections, each of which gives a contribution to the superpotential.  Applying our basic result, we then have
\begin{equation}
W_{Y}(\epsilon)=\mathrm{Res}_{(0,0)}\left[\frac{h_{12}h_{23}h_{31}}{(y)(x(x-\epsilon))}\right]+\mathrm{Res}_{(\epsilon,0)}\left[\frac{h_{12}h_{23}h_{31}}{(y)(x(x-\epsilon))}\right].
\end{equation}
And hence in the notation of \((\ref{wmatrix})\):
\begin{equation}
W_{Y}(\epsilon)(h_{12}=1,h_{23}=x^{k}, h_{31}=x^{j}) = \underbrace{-\left( \begin{array}{cccc}
\frac{1}{\epsilon} & 0 & 0  & \cdots \\
0 & 0 &   \\
0 &  & \ddots  \\
\vdots \\
\end{array} \right) }_{\mathrm{contribution  \ from  \ (0,0)}}+  \underbrace{\left( \begin{array}{cccc}
\frac{1}{\epsilon} & 1 & \epsilon  & \cdots \\
1 & \epsilon &   \\
\epsilon &  & \ddots  \\
\vdots \\
\end{array} \right) }_{\mathrm{contribution \ from \ } (\epsilon,0)}
\end{equation}
Each point gives a Yukawa matrix which is rank one. As can be checked, the sum of the contributions from the two points yields an overall rank two coupling.  Clearly when \(\epsilon \rightarrow 0\) the above converges to the result \((\ref{wmatrix})\) demonstrating the continuity of the coupling in this particular case.  The basic point illustrated by this example is that in order to keep the rank constant under deformations, the two matrices have correlated entries.  Naively one might have expected that in the \(\epsilon \rightarrow 0\) limit the two contributions to the superpotential would align and give a single larger coupling for the \((1,1)\) matrix element of \(W\) and zeros elsewhere.  In fact, however, the \((1,1)\) entry vanishes for arbitrary \(\epsilon\) and the superpotential is always rank two.  The rank of the superpotential is thus a kind of topological invariant of the theory and this makes it possible to determine it in terms of the topological properties of the brane worldvolume \(S\) \cite{Cordova:2009fg, Hayashi:2009bt}.
\section{T-Branes: Basic Examples}
\label{Mono}
In this section we will expand the class of background fields under consideration. In section \(\ref{ReviewB}\) we studied Higgs fields which admit a simple interpretation in terms of intersecting branes.  Such configurations are abelian in nature, being governed by a diagonal matrix.  Here we begin our analysis of \emph{non-diagonalizable} Higgs fields which probe the full non-abelian structure of the theory.  The basic extra ingredient in these solutions is that there are loci in the seven-brane worldvolume where the Higgs field becomes non-zero and nilpotent.  We refer to such brane bound states as T-branes to indicate this triangular structure in the Higgs field.  This section explores such backgrounds by way of the simplest possible examples in \(SU(n)\) gauge theory. Guided by the considerations of the previous section, from now on we will exclusively study the BPS equations of motion locally on \(\mathbb{C}^{2}\) and omit factors of \(dx\wedge dy\) when writing \((2,0)\) forms.
\subsection{Beyond Eigenvalues}
\label{seigenval}
One way to phrase the simplifying assumptions of our previous examples is in terms of the eigenvalues of the Higgs field.  These are encoded in a fundamental gauge invariant observable of the background given by the spectral polynomial of \(\Phi\)
\begin{equation}
P_{\Phi}(z) =  \det \left(z-\Phi \right).
\end{equation}
Since the spectral polynomial is manifestly invariant under complexified gauge transformations it is a holomorphic invariant of the background.  To evaluate \(P_{\Phi}(z)\) we can freely go to a holomorphic gauge where the BPS equation \(\bar{\partial}_{A}\Phi=0\) tells us that the spectral polynomial is a holomorphic function of \((x,y,z)\).\footnote{In the case of a compact brane \(S\), the topological twist implies that \(P_{\Phi}\) is a section of \(K_{S}^{\otimes n}\).}  In the simple abelian examples of section \ref{ReviewB} we have
\begin{equation}
P_{\Phi}(z)=\prod_{i} \left(z-\lambda_{i}\right), \label{factorized}
\end{equation}
with \(\lambda_{i}(x,y)\) the holomorphic eigenvalues of \(\Phi\).  The spectral variable \(z\) then has a geometrical interpretation as a local normal coordinate to the seven-brane stack, and the equation
\begin{equation}
P_{\Phi}(z)=0
\end{equation}
gives the positions of seven-branes in the three-dimensional \((x,y,z)\) space. This is the basic correspondence which has been at the heart of our analysis in section \ref{ReviewB}: via the spectral polynomial, configurations of intersecting branes can be viewed as backgrounds in our original gauge theory.

Once we consider non-diagonalizable Higgs fields, the spectral equation is still an important invariant of a background, but in general it ceases to have such a simple geometric interpretation in terms of intersecting branes.  What is more, while in the diagonal case the whole system was determined by the eigenvalues, for a non-diagonal background the physics knows about much more than just \(P_{\Phi}(z)\).
\subsubsection{A Nilpotent Higgs Field}
\label{NilpH}
An extreme case illustrating the above remarks is a Higgs field which is nilpotent and hence has vanishing eigenvalues and a trivial spectral equation.\footnote{See \cite{ Donagi:2003hh} for additional discussion of nilpotent Higgs fields.} We can construct a simple example with this property in \(SU(2)\) gauge theory following an example due to Hitchin \cite{HitchinSelf}.  Begin in a holomorphic gauge with the desired background Higgs field
 \begin{equation}
  \Phi  =\left(\begin{array}{c c} 0 & 1 \\
 0 & 0
 \end{array}
 \right). \label{nilp}
 \end{equation}

In the holomorphic gauge \(A^{0,1}\) vanishes and the \((0,1)\) part of the covariant derivative is simply \(\bar{\partial}\).  Since the Higgs field \((\ref{nilp})\) is manifestly holomorphic we have solved the F-term equations
\begin{eqnarray}
F^{0,2}_{A} & = & 0, \\
\bar{\partial}_{A} \Phi & = & 0. \nonumber
\end{eqnarray}
To complete the solution it remains to solve the D-term equation
\begin{equation}
\omega \wedge F_{A}+\frac{i}{2}[\Phi^{\dagger}, \Phi]=0. \label{dddterms}
\end{equation}
To this end we must pass from a holomorphic gauge to a unitary gauge.  The procedure we employ is well known: we consider an arbitrary complexified gauge transformation of the Higgs field \((\ref{nilp})\), and we treat the parameters of this gauge transformation as variables to be determined by solving the D-term.  General arguments then imply that somewhere in the complexified gauge orbit of our field configuration there exists a complete solution to the full system of equations of motion which is unique up to unitary gauge transformations. In the case at hand we take our complexified gauge transformation to be of the form
\begin{equation}
g=\left(\begin{array}{c c} e^{f/2} & 0 \\
 0 & e^{-f/2}
 \end{array}
 \right).
\end{equation}
Using the unitary freedom, we can take the function \(f\) above to be real.  On performing the complexified gauge transformation we find that the resulting unitary frame Higgs field and connection are
\begin{equation}
\Phi=\left(\begin{array}{c c} 0 & e^{f} \\
 0 & 0
 \end{array}
 \right), \hspace{1in} A^{0,1}=g \bar{\partial}g^{-1}=\frac{1}{2}\left(\begin{array}{c c} -\bar{\partial}f & 0 \\
 0 & \bar{\partial}f
 \end{array}
 \right). \label{nilpunitary}
\end{equation}
If we again we equip the surface \(\mathbb{C}^{2}\) with a flat K\"{a}hler form
\begin{equation}
\omega=\frac{i}{2}\left(dx \wedge d\bar{x}+ dy \wedge d\bar{y}\right),
\end{equation}
then by evaluating \((\ref{dddterms})\) we see that the D-term is satisfied provided that
\begin{equation}
\Delta f=e^{2f}. \label{liouville}
\end{equation}
This is a generalization to the case of two complex variables of the Liouville equation for the conformal factor of a hermitian metric with uniform negative curvature.  However, since \(f\) is not the conformal factor for the metric on \(\mathbb{C}^{2}\), its interpretation is somewhat different.  Rather, the data of the gauge transformation \(g\) specifies a metric on the \(SU(2)\) gauge bundle \(V\).  As is clear from the diagonal form of the the connection, the solution above has \(V\) split as a direct sum of two line bundles, \(V\cong L\oplus L^{-1}\) and in a holomorphic frame the norm of the basis vector for \(L\), \((1 \  0)^{T}\) is \(e^{f/2}\).  Our technique of construction for the unitary solution is thus quite parallel to the theory of harmonic metrics familiar from the study of Hitchin systems \cite{SimpsonGeneralize}. 

This simple example has an obvious extension to the case of \(SU(n)\) gauge theory.  Possible nilpotent structures of a constant Higgs field are given by a choice of Jordan block decomposition of \(\Phi\).  For the case of a maximal Jordan block, the transformation to unitary gauge is specified by a positive diagonal matrix.  The D-term equation \((\ref{liouville})\) is then replaced by the \(SU(n)\) Toda equation in two complex variables:
\begin{equation}
\Delta f_{a}=C_{ab}e^{f_{b}}, \hspace{.5in} \sum_{a}f_{a}=0,
\end{equation}
with \(C_{ab}\) the Cartan matrix of \(SU(n)\).  For more general Jordan block structures
one still finds Toda-like equations. See \cite{TTSTAR} for more details.

Returning to our simple \(SU(2)\) example, we can see that once we have arrived in unitary gauge the resulting curvature of the gauge bundle is
\begin{equation}
F_{A}=(\bar{\partial}\partial f) (H_{1}-H_{2}). \label{nilpflux}
\end{equation}
Thus the nilpotent Higgs field, in contrast to the diagonal backgrounds studied in section \((\ref{ReviewB})\), depends for its very existence on a non-vanishing gauge field curvature. In the case of diagonal backgrounds the commutator \([\Phi^{\dagger}, \Phi]\) vanishes and the D-term equation (\ref{dddterms}) simplifies to
\begin{equation}
\omega \wedge F_{A}=0
\end{equation}
which along with the conditions $F_{A}^{2,0} = F_{A}^{0,2} = 0$ are the defining equations of an instanton.
The gauge field degrees of freedom are thus largely decoupled from those of the Higgs field and the system reduces to intersecting seven-branes, characterized by the eigenvalues of \(\Phi\), and dissolved three-branes, characterized by gauge instantons. By contrast, for non-diagonalable Higgs fields \([\Phi^{\dagger}, \Phi]\) is not zero and one cannot disentangle the Higgs field from the gauge flux.  In this sense, solutions with non-diagonalizable \(\Phi\) describe non-abelian \emph{bound states} of branes.  For the basic example of the \(SU(2)\) nilpotent Higgs field discussed above, we know that solutions to the Liouville equation are spread over their entire domain of definition, and hence the required flux (\ref{nilpflux}) permeates the entire brane worldvolume.

Once we have the solution in unitary frame, it is straightforward to compute the physical spectrum by solving the linearized BPS equations of motion as in section \(\ref{ReviewMat}\).  If we are just interested in determining the number of fields and their localization properties, we can simplify the computation by proceeding instead in a holomorphic gauge.  As in section \(\ref{MatH}\) the space of gauge inequivalent states is again given by holomorphic matrices modulo those matrices which are commutators with \(\Phi\).  For the nilpotent background in question we can reach a unique gauge where any such state is written as
\begin{equation}
\left(\begin{array}{cc}
0 & 0 \\
h(x,y) & 0
\end{array} \right) \label{nilpfluct}
\end{equation}
with \(h(x,y)\) is an arbitrary holomorphic function.  Since \(h\) depends on both complex coordinates it describes a bulk quantum field which propagates across the whole brane.  Thus this background does not support any localized matter, a fact which should not be surprising given its spatially uniform character in holomorphic gauge.

One feature of the physics for which the unitary gauge solution is essential is in determining the unbroken gauge symmetry.  The relevant terms in the 4\(D\) effective Lagrangian are schematically
\begin{equation}
\mathcal{L}_{4D}=\int_{\mathbb{C}^{2}} |F_{\mu j}|^{2}+|D_{\mu} \Phi|^{2},
\end{equation}
where in the above \(\mu\) denotes a Minkowski space index, and \(j\) an internal index on \(\mathbb{C}^{2}\).  In the presence of a background field configuration \((A,\Phi)\) these terms will give mass to various gauge bosons.  Since both terms are positive, it is clear that to remain unbroken a generator in the Lie algebra must commute with the Higgs field \(\Phi\) everywhere in the internal \(\mathbb{C}^{2}\).  Applying the Jacobi identity to the D-term equation \((\ref{dddterms})\) then shows that any generator which commutes with \(\Phi\) commutes with the gauge field as well.\footnote{We are assuming that the only gauge field we have turned on is the one necessary to satisfy the D-term constraint $\omega \wedge F_{A}^{1,1} + \frac{i}{2} [\Phi^{\dag} , \Phi] = 0$. In additional to this can in principle also consider switching on additional primitive fluxes satisfying $\omega \wedge F_{A}^{1,1\rm{\, extra}} = 0$ associated with a background instanton, or contributions from a flat connection, as would be associated with Wilson lines. As discussed for example in \cite{DWI,BHVI,BHVII,DWII} such contributions will in general induce further symmetry breaking effects.}  Thus the unbroken gauge symmetry is specified by the commutant of \(\Phi\) in a unitary gauge.  In the case at hand it is clear that the unitary frame Higgs field (\ref{nilpunitary}) does not commute with any of the \(SU(2)\) generators and hence the nilpotent Higgs field completely breaks the gauge symmetry.  In four dimensions, the fluctuations \((\ref{nilpfluct})\) are the wavefunctions of chiral multiplets uncharged under any gauge symmetry.  It is interesting to notice that this Higgs mechanism is invisible to both the spectral equation \emph{and} the curvature.  \(F_{A}\) is valued in the \(U(1)\) Cartan subalgebra of \(SU(2)\) and does not itself completely break the symmetry.  This serves to reinforce the basic point of this example: only with a complete knowledge of the BPS solution, that is both the gauge field \(A\) and the full matrix valued Higgs field \(\Phi\) can one completely investigate the physics of any given background.

\subsection{Monodromy Basics}
\label{M101}
While somewhat novel, the nilpotent Higgs field of the previous section is physically boring.  The background supports no unbroken gauge group, no localized matter fields, and there is no hope of generalizing the results of section \ref{ReviewB}.  These deficiencies can be remedied.  The nilpotent Higgs has no localized matter in holomorphic gauge, $\Phi$ is completely uniform over the brane.  To find examples with trapped fluctuations that go beyond the abelian cases studied in section \(\ref{ReviewB}\) we can look for special places in the brane worldvolume by studying the spectral equation
\begin{equation}\label{charpoly}
P_{\Phi}(z)=z^{n}+\sigma_{2}z^{n-2}-\sigma_{3}z^{n-3}+\cdots + (-1)^{n-1}\sigma_{n-1}z+(-1)^{n}\sigma_{n}.
\end{equation}
In the above equation the integer \(n\) indicates that we are now considering an \(SU(n)\) gauge theory.  Each coefficient \(\sigma_{i}\) is a gauge invariant function of the background \(\Phi\) and is the \(i\)-th elementary symmetric polynomial in the eigenvalues of \(\Phi\).\footnote{In our case \(\sigma_{1}\) vanishes since the Higgs field is traceless.}  These symmetric functions are holormorphic functions on the seven-brane and the loci where they vanish are distinguished complex submanifolds where we might expect some physical quantity to reside.

In the case of an abelian background the spectral equation is factorized as in \((\ref{factorized})\) and the data of the symmetric functions \(\sigma_{i}\) can be replaced by the more elementary data of the eigenvalues of the Higgs field.  In general, however, the spectral equation is not factorized and the symmetric functions themselves are the more fundamental gauge invariant data.  This is the basic idea behind the notion of \emph{monodromy} which has already played an important role in many applications of seven-brane gauge theories.  In this context monodromy refers to the behavior of the roots of the spectral polynomial \(P_{\Phi}(z)\) thought of as a polynomial in the spectral variable \(z\) as one traverses the brane worldvolume.  For example, consider the following spectral equation for an \(SU(2)\) Higgs field:
\begin{equation}
P_{\Phi}(z)=z^{2}-x^{m}.
\end{equation}
For \(m\) even the above is factorized into a product of two linear polynomials in \(z\) and thus could be associated to a diagonal \(\Phi\) with eigenvalues \(\pm x^{m/2}\).  By contrast, when \(m\) is odd, the eigenvalues of \(\Phi\) are branched.  As one circles the \(y\) axis they exchange and one says that there is a \(\mathbb{Z}_{2}\) monodromy.  Clearly there is no diagonal holomorphic Higgs field which gives rise to such spectral behavior.  With \(m=1\) the best we can do is to express \(\Phi\) in a holomorphic gauge as
\begin{equation}
\Phi= \left(
\begin{array}{cc}
0 & 1 \\
x & 0
\end{array}
\right). \label{monophiex}
\end{equation}
This Higgs field is something of an intermediate case between the diagonal background of section \ref{ReviewMat} and the nilpotent solution of section \ref{NilpH}.  For \(x\neq0,\) \(\Phi\) has distinct eigenvalues and can be brought to diagonal form by a change of basis.  For \(x=0\), however,
the Higgs field becomes nilpotent.

In general for an \(n \times n\) Higgs field \(\Phi\) the spectral function \(P_{\Phi}(z)\) is a polynomial of degree \(n\) in \(z\), and the monodromy group is defined to be the subgroup of the permutation group on \(n\) letters which acts on the roots as we navigate the brane worldvolume.  In terms of pure mathematics, the monodromy group is the Galois group of the spectral polynomial viewed as a polynomial in \(z\) with coefficients in functions on the worldvolume \(\mathbb{C}^{2}\).  As our \(2\times 2\) example illustrates, the monodromy group in general acts as an obstruction to diagonalizing the Higgs field, with larger monodromy an indication of more non-abelian behavior.
\subsubsection{The $\mathbb{Z}_{2}$ Background}
The most basic case of monodromy to study, and one which suffices for almost all of our applications, is the Higgs field of \((\ref{monophiex})\).  We would like to completely understand this background.  We start, as in our study of the nilpotent background by passing from holomorphic to unitary frame.  This time we parameterize our gauge transformation as
\begin{equation}
g= \left(
\begin{array}{cc}
r^{1/4}e^{f/2} & 0 \\
0 & r^{-1/4}e^{-f/2}
\end{array}
\right).
\end{equation}
Where in the above we have introduced polar coordinates \(x=re^{i\theta}\).  Since the gauge transformation \(g\) must be everywhere non-singular we seek a real solution \(f\) which has a logarithmic singularity at \(r=0\).  We take as an ansatz that \(f\) is independent of \(y\) and \(\theta\).  In that case the same steps that we applied above to the nilpotent Higgs lead to the following D-term equation for \(f\)
\begin{equation}
\left(\frac{d^{2}}{ds^{2}}+\frac{1}{s}\frac{d}{ds}\right)f=\frac{1}{2}\sinh(2f), \label{painleve}
\end{equation}
with \(s=\frac{8}{3}r^{3/2}\).  Equation \((\ref{painleve})\) is a special instance of the Painlev\'{e} III differential equation.\footnote{The same equation \emph{and} boundary condition have previously been studied both in the context of $tt^{\ast}$ metrics for 2\(D\) Landau-Ginzburg models \cite{TTSTAR}, and 4\(D\) wall-crossing \cite{Gaiotto:2009hg}.} A detailed study of its solutions and their boundary behavior can be found in \cite{McCoy:1976cd}. Its solutions have asymptotic behavior specified by an arbitrary constant \(\kappa\).  Near \(s=0\) one has a logarithmic singularity
\begin{equation}
f(s) \rightarrow -\kappa \log(s)+O(1). \label{smallrpainleve}
\end{equation}
While for large \(s\) the solution decays exponentially as
\begin{equation}
f(s)\rightarrow \sqrt{\frac{2}{\pi s}}\sin \left(\frac{\pi \kappa}{2}\right)e^{-s}. \label{largerpainleve}
\end{equation}
In our case we seek a solution for which the resulting gauge transformation \(g\) is non-singular at \(x=0\) and thus we must take \(\kappa =\frac{1}{3}\).  The resulting flux in the unitary gauge is:
\begin{equation}
F_{A}=-2r\sinh(2f)(H_{1}-H_{2})dx \wedge d\bar{x} \longrightarrow \left\{ \begin{array}{c c}  -\frac{3^{8/3}}{\gamma^{2}}(H_{1}-H_{2})dx\wedge d\bar{x} & r <<1 \\  & \\-\sqrt{\frac{3}{\pi}}r^{1/4}e^{-\frac{8}{3}r^{3/2}}(H_{1}-H_{2})dx\wedge d\bar{x} & r >>1 \end{array} \right. .
\end{equation}
Where in the above \(\gamma\) denotes the Euler-Mascheroni constant.

The form of the flux allows us to make precise our previous remark that the \(\mathbb{Z}_{2}\) monodromy background we are studying is an intermediate case between the nilpotent Higgs of section \(\ref{NilpH}\) and the diagonalizable intersecting brane system of section \ref{ReviewMat}.  Like the nilpotent case, as \(x \rightarrow 0\), the flux approaches a non-zero constant matrix valued in the Cartan \(U(1)\subset SU(2)\).  Meanwhile, for large \(r\) the flux decays rapidly; the branch loci of the spectral equation are observable as localized tubes of gauge flux.  However we also see how misleading it would be to try to think of this background purely in terms of a pair intersecting branes: the flux is \emph{always} non-zero.  Only at strictly \(r=\infty\) does this background really approach the simple solutions of section \ref{ReviewB}.

To complete our analysis of the \(\mathbb{Z}_{2}\) monodromy background \((\ref{monophiex})\) it remains to study fluctuations.\footnote{In \cite{Hayashi:2009ge} an analysis of matter fields around a similar background was considered. There, however, it was assumed that one could work in terms of a non-analytic and diagonal Higgs field with explicit branch cuts. There are various subtleties connected with this singular ``branched gauge'' choice. It is therefore important to revisit this example. Indeed, in contrast to what was found in \cite{Hayashi:2009ge}, we find that there is no massless $6D$ localized mode.}  From the form of the unitary solution we can again see that this background completely breaks the \(SU(2)\) gauge symmetry and hence the fluctuations in question are uncharged under any gauge group.  The basic question is whether or not there exists any excitation of this solution which is trapped at \(x=0\).  As usual we can immediately answer this question by looking in the holomorphic gauge.  The space of modes is given by holomorphic matrices modulo commutators with the background \(\Phi\).  If we write a holomorphic matrix as:
\begin{equation}
\left( \begin{array}{cc}
h_{0} & h_{+} \\
h_{-} & -h_{0}
\end{array} \right)
\end{equation}
then under a holomorphic gauge transformation, i.e. a shift by a commutator with the holomorphic gauge \(\Phi\), we have:
\begin{eqnarray}
h_{0} & \rightarrow & h_{0}+ \alpha  \\
h_{+} & \rightarrow & h_{+} + \beta \nonumber \\
h_{-} & \rightarrow & h_{-}-\beta x \nonumber
\end{eqnarray}
with \(\alpha, \beta\) arbitrary holomorphic functions.  Via such a gauge transformation we can set the diagonal term \(h_{0}\) to zero.  Meanwhile for the off-diagonal perturbations, gauge transformations shift both components simultaneously.  Again using \(\mathcal{O}\) to denote the local ring of holomorphic functions, we can write the space of equivalence classes as a doublet
\begin{equation}
[h_{+}, h_{-}] \in \frac{\mathcal{O} \oplus \mathcal{O}}{\langle(1,-x)\rangle}. \label{mixedhpm}
\end{equation}
To investigate the issue of localized matter we can use our gauge freedom to set \(h_{+}\) to zero.  The holomorphic fluctuation is then specified by the arbitrary function \(h_{-}(x,y)\) and since this depends on two complex coordinates we reach the conclusion that the only excitations are bulk fields spread over the entire \(\mathbb{C}^{2}\) worldvolume, and there is no massless fluctuation trapped at \(x=0\).

To further bolster our conclusion that there is no massless 6$D$ mode, we now recover the same result in the physical unitary frame.  We start in holomorphic gauge with a perturbation matrix of the form
\begin{equation}
\left(\begin{array}{cc}
0 & 0 \\
h & 0
\end{array}\right) \label{monoperturb}
\end{equation}
and we ask for the unitary gauge wavefunction of this mode.  As a representative example let us study the case where the holomorphic perturbation \(h\) is a real constant, independent of both complex coordinates \((x,y)\).  In that case we simply note that the addition of \((\ref{monoperturb})\) to the original Higgs field background \((\ref{monophiex})\) can be interpreted as a change of coordinates \(x\rightarrow x +h\).  Thus our construction of the background itself suffices to determine the resulting perturbation, and we need only perform a Taylor expansion.  If we denote by \(\varphi\) the resulting Higgs field perturbation in the physical, unitary gauge then a short calculation reveals that
\begin{equation}
\varphi=h \cos(\theta)\left(\frac{1}{2r}+\frac{df}{dr}\right)\left( \begin{array}{cc}
0 & \sqrt{r}e^{f} \\
-\sqrt{r}e^{-f+i\theta} & 0
\end{array}
\right)+
h \left(\begin{array}{cc}
0 & 0 \\
\frac{e^{-f}}{\sqrt{r}} & 0
\end{array} \right).
\label{bulkfull}
\end{equation}
One can see that these are indeed bulk modes by examining their asymptotic behavior for large \(r\).  If we neglect terms that vanish exponentially fast then we find that
\begin{equation}
\varphi \rightarrow \frac{h}{2\sqrt{r}} \left( \begin{array}{cc}
0 & \cos(\theta) \\
2-e^{i\theta}\cos(\theta) & 0
\end{array}
\right). \label{bulkfar}
\end{equation}
In contrast to the unitary gauge modes on matter curves of section \(\ref{MatU}\) these fields decay as a power law for large \(r\).  This is the basic difference between bulk fields and localized fields.  A mode localized on a curve has a finite norm per unit-length along the curve; if the matter curve is compact then the field is normalizable independent of the compactness of the transverse direction.  Meanwhile, the wavefunction of a bulk field is spread over the whole worldvolume, its normalizability is only possible if \(S\) is compact.  For the fluctuation (\ref{bulkfar}) we can see that for large \(r\):
\begin{equation}
||\varphi||^{2}=\mathrm{Tr}\left(\varphi \varphi^{\dagger}\right)\sim \frac{1}{r}
\end{equation}
and thus the mode is not normalizable in the \(x\)-plane: it is a bulk mode.  This does not mean that the physical wavefunction is completely uniform over the brane.  In fact one can see that the wavefunction of this mode does decay for large \(r\), and in this sense is a unique  element of the spectrum of bulk modes.  The norm of a unitary wavefunction of a general perturbation \(h\) which is not-necessarily constant in \(x\) behaves at large radius like \(\frac{|h|^{2}}{r}\) and hence aside from the case studied above of constant \(h\) all such modes \emph{grow} at infinity.  For these bulk fields one then expects the local picture developed here to be quite inaccurate as in general their wavefunctions are mostly supported outside the patch we have focused on.
By contrast, in the case where $h$ is constant the local picture still captures the relevant behavior.

\begin{figure}
\begin{center}
\framebox{
\includegraphics[totalheight=0.34\textwidth]{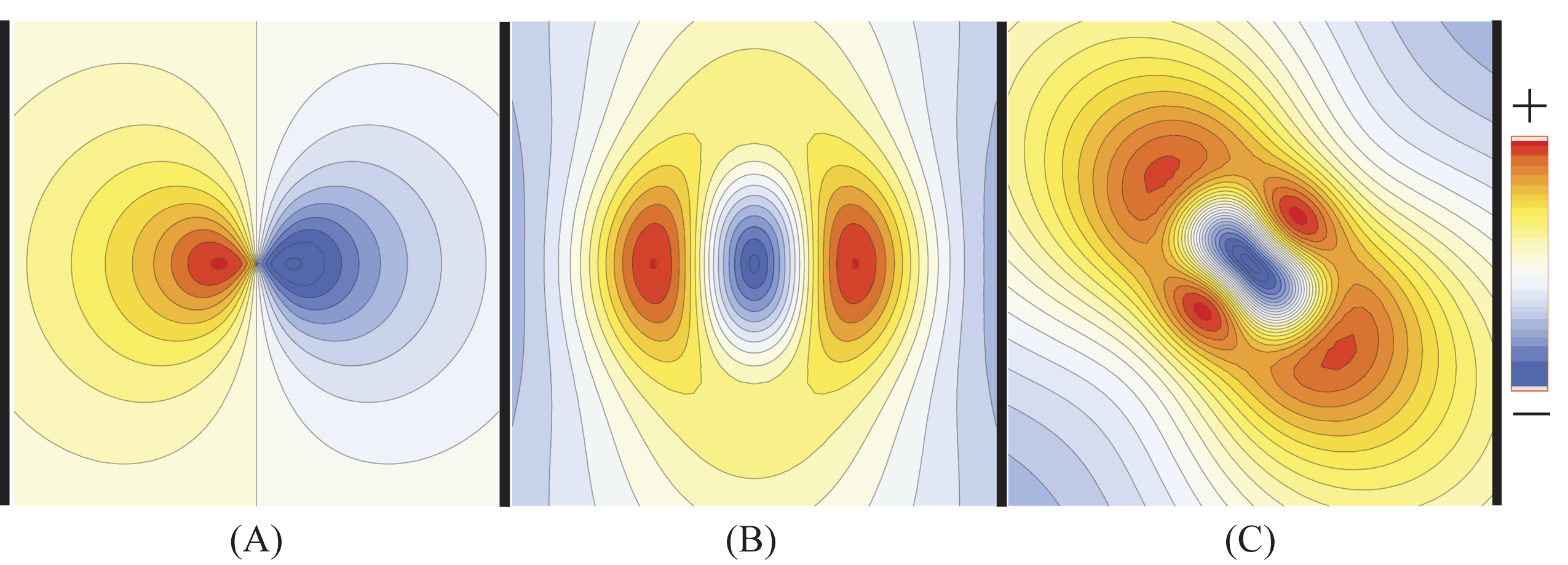}
}
\caption{Contour plots of the distorted bulk mode \(\varphi\) in the background \((\ref{monophiex})\) with nontrivial \(\mathbb{Z}_{2}\) monodromy.  The images show the profile of the zero modes in the complex \(x\) plane centered on the branch locus.  The complex coordinate \(y\) is suppressed.  Figure (A) shows the upper-right entry of \(\varphi\) while (B) and (C) illustrate the real and imaginary parts of the lower-left entry of \(\varphi\). }
\label{fig:bulk}
\end{center}
\end{figure}

Finally, let us note that although there are no localized \textit{massless} 6\(D\) fields, there will still be massive modes which are ``quasi-localized'' around $x = 0$. In this sense, there may still be an effective notion of a ``matter curve'' visible at high-energies.  In the Kaluza-Klein Majorana neutrino scenario considered in \cite{BHSV}, similar quasi-localized modes were identified with right-handed Majorana neutrinos.  It would be interesting to undertake a detailed study of Kaluza-Klein modes in this backgrounds and determine the precise correspondence with the ideas presented in \cite{BHSV}.

\subsubsection{Charged Matter}
\label{MatC}
If the stated goal of section \ref{M101} was to generalize the study of localized modes and couplings to the case of non-diagonal backgrounds then clearly section \ref{M101} has ended in failure.  We started with the simplest possible spectral equation exhibiting monodromy, \(P_{\Phi}(z)=z^{2}-x\), and we found that although the branch locus \(x=0\) supports a concentrated gauge flux tube, it does not trap any modes.  This example is typical of backgrounds which break all of the gauge symmetry. Referring back to the form of the spectral equation \eqref{charpoly}, at what appear to be be special places in the geometry, the symmetric functions \(\sigma_{i}\) defining $P_{\Phi}$ will vanish, and the Higgs field in general will approach a non-zero \emph{nilpotent} matrix with associated flux but no trapped perturbations.  The situation is more interesting for backgrounds that leave unbroken a subgroup of the gauge symmetry.  In this case we will indeed find localized matter, and in this section we analyze an example of this phenomenon.
\subsubsection{Holomorphic Gauge}
\label{localizedcharged}
The basic example to study is a breaking from \(SU(3)\rightarrow U(1)\) described by the holomorphic Higgs field
\begin{equation}
\Phi = \left( \begin{array}{cc|c}
0 & 1 & 0 \\
x & 0 & 0 \\
\hline
0 & 0 & 0
\end{array}
\right)=\left( \begin{array}{c|c}
\Psi & \begin{array}{c} 0 \\ 0 \end{array} \\
\hline
\begin{array}{c c} 0 & 0 \end{array} & 0
\end{array}
\right).
\end{equation}
Where \(\Psi\) denotes the \(2\times 2\) non-trivial block studied in the previous section.  This is a situation with both an unbroken gauge group and monodromy.  If we look at perturbations around this background, the new feature is the existence of bifundamental modes charged under both the unbroken \(U(1)\) and the broken \(SU(2)\) subgroup.  Embedded in the adjoint of \(SU(3)\) these modes are
\begin{equation}
 \left( \begin{array}{cc|c}
0 & 0 & h_{+} \\
0 & 0 & h_{-} \\
\hline
0 & 0 & 0
\end{array}
\right) \label{hdoublet}
\end{equation}
together with the associated transpose degrees of freedom which have opposite \(U(1)\) charge.

Following the usual method we find that the space of gauge inequivalent perturbations is
\begin{equation}
[h_{+}, h_{-}] \in \frac{\mathcal{O}\oplus \mathcal{O}}{\langle(1,0), (0,x)\rangle}.
\end{equation}
We can use this gauge freedom to eliminate \(h_{+}\) and if we do so then we find that the remaining perturbation \(h_{-}\) is valued in
\begin{equation}
h_{-}\in \frac{\mathcal{O}}{\langle x\rangle}.
\end{equation}
The bifundamental modes can thus be thought of as residing on the matter curve \(x=0\).  It is exactly in this situation that we expect to find localized perturbations concentrated sharply around this curve.  As usual, it is instructive to study the solutions to the equations of motion both from the holomorphic and unitary perspective.  In the holomorphic setting our goal is to construct the non-abelian version of a localized gauge where the perturbation vanishes outside of a tiny tube around the matter curve.  In the unitary frame our goal is to explicitly solve the equations of motion and determine the behavior of fluctuations both near the matter curve and at asymptotically long distances.

In the holomorphic gauge all information is encoded in \(\Phi\) and we seek to solve the linearized F-term equations:
\begin{eqnarray}
\bar{\partial}a & = & 0 \\
\bar{\partial}\varphi &= & [\Phi, a].  \label{commutator}
\end{eqnarray}
We take the matrix polarizations of \(a,\varphi\) to be identical to \((\ref{hdoublet})\) so that the commutator in the above is simply \(\Psi\) acting on the doublet \(a\) in the fundamental representation.  The formal solution to these equations proceeds exactly as in section \(\ref{MatH}\).  We introduce a doublet \(\xi\) which is a \(\bar{\partial}\) anti-derivative to \(a\) together with a holomorphic doublet \(h\) and write
\begin{eqnarray}
a & = & \bar{\partial}\xi \label{formalint1} \\
\varphi & = & \Psi \xi+h. \label{formalint}
\end{eqnarray}
The key observation is to consider the spectral equation for the \(2\times 2\) background \(\Psi\).  According to the Cayley-Hamilton theorem of linear algebra the matrix \(\Psi\) satisfies its own spectral equation.  In other words
\begin{equation}
\Psi^{2}=x\ident _{2},
\end{equation}
with $\ident _{2}$ the \(2\times 2\) identity matrix.  Thus although the matrix \(\Psi\) has a non-diagonal action on the doublets, the matrix \(\Psi^{2}\) acts trivially by multiplication by \(x\).  This allows us to invert the formal solution \((\ref{formalint})\) to write
\begin{equation}
\xi=\frac{\Psi(\varphi-h)}{x}.
\end{equation}

Now we can easily deduce the correct notion of localization; the important quantity is not the behavior of \(\varphi\) near the matter curve, but rather \(\Psi\varphi\).  In particular given any tube \(T_{\epsilon}\) centered on the matter curve \(x=0\) and any doublet solution \(\varphi\) we can define another smooth \((2,0)\) form \(\varphi'\) with the property
\begin{equation}
\Psi\varphi'=\left \{ \begin{array}{cc} \Psi\varphi  & \mathrm{inside}  \ T_{\epsilon} \\
0 & \mathrm{outside} \  T_{2\epsilon}
\end{array}
\right. .
\end{equation}
By construction then a complexified gauge transformation with parameter
\begin{equation}
\chi = \frac{\Psi(\varphi'-\varphi)}{x}
\end{equation}
shows that the two solutions \(\varphi\) and \(\varphi'\) are gauge equivalent.  In particular we see that given any solution, we can find a localized gauge where \(\Psi\varphi\) vanishes outside an arbitrarily small neighborhood of the matter curve.  If we explicitly write out the doublet mode then we find that at the matter curve
\begin{equation}
\Psi \varphi\rightarrow \Psi h \rightarrow \left(\begin{array}{c} h_{-}(0,y) \\ 0 \end{array} \right).
\end{equation}
This is exactly what we expect from the characterization of the excitations as the set of holomorphic matrices modulo commutators with the background.  There we saw that all the data of the doublet perturbation should be encoded in the single holomorphic function \(h_{-}\) along the matter curve \(x=0\), and here we recover this fact from the localized gauge construction.
\subsubsection{Unitary Gauge}
An alternative perspective on the doublet fluctuations is to solve the equations of motion directly in a unitary gauge.  Since we have already studied the nontrivial \(SU(2)\) background \(\Psi\) in the previous section, this is a calculation which we are prepared to undertake.   The complexified gauge transformation which takes the full \(SU(3)\) background Higgs field from holomorphic to unitary gauge is simply
\begin{equation}
g=\left(\begin{array}{cc|c}
r^{1/4}e^{f/2} & 0 & 0 \\
0 & r^{-1/4}e^{-f/2} & 0 \\
\hline
0 & 0 & 1
\end{array}
\right).
\end{equation}
Where \(f\) denotes the Painlev\'{e} transcendent introduced in \((\ref{painleve})\).  Since the F-term equations are invariant under complexified gauge transformations solving them in unitary frame is trivial; we simply perform a gauge transformation by \(g\) on the solution \((\ref{formalint1})-(\ref{formalint})\).  Further we know from our analysis in holomorphic gauge that we should be able to find solutions with \(h_{+}=0\) and then the equations of motion fix \(\xi_{-}=0\).  Hence in unitary gauge we can write:
\begin{equation}
a =\left( \begin{array}{c}
r^{1/4}e^{f/2}\bar{\partial}\xi \\
0
\end{array}
\right)
\hspace{1in}
\varphi =\left( \begin{array}{c}
0 \\
r^{-1/4}e^{-f/2}( x\xi+h)
\end{array}
\right).
\end{equation}
Now we need to solve the D-term
\begin{equation}
\omega \wedge \partial_{A} a+\frac{i}{2}\Psi^{\dagger}\varphi=0. \label{crazyd}
\end{equation}
This implies a single remaining equation
\begin{equation}
\left[\Delta+\left(\frac{1}{4x}+\partial_{x}f\right)\bar{\partial}_{\bar{x}}-|x|e^{-2f}\right]\xi=\frac{|x|}{x}e^{-2f}h. \label{singled}
\end{equation}
We can assume that the holomorphic function \(h\) depends only on the coordinate \(y\) along the matter curve and that \(\xi\) is independent of \(\bar{y}\).  To solve this equation we proceed as follows.  First observe that the differential operator in brackets above conserves angular momentum in the \(x\) plane.  Thus it is natural to make the substitution:
\begin{equation}
\xi=\frac{\rho(r)-h}{x} \label{rhodef}
\end{equation}
and to solve a radial equation for \(\rho\).  One can see that in terms of the original doublets \(a\) and \(\varphi\)
\begin{equation}
a=\frac{1}{2}\left(\begin{array}{c} r^{-3/4}e^{f/2}\frac{d\rho}{dr} \\ 0  \end{array}\right), \hspace{1in}\varphi=\left(\begin{array}{c} 0 \\ r^{-1/4} e^{-f/2} \rho(r)  \end{array}\right),
\end{equation}
and thus we expect to find a solution \(\rho(r)\) which is localized near the origin.  On substituting \((\ref{rhodef})\) into \((\ref{singled})\) we find
\begin{equation}
\left[\frac{d^{2}}{dr^{2}}+\left(\frac{df}{dr}-\frac{1}{2r}\right)\frac{d}{dr}-4re^{-2f}\right]\rho(r)=0. \label{rhoeqn}
\end{equation}
Since the function \(f\) is so complicated we do not expect that this equation can be solved analytically.  However we can construct an approximate solution by solving the equation for small and large \(r\) where \(f\) simplifies and then matching the solutions at an intermediate radius, and this technique suffices for seeing the existence of a localized mode.

For \(r\) very small we use the asymptotic behavior (\ref{smallrpainleve}) to simplify equation (\ref{rhoeqn}):
\begin{equation}
\left[\frac{d^{2}}{dr^{2}}-\frac{1}{r}\frac{d}{dr}-4r^{2}\right]\rho(r)=0.
\end{equation}
The solution to this is
\begin{equation}
\rho(r)=Be^{r^{2}}+Ce^{-r^{2}}. \label{smallrrhosol}
\end{equation}
Meanwhile according to (\ref{largerpainleve}) for large \(r\) the Painlev\'{e} transcendent \(f\) vanishes exponentially fast and hence (\ref{rhoeqn}) simplifies to
\begin{equation}
\left[\frac{d^{2}}{dr^{2}}-\frac{1}{2r}\frac{d}{dr}-4r\right]\rho(r)=0.
\end{equation}
The general solution to this is:
\begin{equation}
\rho(r)=De^{\frac{4}{3}r^{3/2}}+Ee^{-\frac{4}{3}r^{3/2}}. \label{largerrhosol}
\end{equation}
In the solutions (\ref{smallrrhosol}) and (\ref{largerrhosol}) \((B,C,D,E)\) denote holomorphic functions of \(y\) which must be determined by boundary conditions.  They are fixed as follows:
\begin{itemize}
\item Regularity of \(\xi\) at \(r=0\) fixes \(B+C=h\).
\item Normalizability for large \(r\) fixes \(D=0\).
\item Matching the zeroth and first derivatives of the near and far solution.
\end{itemize}
One can see that independent of where the matching takes place, the constant \(B\) must be very nearly zero.  Thus the solution is peaked at \(r=0\) with an amplitude fixed by \(h\), and for small \(r\) it decays as a Gaussian.  Meanwhile for large \(r\) the decay changes from the standard Gaussian to \(e^{-\frac{4}{3}r^{3/2}}\) characteristic of solutions to Painlev\'{e} III.  Glueing these behaviors together, we obtain a single normalizable charged \(6D\) field localized on the matter curve \(x=0\).
\section{Localized Modes and Their Couplings}
\label{formalism}

The previous section illustrates some of the most primitive examples of T-branes and their associated intricate physics of symmetry breaking, flux tubes, and localized charged matter.  In general an arbitrary Higgs background can exhibit a wide variety of exotic and novel phenomena and it is beyond the scope of this work to classify completely all such behavior.  Nevertheless we can introduce a setup which extends the holomorphic intersecting brane techniques of section \ref{ReviewB} and allows us to study any given example.  We generalize our theory to the case of an arbitrary gauge group with Lie algebra \(\mathfrak{g}\) and we write \(\mathrm{ad}_{\Phi}(M)\) to denote the adjoint action of \(\Phi\) on a matrix \(M\in \mathfrak{g}\).

As the examples of the previous section should make clear, keeping track of all of the D-term data of such configurations is quite cumbersome.  However we have seen that for intrinsically holomorphic questions, such as the spectrum of localized modes and their superpotential couplings, this D-term data is also unnecessary. For this reason from now on we work exclusively with the complexified gauge group and neglect the D-term equation of motion \eqref{dterm11}.  The basic equations in a holomorphic gauge for the background are identical to those given in subsection \ref{MatH}. Namely, the gauge field satisfies
\begin{equation}
\bar{\partial}_{A}=\bar{\partial}+A^{0,1}=\bar{\partial}
\end{equation}
and the matter field fluctuations satisfy:
\begin{equation}\label{modesrepeat}
a =\bar{\partial}\xi, \hspace{1in} \varphi = \mathrm{ad}_{\Phi}(\xi)+h,
\end{equation}
with \(\xi\) an arbitrary complex matrix in \(\mathfrak{g}\).

The analysis of the Yukawa proceeds much as for diagonalizable backgrounds. Starting from the bulk integral:
\begin{equation}
W_{Y} = \int_{S} \Tr(a \wedge a \wedge \varphi)
\end{equation}
we reduce this to a residue integral by noting that the $a$ modes satisfy $a = \bar{\partial}\xi$ and so can be integrated to boundary terms, and evaluated as a residue. To compute the coupling, it is therefore enough to track the boundary behavior of $\xi$. In practice, this involves formally solving for $\xi$ in equation \eqref{modesrepeat}:
\begin{equation}
\xi = \mathrm{ad}_{\Phi}^{-1}(\varphi - h)
\end{equation}
where here, ``$\mathrm{ad}_{\Phi}^{-1}$'' denotes a formal inversion of the adjoint map \(\mathrm{ad}_{\Phi}(\cdot)\).  Roughly speaking we expect that ``$\mathrm{ad}_{\Phi}^{-1}$'' is invertible away from some loci defining the matter curves \(f=0\).  In this case we may write 
\begin{equation}
\xi = \mathrm{ad}_{\Phi}^{-1}(\varphi - h) \equiv \frac{\eta}{f}
\end{equation}
Assuming $\varphi$ falls off rapidly at the boundary, the computation of the residue depends only on $h$. The residue integral then becomes:
\begin{equation}
W_{Y}=\mathrm{Res}_{(0,0)}\left[\frac{\mathrm{Tr}\left([\eta_{1}, \eta_{2}]\varphi_{3}\right)}{f_{1}f_{2}}+\frac{\mathrm{Tr}\left([\eta_{2}, \eta_{3}]\varphi_{1}\right)}{f_{2}f_{3}}+\frac{\mathrm{Tr}\left([\eta_{3}, \eta_{1}]\varphi_{2}\right)}{f_{3}f_{1}}\right].
\end{equation}

Though the above description provides a rough guide to matter fields and their couplings, making sense of the intuitive formal manipulations presented above requires being more careful with our notions of ``localized modes'' and wave function overlap integrals. In this section we develop in more precise terms the necessary ingredients to treat a general class of localized modes and their superpotential couplings.

The rest of this section is organized as follows. First we clarify the precise meaning of a localized mode, and describe explicitly how a localized solution to the 8\(D\) fluctuation equations \((\ref{eomda})-(\ref{oeom})\) gives rise to a 6\(D\) field living on a matter curve.  Following this identification, we write the general explicit form of the superpotential interactions of such modes, both for their \(6D\) kinetic superpotential and their \(4D\) localized Yukawa couplings. Additional technical details for the diligent reader are given in the Appendices.

\subsection{Localized 6$D$ Fields}
\label{local6d}

We will continue to denote the ring of holomorphic power series in two complex variables by \(\mathcal{O}\).  We work in a holomorphic gauge so that \(A^{0,1}=a=0\).  Then all information of the background is contained in the Higgs field
\begin{equation}
\Phi \in \mathfrak{g}\otimes \mathcal{O}.
\end{equation}
It is easy to characterize the space of physically distinct modes in the given background.  Working holomorphically, a perturbation \(\varphi\) is gauge equivalent to zero if there exists a holomorphic \(\chi\) such that
\begin{equation}
\varphi=\mathrm{ad}_{\Phi}(\chi).
\end{equation}
Thus in the presence of the background \(\Phi\) the space of physically distinct modes is given by the quotient space
\begin{equation}
\frac{\mathfrak{g}\otimes \mathcal{O}}{\mathrm{ad}_{\Phi}\left(\mathfrak{g}\otimes \mathcal{O}\right)}. \label{quo}
\end{equation}
This abstract expression for the space of modes already illustrates one of the primary advantages of working in a holomorphic gauge: we have translated the problem of solving the fluctuation equations \((\ref{eomda})-(\ref{oeom})\), a priori one which involves solving \emph{differential} equations, to a purely \emph{algebraic} problem of determining the quotient space (\ref{quo}).

Now, the space of modes is populated by both bulk fields, whose unitary wavefunctions permeate the entire brane, and localized modes, whose unitary wavefunctions are concentrated along matter curves.  Our primary interest in this section is in the latter and we seek to extract these from the general quotient space (\ref{quo}).  Let \(\Sigma\subset \mathbb{C}^{2}\) be any matter curve.  Locally this curve is defined by the vanishing a single holomorphic function \(f\).\footnote{We make the assumption that this matter curve is irreducible so that \(f\) does not factor.}  The basic observation is that if \(\varphi\) is a localized mode on \(\Sigma\) then all the gauge invariant data of this mode should be contained inside an arbitrarily small neighborhood of \(\Sigma\).  It follows that if we consider the restriction of the mode to the complement of the matter curve then a localized mode \(\varphi\) is gauge equivalent to zero.  This means that there exists a holomorphic \(\eta \in \mathfrak{g}\otimes \mathcal{O}\) and a positive integer \(m\) such that
\begin{equation}
\varphi =\mathrm{ad}_{\Phi}\left(\frac{\eta}{f^{m}}\right). \label{etadef}
\end{equation}
The function \(f\) has is zero on \(\Sigma\) and thus while \(\varphi\) is gauge equivalent to zero off the matter curve, \(\eta/f^{m}\) cannot be extended over all of \(\mathbb{C}^{2}\) and hence \(\varphi\) is not globally trivial.

This definition of localized modes has a natural mathematical interpretation in terms of sheaf theory.  We are studying the space of modes (\ref{quo}) which is a coherent sheaf on the brane worldvolume \(\mathbb{C}^{2}\), and the mode \(\varphi\) is a holomorphic section of this sheaf.  The prescription \((\ref{etadef})\) says that the localized modes are exactly those sections for which there exists a holomorphic function \(f\) with
\begin{equation}
f^{m} \varphi = 0 \in \frac{\mathfrak{g}\otimes \mathcal{O}}{\mathrm{ad}_{\Phi}\left(\mathfrak{g}\otimes \mathcal{O}\right)}. \label{torsionr}
\end{equation}
These are the \emph{torsion} elements of the sheaf.  In the mathematical language, the function \(f^{m}\) is the \emph{annihilator} of \(\varphi\).

From a physical perspective we can motivate our definition of localized modes by considering the potential induced for the adjoint scalars in the presence of a background Higgs field \(\Phi\).  If we write \(g\) for the complexified gauge transformation which takes us from holomorphic to unitary gauge, then the physical potential takes the form
\begin{equation}
V(\eta)=\left|\left|g ( \mathrm{ad}_{\Phi}(\eta))g^{-1}\right|\right|^{2}
\end{equation}
For a general \(\eta\) this potential is a non-zero mass term and lifts these matrices from the low-energy spectrum.  Meanwhile for those matrices which commute with \(\Phi\) along the entire brane, the induced mass is zero and such \(\eta\) give rise to bulk modes.  Finally there are the localized modes.  According to the definition \((\ref{etadef})\) these are given by \(\eta\)'s which commute with the background only along the matter curve \(\Sigma\).  On this curve the mass induced by the potential \(V\) vanishes and this has the effect of confining the mode to the curve.

We can obtain further intuition about localized modes by thinking about them in terms of partial symmetry restoration or \emph{unHiggsing}.  The bulk fields are given by the commutant of \(\Phi\).  Along matter curves, new matrices commute with \(\Phi\) and give rise to localized modes.  In this way matter curves can be seen as the loci where a symmetry is restored.  This point of view also sheds light on the basic difference between a diagonal intersecting brane background and a T-brane.  Since seven-branes are complex codimension one objects, the Higgs field \(\Phi\) is complex and thus so are the fluctuations \(\eta\).  At each point \(p\) in the brane the relevant space where possible \(\eta\)'s are valued is the complexified Lie algebra \(\mathfrak{g}_{\mathbb{C}}\).  Now let us consider labeling points \(p\) by the locally restored symmetry algebra given by the commutant of \(\Phi(p)\).  In the case of a diagonal background, \(\Phi\) is valued in the Cartan subalgebra and the local commutant at each \(p\) is determined by setting to zero some number of roots.  It follows from this that the local commutant will always be the complexification of a semi-simple Lie algebra.  Thus we can label points by an associated \emph{real} semi-simple Lie algebra and think of this real Lie algebra as a local gauge group which has been unHiggsed at the given point.  This terminology permeates much of the current F-theory literature: seven-branes, matter curves, and Yukawa points are typically denoted by compact real Lie groups.  From this perspective the interesting feature of T-brane backgrounds is then that the local symmetry algebra need not be the complexification of a semi-simple Lie algebra.  Since the Higgs field is now a general non-diagonal matrix, the local commutants are general \emph{complex} Lie subalgebras of \(\mathfrak{g}_{\mathbb{C}}\).  If one wanted to continue to phrase the discussion in terms of local Higgsing and unHiggsing, then the relevant ``gauge group'' to consider is the complexified one with Lie algebra \(\mathfrak{g}_{\mathbb{C}}\).  Localized matter occurs exactly when this complexified gauge group is partially unHiggsed.

Having identified the localized modes as torsion elements in the space of modes, we now assert that the correspondence between an \(8D\) field \(\varphi\) which represents a localized mode and satisfies (\ref{etadef}), and the on-shell \(6D\) representative which naturally resides at the matter curve is given by passing from \(\varphi\) to the \(\eta\) restricted to the curve.  More precisely, any \(\eta\) satisfying \eqref{etadef} is ambiguous up to an element in the kernel of the adjoint map \(\mathrm{ad}_{\Phi}(\cdot)\).  Thus the space of possible \(\eta\)'s is naturally
\begin{equation}
\frac{\mathfrak{g}\otimes \mathcal{O}}{\mathrm{ker}(\mathrm{ad}_{\Phi})}. \label{etaspace}
\end{equation}
Then the map from \(8D\) fields \(\varphi\) to \(6D\) fields is given by choosing any representative \(\eta\) which solves \eqref{etadef} and taking the residue class of \(\eta\) in \(\mathcal{O}/\langle f^{m} \rangle\).
\begin{equation}
\varphi \mapsto [\eta]\in \frac{\mathfrak{g}\otimes \mathcal{O}/\langle f^{m}\rangle}{\mathrm{ker}(\mathrm{ad}_{\Phi})} \label{6dmap}
\end{equation}
We make several comments justifying and explaining this identification:
\begin{itemize}
\item The assignment \((\ref{6dmap})\) is gauge invariant.  If we modify \(\varphi\) by a gauge transformation:
\begin{equation}
\varphi \longrightarrow \varphi +\mathrm{ad}_{\Phi}(\chi)
\end{equation}
then we have
\begin{equation}
\eta \longrightarrow \eta + f^{m}\chi
\end{equation}
and hence the residue class \([\eta]\) is unchanged.
\item All the gauge invariant data of an \(8D\) localized field \(\varphi\) is contained in the class \([\eta]\).  If \(\varphi\) and \(\varphi'\) are two solutions with associated \(\eta\) and \(\eta'\) and \([\eta]=[\eta']\), then we can define a holomorphic gauge parameter
\begin{equation}
\chi = \frac{\eta-\eta'}{f^{m}}.
\end{equation}
A gauge transformation with parameter \(\chi\) shows that \(\varphi\) and \(\varphi'\) are holomorphically equivalent.
\item The correspondence \((\ref{6dmap})\) makes precise our intuition that a localized field should be one whose wavefunction in holomorphic gauge depends only on the coordinate along the matter curve.  The gauge invariant class \([\eta]\) is a matrix with entries valued \(\mathcal{O}/\langle f^{m}\rangle\) and \(f\) vanishes along the curve.
\item The identification \((\ref{6dmap})\) allows us to construct a localized gauge.  Given an 8\(D\) field \(\varphi\) with associated residue class \([\eta]\) we define a new non-holomorphic (smooth) function \(\eta'\) which agrees with \(\eta\) in an epsilon neighborhood of the matter curve and which vanishes outside a slightly larger  neighborhood.  Then since all physical information is contained in the behavior of \(\eta'\) near the matter curve we may as well replace \(\eta\) with \(\eta'\) and hence \(\varphi\) with the pair $(a^\prime,\:\varphi^\prime)$
\begin{equation}
\varphi'=\mathrm{ad}_{\Phi}\left(\frac{\eta'}{f^{m}}\right) \quad\text{and}\quad a^\prime=\overline{\partial}\,\left(\frac{\eta'}{f^{m}}\right)
\end{equation}
which vanishes outside an arbitrarily small tube surrounding \(\Sigma\).
\end{itemize}
These definitions can be made more concrete by seeing explicitly how they work in the two cases of localized matter we have studied thus far.  In the first case of a diagonal background in \(SU(2)\) gauge theory
\begin{equation}
\Phi=\left(\begin{array}{cc}x/2& 0 \\ 0 & -x/2  \end{array}\right),
\end{equation}
and we have a localized mode corresponding to the root \(R_{12}\)
\begin{equation}
x \left(\begin{array}{cc}0 & 1 \\0 & 0\end{array}\right)=\mathrm{ad}_{\Phi}\left(\begin{array}{cc}0 & 1 \\0 & 0\end{array}\right).
\end{equation}
This takes the form of our general expression (\ref{etadef}) with annihilator \(f=x\) and
\begin{equation}
\varphi=\eta= \left(\begin{array}{cc}0 & 1 \\0 & 0\end{array}\right).
\end{equation}
Meanwhile in the more interesting case of the background with monodromy and localized charged matter of section \ref{localizedcharged} we have
\begin{equation}
\Phi=\left(\begin{array}{cc|c}
0 & 1 & 0 \\
x & 0 & 0 \\
\hline
0 & 0 & 0
\end{array}\right) \label{phimonoeta}
\end{equation}
The localized doublet mode satisfies the equation
\begin{equation}
x\left(\begin{array}{c}\varphi_{+} \\ \varphi_{-}\end{array}\right)=\mathrm{ad}_{\Phi} \cdot \left(\begin{array}{c}\varphi_{-} \\ x\varphi_{+}\end{array}\right) \label{nmonoetadef}
\end{equation}
Again this is of the general form (\ref{etadef}) with annihilator \(f=x\) but now with a non-trivial relationship between the 8\(D\) field \(\varphi\) and the 6\(D\) field \(\eta\).  In particular the gauge invariant residue class \([\eta]\) in this case is given by
\begin{equation}
[\eta]=\eta|_{x=0}=\left(\begin{array}{c}\varphi_{-}(y) \\ 0\end{array}\right),
\end{equation}
exactly as we found in section \ref{localizedcharged}.  From these two examples one can see that part of the interesting structure that distinguishes diagonal backgrounds from more general T-brane configurations is that in the latter case the map between 8\(D\) and 6\(D\) fields can be non-trivial.  One of the most important features of our general formalism is that it exhibits the precise relationship between these fields.

\subsection{Superpotentials}
\label{supergen}
\subsubsection{The Matter Curve Action}\label{TheMatterCurveAction}
Now that we have identified the on-shell \(6D\) fields it is straightforward to compute the general formulas governing their superpotential couplings.  The first step is to extend our notion of \(6D\) fields off-shell.  To do this we proceed as in the diagonal examples of section \ref{ReviewB}.  An off-shell 8\(D\) field is one which satisfies the equations of motion only in the transverse direction to the matter curve, but not in the parallel direction.  If we use the map to \(6D\) this implies that an off-shell \(6D\) field is given by an \(\eta\) with the property that \(\eta|_{\Sigma}\) is not necessarily holomorphic.  Putting the \(6D\) modes on-shell then simply corresponds to enforcing the holomorphy constraint.

The assertions of the previous paragraph can be rigorously derived by evaluating the \(8D\) quadratic superpotential on the off-shell 6\(D\) fields.  We consider a smooth matter curve \(\Sigma\) which we may as well approximate by the \(y\)-axis, so that \(\Sigma\) is defined by \(x=0\).  On this matter curve propagate \(k\) distinct \(6D\) hypermultiplets \(\eta_{i}\).  These fields are related to the holomorphic \(8D\) fluctuation fields \(\varphi_{i}\) by the basic equation torsion condition for a localized mode (\ref{etadef})
\begin{equation}\label{basic2}
x^{m_{i}} \varphi_{i}=\mathrm{ad}_{\Phi}(\eta_{i}).
\end{equation}
The off-shell extension of these modes is achieved by relaxing holomorphy in the directions along the matter curve.  Thus off-shell fields are holomorphic functions of \(x\) but arbitrary smooth functions of the matter curve coordinates \((y,\bar{y})\).

Next we want to evaluate the \(8D\) superpotential on these 6\(D\) off-shell fields.  The same steps and localization techniques of section \ref{mcact} immediately lead us to the form
\begin{equation}
W_{6D}  =  \sum_{i,j}\int _{\Sigma}dy\wedge d\bar{y}\left[\frac{1}{2\pi i}\oint_{|x|=\epsilon} \frac{\mathrm{Tr}\left(\eta_{i} \ \bar{\partial}_{\bar{y}}\varphi_{j}\right)}{x^{m_{i}}}dx\right] \label{w6dgen}
\end{equation}
The expression in brackets above has the desired form of a pairing between the distinct \(6D\) fields.  The residue integral extracts the behavior of the product \(\mathrm{Tr}(\eta_{i}\bar{\partial}_{\bar{y}}\varphi_{j})\) along the matter curve \(x=0\).  This makes gauge invariance obvious: a change of gauge on \(\varphi\) or change in representative of the residue class \([\eta]\) has the effect of shifting the product \(\mathrm{Tr}(\eta_{i}\bar{\partial}_{\bar{y}}\varphi_{j})\) by a quantity which does not contribute to the residue and hence leaves the superpotential invariant.  Notice also that as a consequence the basic equation (\ref{etadef}) we can alternatively write \(W_{6D}\) as
\begin{equation}
W_{6D} =\sum_{i,j}\int _{\Sigma}dy\wedge d\bar{y}\left[\frac{1}{2\pi i}\oint_{|x|=\epsilon} \frac{\mathrm{Tr}\left(\eta_{i} \ \mathrm{ad}_{\Phi}(\bar{\partial}_{\bar{y}}\eta_{j})\right)}{x^{m_{i}+m_{j}}}dx\right].
\end{equation}
In particular we see from this that the contribution to the integral over the complex \(y\) plane from any pair of indices \((i,j)\) is \emph{symmetric} under the exchange \(i\leftrightarrow j\).

Now we come to the main point.  The 8\(D\) on-shell fields are holomorphic even on the matter curve and hence so are their on-shell \(6D\) representatives.  This means that the minimization of the \(6D\) superpotential (\ref{w6dgen}) must enforce the BPS equations of motion
\begin{equation}
\bar{\partial}_{\bar{y}}\eta_{i}=0.
\end{equation}
We would like to see this directly from \(W_{6D}\).  Varying \((\ref{w6dgen})\) with respect to \(\varphi_{j}\) one sees that this will be so provided that the skew-symmetric matrix \(\Omega^{ij}\) defined via
\begin{equation}
\Omega^{ij}\: \eta_i \,\bar{\partial}_{\bar y}\,\eta_j =\sum_{i,j}\oint_{|x|=\epsilon} \frac{\mathrm{Tr}\left(\eta_{i} \ \mathrm{ad}_{\Phi}(\bar{\partial}_{\bar{y}}\eta_{j})\right)}{x^{m_{i}+m_{j}}}dx,
\end{equation}
is non-degenerate.  The general techniques required to prove the non-degeneracy of this pairing, while interesting, are somewhat orthogonal to the main thrust of our work and are confined to Appendix \ref{pairing}.  There we develop the local structure of modes on curves in more detail and use this to prove that the \(6D\) superpotential is extremized when all \(6D\) fields are holomorphic.  As in section \ref{ReviewB}, one can view this result as a reflection of the fact that six-dimensional matter must come in \emph{hypermultiplets}.  The non-degeneracy result then shows that given any field \(\eta\) there exists a conjugate field \(\eta^{c}\) which lives on the same matter curve and pairs canonically with \(\eta\) in the \(6D\) superpotential.

It is illuminating to see the \(6D\) superpotential worked out for the simplest non-trivial example of localized matter in a T-brane background.  We again take \(\Phi\) to be of the form \eqref{phimonoeta}.  This background supports two localized charged fields on the matter curve \(x=0\).  Up to gauge transformations we may write the off-shell fields as in equation \eqref{nmonoetadef}
\begin{eqnarray}
\varphi(x,y,\bar{y}) & = &\left(
\begin{array}{cc|c}
0 & 0 & 0 \\
0 & 0 & \varphi_{1}(x,y,\bar{y})\\
\hline
\varphi_{2}(x,y,\bar{y}) & 0 & 0
\end{array}
\right),
\\
\eta(x,y,\bar{y}) &= &\left(
\begin{array}{cc|c}
0 & 0 & \varphi_{1}(x,y,\bar{y}) \\
0 & 0 & 0\\
\hline
0 & -\varphi_{2}(x,y,\bar{y}) & 0
\end{array}
\right).
\end{eqnarray}
According to the general result \eqref{w6dgen} the \(6D\) superpotential is
\begin{eqnarray}
W_{6D} & = & \int_{\Sigma} dy \wedge d\bar{y}\left(\frac{1}{2\pi i}\oint_{|x|=\epsilon}\frac{\mathrm{Tr}(\eta \ \overline{\partial}_{\bar{y}}\varphi)}{x} dx\right)\\
& = &\int_{\Sigma} dy \wedge d\bar{y} \left(\epsilon^{ij}\varphi_{i}(0,y,\bar{y})\overline{\partial}_{\bar{y}}\varphi_{j}(0,y,\bar{y})\right).
\end{eqnarray}
This illustrates the symplectic pairing at work.  The modes \(\eta\) and \(\varphi\) pair up to form a canonical quadratic action for the \(6D\) fields.
\subsubsection{Yukawa Couplings}
\label{yukgens}
The final piece of formalism we must develop is a general expression for the localized Yukawa coupling for fields on matter curves.  We will confine the general derivation to Appendix \ref{resformgen}.  Armed with the general notion of \(6D\) matter and localized gauges its proof is a straightforward application of multidimensional residue theory.  The final result can be phrased elegantly in terms of the \(6D\) fields \(\eta_{i}\) which reside on a matter curve.  We consider three localized modes
\begin{equation}
f_{i}\varphi_{i}=\mathrm{ad}_{\Phi}(\eta_{i}), \hspace{1in} i=1,2,3.
\end{equation}
We assume that these matter curves have an isolated intersection at the origin
\begin{equation}
f_{1}(x,y)=f_{2}(x,y)=f_{3}(x,y)=0 \Longrightarrow (x,y)=(0,0).
\end{equation}
Then the universal localized Yukawa for the three modes is given by
\begin{equation}\label{resideufromfin}
W_{Y}=\mathrm{Res}_{(0,0)}\left[\frac{\mathrm{Tr}\left([\eta_{1}, \eta_{2}]\varphi_{3}\right)}{f_{1}f_{2}}+\frac{\mathrm{Tr}\left([\eta_{2}, \eta_{3}]\varphi_{1}\right)}{f_{2}f_{3}}+\frac{\mathrm{Tr}\left([\eta_{3}, \eta_{1}]\varphi_{2}\right)}{f_{3}f_{1}}\right].
\end{equation}
There are four important consistency checks on this coupling.
\begin{itemize}
\item It is gauge invariant.  Because the coupling is written as a residue it is sensitive only to the gauge invariant residue classes \([\eta_{i}]\).  Further, as a consequence of the basic definition \((\ref{etadef})\) of a localized mode it is unchanged under any change of gauge on the modes \(\varphi_{i}\).
\item The formula can be phrased entirely in terms of holomorphic quantities and hence it is manifestly independent of K\"ahler and flux data.
\item The residue formula is manifestly symmetric in the indices \(1,2,3\) of the localized modes.\footnote{\ Recall that the residue carries a sign arising from the orientation of $\mathbb{C}^2$. The three terms in the right-hand-side of equation \eqref{resideufromfin} are meant to be defined with respect to the orientations $df_1\wedge df_2$,
$df_2\wedge df_3$, and $df_3\wedge df_1$, respectively.}
\item This coupling can readily be seen to reduce to the Yukawa derived in section \ref{ReviewB} for the case of intersecting branes: in that case \(\eta_{i}=\varphi_{i}\) reduces to the holomorphic wavefunction \(h\) and we recover our previous result.
\end{itemize}
This completes the technological developments of this section for the case of trivial worldvolume \(S=\mathbb{C}^{2}\).  In sections \ref{Higgsing}-\ref{examples} of this paper we will apply these results to compute the localized matter spectrum and interactions in a variety of examples.  As explained in detail in section \ref{ReviewB} the restriction to trivial worldvolumes is appropriate when one studies the universal localized contributions to the superpotential for localized matter fields.  By contrast, the study of bulk modes, in particular their existence or lack thereof can only be deduced once a compact brane worldvolume is specified.  In Appendix \ref{gens} we briefly indicate what is required to generalize the results of this section to an arbitrary brane worldvolume \(S\), gauge bundle \(\mathrm{ad}(P)\), and matter curve \(\Sigma\).  Aside from this appendix, all remaining analysis will be carried out locally on \(S=\mathbb{C}^{2}\).

\section{Brane Recombination}
\label{Higgsing}

The techniques of the previous section provide tools to analyze the holomorphic sector of any given background Higgs field.  In this section we apply these ideas to reinterpret the spectrum of a wide class of backgrounds in terms of brane recombination. The basic intuitive picture is that if one starts with a pair of intersecting branes and allows the localized charged field at the intersection to condense then recombination occurs.  On the other hand, from our study of intersecting branes in section \ref{ReviewB}, we know that the localized modes correspond to off-diagonal perturbations of \(\Phi\).  Thus condensation of the bifundamental matter field results in a new non-diagonal backgroud.  This suggests that some T-branes have a simple interpretation in terms of recombined branes.  This gives an alternative method for calculating the spectrum in such examples: work in the recombined frame.  Comparing this recombined picture with our general formalism then gives an elegant and successful check on our techniques.
\subsection{Reconstructible Higgs Fields}
\label{generic}

As we have repeatedly stressed, a T-brane configuration is specified by $\Phi$ rather than its spectral equation. In this section we study a special class of examples where the spectral equation is enough to determine $\Phi$. For simplicity, we restrict our analysis to a unitary gauge group $U(n)$. Up to an overall change in the decoupled \(U(1)\) center of mass, we can always assume that the background Higgs field $\Phi$ is traceless. The backgrounds we consider are those which are non-singular in a suitable sense. We say that a Higgs field is ``reconstructible'' when its defining spectral surface
\begin{equation}
P_{\Phi}(x,y,z)=z^{n}+\sigma_{2}(x,y)z^{n-2}-\sigma_{3}(x,y)z^{n-3}+\cdots +(-1)^{n}\sigma_{n}(x,y)=0
\end{equation}
is non-singular in the three dimensional \((x,y,z)\) space.  Since we are studying the problem locally, the only singularities we are concerned with reside at the origin \((x,y,z)=(0,0,0)\).  The case of maximal interest is when all the eigenvalues of the Higgs field vanish at the origin and from now on we assume this is the case.  Then non-singularity of the spectral surface is determined completely in terms of
\begin{equation}
\sigma_{n}(x,y)=\det(\Phi).
\end{equation}
A reconstructible Higgs field is one for which \(\det(\Phi)\) vanishes to exactly first order at the origin.  As we will see momentarily the terminology ``reconstructible'' is motivated by the fact that such a background is completely determined by its spectral equation, and hence the matrix valued Higgs field can be reconstructed just from the behavior of the eigenvalues.

In terms of the examples we have studied thus far, the intersecting brane configurations of section \ref{ReviewB} are \emph{not} reconstructible.  Their spectral polynomials are factorized as in equation (\ref{factorized}) and hence are singular when the branes intersect.  Meanwhile the basic example of monodromy studied in section \ref{M101} is reconstructible: its spectral polynomial is \(z^{2}-x\) and this cuts out a smooth surface.  One significant consequence of this definition is that a reconstructible Higgs field must break all of the gauge symmetry except the overall center of mass \(U(1)\subset U(n)\) which we have thus far ignored.  Any larger unbroken gauge group would imply a non-trivial commutant of the background \(\Phi\), and thus as in sections \ref{ReviewB} and \ref{MatC} the background could be written in a block diagonal form.  The spectral surface then factorizes and hence is singular when the two components collide.  The fact that reconstructible Higgs fields preserve such a small symmetry group is one of the principle reasons we are able to give a complete picture of their physics.
\subsubsection{Total Recombination}
\label{recombine}
The basic fact that we want to prove in this section is that a \(U(n)\) seven-brane gauge theory deformed by a reconstructible Higgs field is just a \(U(1)\) brane in disguise.  Geometrically this statement is easy to understand.  The correspondence we have used to describe intersecting branes in section \(\ref{ReviewB}\) is that a background Higgs field deforms the stack of \(n\) seven-branes from the \((x,y)\) plane to the spectral surface
\begin{equation}
z^{n}=0\longrightarrow P_{\Phi}(z)=0.
\end{equation}
This fact continues to be true for non-diagonal backgrounds.  Our procedure in the previous sections of this paper has been to view the resulting gauge theory after deformation from the point of view the original gauge theory on the \(z=0\) plane.  Of course absolutely nothing forces us to do this.  After turning on the background it may be more convenient to describe the physics from the point of view of the resulting branes.  At the level of equations this means that we may change coordinates from \((x,y,z)\) to some new more suitable variables.

Now we come to the key observation: for a reconstructible Higgs field the spectral surface is non-singular.  Therefore up to a change of variables on the \(z=0\) plane, we may as well assume that
\begin{equation}
\det(\Phi)= \pm x.
\end{equation}
To describe the resulting branes after deformation we introduce a new system of coordinates where \(x\) has been eliminated
\begin{equation}
(\tilde{x}, \tilde{y}, \tilde{z})=(z, y, P_{\Phi}(z)).
\end{equation}
If we view the ambient space \(\mathbb{C}^{3}\) as being described by the tilded coordinates then the result of deformation by a reconstructible Higgs field is that the brane is now described by \(\tilde{z}=0\) with worldvolume coordinates \((\tilde{x}, \tilde{y})\).  In these coordinates the resulting brane can be identified with an ordinary D7 brane.  This gives a very concrete physical meaning to our non-singularity condition on Higgs fields.  A reconstructible background is one which causes our original brane stack to totally recombine into a single smooth seven-brane.

At the abstract level of the previous paragraphs the claim that a reconstructible Higgs field describes a \(U(1)\) gauge theory is obvious.  However from the point of view of the gauge theory it is a remarkable statement.   We have seen in section \(\ref{M101}\) that even the simplest example of monodromy
\begin{equation}
\Phi=\left(
\begin{array}{cc}
0 & 1 \\
x & 0
\end{array}
\right), \label{nnnmonophi}
\end{equation}
involves an intricate flux tube determined by the highly non-trivial Painlev\'{e} III differential equation, and we are asserting that this solution is simply a complicated perspective for a D7 brane.  We can see a simple check of this claim by looking at the massless spectrum of both theories.  A \(U(1)\) gauge theory has only the bulk massless fluctuations of the free abelian gauge multiplet.  Meanwhile the \(\mathbb{Z}_{2}\) monodromy background naively has \emph{two} bulk fields given by the overall trace and the bulk fluctuation studied in detail in section \ref{M101}.  Written in holomorphic gauge as perturbations of \(\Phi\) these are
\begin{equation}
\left(\begin{array}{cc}
\frac{\alpha(x,y)}{2} & 0 \\
0 & \frac{\alpha(x,y)}{2}
\end{array}\right)
\hspace{.5in} \mathrm{and}
\hspace{.5in}
\left(\begin{array}{cc}
0 & 0 \\
\beta(x,y) & 0
\end{array}\right). \label{u2perturb}
\end{equation}
The astute reader might claim that the massless spectra of the two theories in question, an ordinary \(U(1)\) D7 brane, and an \(U(2)\) seven-brane gauge theory deformed by the monodromy background (\ref{monophiex}), are different and hence these theories cannot possibly be the same.  In fact however, when properly interpreted the two fields (\ref{u2perturb}) are simply two pieces of a single field in the recombined theory.  To see this, we start in the \(U(2)\) gauge theory and we consider the geometric result of activating the first order perturbations (\ref{u2perturb}).  The spectral equation is deformed as
\begin{equation}
z^{2}-x\longrightarrow(z-\frac{1}{2}\alpha(x,y))^{2}-(x+\beta(x,y)) .
\end{equation}
To determine the effect of the deformation from the point of view of the recombined brane we must now change variables to the tilded coordinates.  We want to view the perturbation fields \(\alpha, \beta\) as functions on the new brane worldvolume \(\tilde{z}=0\).  The spectral equation then becomes, at leading order
\begin{equation}
\tilde{z}-(\tilde{x}\alpha(\tilde{x}^{2},\tilde{y})+\beta(\tilde{x}^{2},\tilde{y})) = 0.
\end{equation}
Thus on the new brane worldvolume the two perturbations \(\alpha\) and \(\beta\) are simply the even and odd parts of a single function of \(\tilde{x}\).  They combine to one bulk field which is identified with the \(U(1)\) bulk center of mass field of a D7 brane.

\begin{figure}
\begin{center}
\framebox{
\includegraphics[width=0.7\textheight]{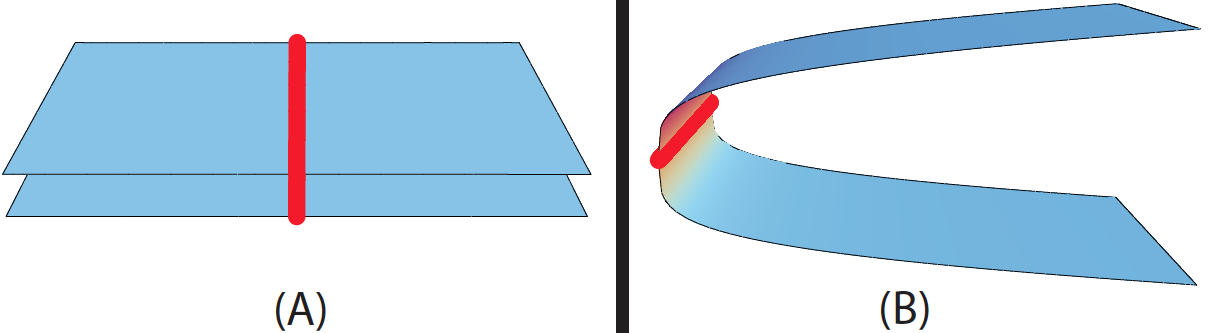}
}
\caption{A schematic cartoon of brane recombination.  In (A) we have an $SU(2)$ gauge theory in the presence of the $\mathbb{Z}_{2}$ monodromy background \eqref{nnnmonophi}.  Along the branch locus of the spectral equation, \(x=0\), this solution has a concentrated gauge flux tube illustrated in red.  In (B) the same system is described from the perspective of a single recombined brane whose worldvolume is a double cover of the original brane stack.  The data of the brane flux in the original \(SU(2)\) theory is carried in the recombined picture by a concentrated worldvolume curvature at the branch locus of the cover.}
\label{fig:recombine}
\end{center}
\end{figure}

One can make the match between the \(U(2)\) monodromy background and the D7 recombined picture sharper still.  In section \ref{M101} we found that in unitary gauge there was a single bulk field whose wavefunction decayed far from the branch locus.  How do we explain such a mode in the recombined picture?  The answer is that although holomorphically the recombined brane is described by the simple equation \(\tilde{z}=0\), if we keep track of the additional real data then there is more to the story.  Locally we approximated the \(U(2)\) brane as flat with canonical K\"{a}hler form.  If we further approximate the normal direction as flat then it follows that on the recombined brane the metric is specified by the K\"{a}hler form:
\begin{equation}
\omega=\frac{i}{2}\left((1+4|\tilde{x}|^{2})d\tilde{x}\wedge d\bar{\tilde{x}}+d\tilde{y}\wedge d\bar{\tilde{y}}\right).
\end{equation}
The recombined brane, illustrated schematically in Figure \ref{fig:recombine}, is therefore \emph{curved}.  The unitary wavefunctions are explicitly sensitive to this curvature and by solving the wave equation on the recombined brane we recover a mode localized on its worldvolume.  

Thus the recombined picture offers an a posteriori explanation of the results of section \ref{M101}.  There are no localized modes in the monodromic background \eqref{nnnmonophi} because there are no localized modes for an isolated D7 brane.  Meanwhile the Painlev\'{e} flux trapped at the branch locus which gives rise to the decaying bulk field is captured in the recombined picture by a non-trivial worldvolume curvature.

The argument above that reconstructible Higgs fields result in total brane recombination can be extended to the case of \(U(n)\) gauge theory.  To prove this the most significant point we need to address is the following.  Holomorphically, a \(U(1)\) gauge theory on a smooth isolated D7 brane is a completely unique theory.  To fix the configuration one simply chooses a location in space for the brane, modeled in the gauge theory by a background for the Higgs field in the free \(U(1)\) gauge multiplet.  Meanwhile the \(U(n)\) gauge theory on \(\mathbb{C}^{2}\) has a huge moduli space of backgrounds described by choosing an arbitrary \(n \times n\) holomorphic matrix \(\Phi\) up to holomorphic gauge transformations.  Given any such background \(\Phi\) we know that the spectral polynomial \(P_{\Phi}(z)\) is one piece of gauge invariant data but in general it does not carry complete information.  The nilpotent background of section \ref{NilpH} serves to illustrate this point.  Nevertheless, in the correspondence between \(U(n)\) seven-brane gauge theories in a generic background \(\Phi\) we are asserting that \(P_{\Phi}(z)=0\) describes the new position of an isolated D7 brane.  Since the holomorphic sector of the latter theory is now completely fixed it must be that for a reconstructible Higgs field the spectral polynomial yields complete information.  Thus we are led to a sharp mathematical consequence of our claim.   \emph{Any two reconstructible Higgs fields with the same spectral polynomial are gauge equivalent.}  In other words, the terminology ``reconstructible'' is justified: such Higgs fields can be extracted from the behavior of their eigenvalues.

In Appendix \ref{classgen} of this paper we prove this claim directly.  The result is that if \(\Phi\) is a reconstructible \(SU(n)\) Higgs field and has spectral polynomial
\begin{equation}
P_{\Phi}(z)=z^{n}+\sigma_{2}z^{n-2}-\sigma_{3}z^{n-3}+\cdots+(-1)^{n}\sigma_{n}
\end{equation}
then we may reach a gauge where
\begin{equation}
\Phi=\left(
\begin{array}{cccccc}
0 & 1 & 0 & \cdots & 0 & 0 \\
0 & 0 & 1& \cdots & 0 & 0 \\
\vdots & \vdots & \vdots & \ddots & \vdots & \vdots \\
0 & 0 & 0 & \cdots & 0 & 1 \\
(-1)^{n-1}\sigma_{n} & (-1)^{n-2}\sigma_{n-1} & (-1)^{n-3}\sigma_{n-2} & \cdots & -\sigma_{2} & 0
\end{array}
\right). \label{explicitgeneric}
\end{equation}
A further consequence of the theorem proved in the Appendix is that the spectrum in the background (\ref{explicitgeneric}) is given by \(n-1\) bulk fields which are fluctuations in the spectral coefficients \(\sigma_{i}\) as well as the overall \(U(1)\) trace.  In particular there are no localized fields.  To calculate the bulk spectrum and match to the D7 theory we now follow the same argument as above for the simple \(2\times2\) monodromy background.  The recombined brane defined by
\begin{equation}
z^{n}+\sigma_{2}z^{n-2}-\sigma_{3}z^{n-3}+\cdots+(-1)^{n}\sigma_{n}=0
\end{equation}
is a smooth \(n\)-sheeted cover of the original seven-brane stack at \(z=0\).  Since the Higgs field on the recombined brane is a generic holomorphic function this means that we require \(n\) holomorphic functions on the original brane stack \(z=0\) to reproduce the single Higgs field on the recombined brane.  The bulk fields discussed above are exactly these modes.

Thus at the level of the massless spectrum we have explicitly demonstrated that \(U(n)\) seven-brane gauge theory deformed by a generic Higgs field results in complete brane recombination into a single smooth isolated D7 brane.  For such backgrounds there are no matter curves or localized fields because there are none for the D7.  For comparison with previous results studied in the literature, what we have derived is essentially the conditions under which the spectral equation technology used in references \cite{Hayashi:2009ge, DWIII} is applicable.\footnote{Let us stress that in abstract terms, the spectral \emph{cover}, as opposed to the spectral \emph{equation}, contains more information than just the eigenvalues of $\Phi$.} The method used there is based on the idea that for reconstructible Higgs fields the spectral equation gives complete information and hence the physics of the gauge theory can be extracted from the geometry of the smooth spectral surface cut out by
\begin{equation}
P_{\Phi}(z)=0.
\end{equation}
By contrast when the spectral equation is singular, one needs to know the actual Higgs field.  Again the most basic example of this is the nilpotent background of section \ref{NilpH}.  The spectral surface is cut out by \(P_{\Phi}(z)=z^{2}=0\) and this is manifestly singular.  In this case the theory is not uniquely determined by the spectral polynomial and the set of physically meaningful ways of resolving the singularity is given by possible Higgs fields with the given \(P_{\Phi}\).

Given that the non-reconstructible backgrounds are in a sense singular, one might wonder if we can simplify our lives and ignore these cases.  Even at the level of phenomenological applications this is decidedly not so.  Indeed, the case of maximal interest for various model building applications in F-theory GUTs occurs precisely when the spectral surface is singular! In such cases one should not expect to be able to fully specify a model with just the eigenvalues of $\Phi$. One major conceptual point of our work is that although in such cases the spectral equation does not completely characterize the background, one can utilize the gauge theory techniques of section \ref{formalism} to directly analyze the physics of any given example.

\subsection{Intersecting Recombined Branes }
\label{spectraltrick}
Having classified reconstructible backgrounds we can now study a simple restricted class of backgrounds which involve both brane recombination and intersecting branes.  We consider a Higgsing process \(SU(k_{1}+k_{2}+\cdots k_{j}+n)\) broken to \(U(1)^{j}\times SU(n)\) described in holomorphic gauge by a block diagonal Higgs field
\begin{equation}
\Phi=\left(
\begin{array}{c|c|c|c|c}
\Psi_{1} & 0 & \cdots & 0 & 0 \\
\hline
0 & \Psi_{2} & \cdots & 0 & 0 \\
\hline
\vdots & \vdots & \ddots & \vdots & \vdots \\
\hline
0 & 0 & \cdots & \Psi_{j} & 0 \\
\hline
0 & 0 & \cdots & 0 & 0
\end{array}
\right). \label{generalhiggs}
\end{equation}
This background has a factorized spectral polynomial and thus is not reconstructible.  To constrain the problem, we restrict ourselves to the case where each block \(\Psi_{i}\) is itself a reconstructible \(k_{i}\times k_{i}\) Higgs field and we assume that \(P_{\Psi_{i}}(z)\neq P_{\Psi_{j}}(z)\) for \(i\neq j\).  Given the discussion of the previous section one expects that this background can be interpreted as a stack of \(n\) \(D7\) branes which support the non-abelian \(SU(n)\) gauge group and which intersect with \(j\) distinct smooth \(D7\) branes.  Each of these smooth \(D7\) branes has a worldvolume given by the vanishing locus of the spectral equation of \(\Psi_{i}\).  These meet the \(z=0\) plane along the determinant loci
\begin{equation}
\det(\Psi_{i})=0,
\end{equation}
and hence we expect that these are the matter curves.  For simplicity we will assume that all of these determinant curves have generic intersections, so this should describe the basic example of transversally intersecting branes studied in detail in section \ref{ReviewB}.
In this section we will provide evidence that this is the case by comparing the massless spectra of these theories.

For simplicity we confine our attention to the localized spectrum charged under the non-abelian part of the unbroken gauge group \(SU(n)\).  The modes are then grouped into \(2j\) groups \(B_{i}\) and \(B^{c}_{i}\) for \(i=1, \cdots, j\) embedded in the adjoint of \(SU(k_{1}+k_{2}+\cdots k_{j}+n)\) in the block diagonal notation of equation (\ref{generalhiggs}) as
\begin{equation}
\left(
\begin{array}{c|c|c|c|c}
0 & 0 & \cdots & 0 & B_{1} \\
\hline
0 & 0 & \cdots & 0 & B_{2} \\
\hline
\vdots & \vdots & \ddots & \vdots & \vdots \\
\hline
0 & 0 & \cdots & 0 & B_{j} \\
\hline
B^{c}_{1} & B^{c}_{2} & \cdots & B^{c}_{j} & 0
\end{array}
\right). \label{bigbdef}
\end{equation}
To study the existence of localized modes in the block \(B_{i}\) we follow the general procedure of section \ref{formalism} and consider the torsion equation \eqref{etadef} for a localized mode
\begin{equation}
f_{i}\varphi_{i}=\mathrm{ad}_{\Phi}(\eta_{i})=\Psi_{i} \eta_{i}.
\end{equation}
To solve this we observe that the matrix \(\Psi_{i}\) is reconstructible and hence invertible away from the curve defined by the vanishing of its determinant.  Thus the only possible matter curve in the block \(B_{i}\) has
\begin{equation}
f_{i}=\det(\Psi_{i}).
\end{equation}
Let \(\mathcal{A}_{i}\) denote the adjugate matrix to \(\Psi_{i}\).  If the spectral equation for \(\Psi_{i}\) is
\begin{equation}
P_{\Psi_{i}}(z)=z^{k_{i}}-\sigma_{1}z^{k_{i}-1}+\cdots+(-1)^{k_{i}}\sigma_{k_{i}},
\end{equation}
Then the adjugate is given by
\begin{equation}
\mathcal{A}_{i}=(-1)^{k_{i}+1}\left(\Psi_{i}^{k_{i}-1}-\sigma_{1}\Psi^{k_{i}-2}+\cdots+(-1)^{k_{i}-1}\sigma_{k_{i}-1}\right)
\end{equation}
According to the Cayley-Hamilton theorem of linear algebra the matrix \(\Psi_{i}\) satisfies
its own spectral equation and this implies the matrix equation
\begin{equation}
\mathcal{A}_{i}\Psi_{i}=\Psi_{i}\mathcal{A}_{i}=\det(\Psi_{i}) \mathbbm{1}. \label{adjdet}
\end{equation}
Thus up to a factor the determinant, the adjugate \(\mathcal{A}_{i}\) is just the inverse \(\Psi_{i}^{-1}\).  However unlike the inverse, which ceases to exist along the curve \(\det(\Psi_{i})=0\), the adjugate exists everywhere.  Now act with the adjugate on \(\varphi_{i}\) to obtain
\begin{equation}
\mathcal{A}_{i}\varphi_{i}=\eta_{i}. \label{adjeta}
\end{equation}
Thus the adjugate matrix allows us to pass from the \(8D\) fields to their \(6D\) representatives.

To determine how many localized modes exist in the block \(B_{i}\) we must now deduce how many distinct residue classes \([\eta]\) exist i.e. we must calculate the dimension of the space of solutions to \eqref{adjeta} once we restrict to the matter curve locus defined by \(\det(\Psi_{i})=0\).  This is an elementary problem in linear algebra.  The reconstructible matrix \(\Psi_{i}\) can be written in the general form \eqref{explicitgeneric}, and hence where \(\det(\Psi_{i})=0\), the matrix \(\Psi_{i}\) has rank one less than maximal, namely \(k_{i}-1\).  From the matrix relation \eqref{adjdet} we then deduce that on the matter curve the kernel of \(\mathcal{A}_{i}\) has dimension at least \(k_{i}-1\).  On the other hand, the adjugate matrix \(\mathcal{A}_{i}\) does not vanish on the matter curve.  By Cramer's rule the matrix entries of \(\mathcal{A}_{i}\) are given by minors of \(\Psi_{i}\) and at least one such minor, the one corresponding to the maximal Jordan block, never vanishes.  It follows that the kernel of \(\mathcal{A}_{i}\) has dimension exactly \(k_{i}-1\) on the matter curve and by the rank nullity theorem the image of \(\mathcal{A}_{i}\) is dimension \(1\) on the matter curve.  This implies that the block \(B_{i}\) supports exactly one localized field.  This mode is a bifundamental transforming with charge \(+1\) under the \(U(1)_{i}\) and as an antifundamental under \(SU(n)\).  That there is one such mode is exactly what we expect from the picture of the spectrum in terms of intersecting recombined branes.
\subsubsection{Yukawa Couplings for Intersecting Recombined Branes}\label{YukRec}
To complete our analysis of brane recombination we want to match the Yukawa couplings of section \ref{ReviewB} with the more general abstract formulas of section (\ref{yukgens}) in the examples of the previous section.  For the block diagonal Higgs backgrounds which describe recombining branes the basic structure for studying a Yukawa is a \(3\times 3\) block matrix.  Thus without loss of generality we take
\begin{equation}\label{Psithrblock}
\Phi=\left(
\begin{array}{c|c|c}
\Psi_{1} & 0 & 0 \\
\hline
0 & \Psi_{2} & 0 \\
\hline
0 & 0 & 0
\end{array}
\right).
\end{equation}
As in the previous section we take the \(\Psi_{i}\) to be reconstructible \(k_{i}\times k_{i}\) matrices.  The vanishing lower-right block of the background implies an unbroken \(SU(n)\) gauge symmetry.

The localized matter in this background is naturally decomposed into blocks as
\begin{equation}
\varphi=\left(
\begin{array}{c|c|c}
0 & \varphi_{12} & 0 \\
\hline
0 & 0 & \varphi_{23} \\
\hline
\varphi_{31} & 0 & 0
\end{array}
\right).
\end{equation}
Clearly a gauge invariant superpotential is possible between these modes and we aim to compute it.  For the modes \(\varphi_{23}\) and \(\varphi_{31}\) which are charged under the non-abelian gauge group we write the localized \(6D\) fields as in the previous section.  The matter curves are the determinant loci for the matrices \(\Psi_{i}\), and upon introducing the adjugate matrices \(\mathcal{A}_{i}\) we have
\begin{eqnarray}
\mathcal{A}_{2}\varphi_{23} & = & \eta_{23}, \label{adjloc23}\\
-\varphi_{31}\mathcal{A}_{1} & = & \eta_{31}. \label{adjloc31}
\end{eqnarray}
Meanwhile for the modes \(\varphi_{12}\) the localization equation takes the general form \eqref{etadef} for a matter curve defined by \(f=0\)
\begin{equation}
f \varphi_{12}=\mathrm{ad}_{\Phi}(\eta_{12})=\Psi_{1}\eta_{12}-\eta_{12}\Psi_{2}. \label{wierdpsiact}
\end{equation}
In principle one can introduce an adjugate for the combined action of \(\Psi_{1}\) and \(\Psi_{2}\) on \(\eta_{12}\) which appears on the right-hand-side of the above equation.  Using this adjugate one could then solve for the matter curve \(f\) and the localized field \(\eta_{12}\).  However, our interest is in computing a Yukawa coupling and all that is required for this is the general structure of \eqref{wierdpsiact}.

Now according to our results from section \ref{yukgens}, for the holomorphic interactions of localized fields, the Yukawa coupling takes the form
\begin{equation}
W_{Y}=\mathrm{Res}_{(0,0)}\left[\frac{\mathrm{Tr}\left(\eta_{12}\eta_{23}\varphi_{31}\right)}{f \det(\Psi_{2})}+\frac{\mathrm{Tr}\left(\eta_{31}\eta_{12}\varphi_{23}\right)}{ \det(\Psi_{1})f}+\frac{\mathrm{Tr}\left(\eta_{23}\eta_{31}\varphi_{12}\right)}{\det(\Psi_{2})\det(\Psi_{1})}\right]
\end{equation}
Again in principle, one might think that to evaluate the above requires knowledge of \(\eta_{12}\) and \(f\), but this is not so.  Elementary manipulations using only equations \eqref{adjloc23}-\eqref{wierdpsiact} imply that the residue contributions of the first two terms in brackets above cancel.  The final answer, expressed in terms of the \(\varphi_{ij}\) fields takes the form
\begin{equation} \label{shabadoo}
W_{Y}  =  \mathrm{Res}_{(0,0)}\left[\frac{\mathrm{Tr}\left(\varphi_{31}\mathcal{A}_{1}\varphi_{12}\mathcal{A}_{2}\varphi_{23}\right)}{\det(\Psi_{1})\det(\Psi_{2})}\right]. 
\end{equation}
As a simple consistency check on our interpretation in terms of intersecting branes, we note that since the matrices \(\Psi_{i}\) are reconstructible and distinct, the two matter curves which appear in the denominator of the residue formula meet transversally at the origin.   The local duality theory of residues \cite{Griffiths} then implies that the rank of this Yukawa coupling is one.  This is in agreement with our general picture of block reconstructible T-brane configurations describing transversally intersecting recombined branes.

\subsubsection{An Example: $\mathbb{Z}_{k_{1}}\times \mathbb{Z}_{k_{2}}$ Monodromy}
\label{zmznex}
As a simple sample application of the techniques of this section let us conclude with an explicit example.
We consider a block diagonal Higgs field of the form \eqref{Psithrblock} with
\begin{equation}
\Psi_{1}=\left(
\begin{array}{ccccc}
0 & 1 & 0 & \cdots & 0 \\
0 & 0 & 1 & \cdots & 0 \\
\vdots & \vdots & \vdots & \ddots & \vdots \\
0 & 0 & 0 & \cdots & 1\\
x & 0 & 0 &\cdots & 0
\end{array}
\right), \hspace{.5in} \Psi_{2}=\left(
\begin{array}{ccccc}
0 & 1 & 0 & \cdots & 0 \\
0 & 0 & 1 & \cdots & 0 \\
\vdots & \vdots & \vdots & \ddots & \vdots \\
0 & 0 & 0 & \cdots & 1\\
y & 0 & 0 &\cdots & 0
\end{array}
\right).
\end{equation}
\(\Psi_{i}\) is a \(k_{i}\times k_{i}\) reconstructible matrix and hence this example fits into the paradigm of intersecting recombined branes.  The spectral polynomials of these matrices are
\begin{equation}
P_{\Psi_{1}}(z)=z^{k_{1}}-x, \hspace{1in} P_{\Psi_{2}}(z)=z^{k_{2}}-y.
\end{equation}
Thus the roots of the spectral equation for \(\Psi_{1}\) are exactly the \(k_{1}\)-th roots of \(x\) and hence this block realizes a cyclic \(\mathbb{Z}_{k_{1}}\) monodromy group.  Similarly, the block \(\Psi_{2}\) realizes a \(\mathbb{Z}_{k_{2}}\) monodromy group so that the full \(\Phi\) background has a product \(\mathbb{Z}_{k_{1}}\times \mathbb{Z}_{k_{2}}\) monodromy group.

Now we know from our previous analysis that the matter curves for the blocks \(\varphi_{13}\) and \(\varphi_{23}\) are given by the determinant loci of the \(\Psi_{i}\) that is respectively \(x=0\) and \(y=0\).  The adjugate matrices along these loci take the very simple form
\begin{equation}
\mathcal{A}_{1}|_{x=0}=\left(
\begin{array}{cccc}
0 & 0  & \cdots & (-1)^{k_{1}-1} \\
0 & 0 &  \cdots & 0 \\
\vdots & \vdots & \ddots & \vdots \\
0 & 0 &  \cdots & 0\\
\end{array}
\right), \hspace{.5in}\mathcal{A}_{2}|_{y=0}=\left(
\begin{array}{cccc}
0 & 0  & \cdots & (-1)^{k_{2}-1} \\
0 & 0 &  \cdots & 0 \\
\vdots & \vdots & \ddots & \vdots \\
0 & 0 &  \cdots & 0\\
\end{array}
\right).
\end{equation}
If we represent \(\varphi_{23}\) as a column vector with \(k_{2}\) rows each of which is an \(\overline{n}\)  under the unbroken \(SU(n)\),  Then we can write the gauge invariant residue classes as in \eqref{adjloc23}-\eqref{adjloc31}
\begin{equation}
\varphi_{23}(x,y)=\left(\begin{array}{c}
\alpha_{1}(x,y) \\
\alpha_{2} (x,y)\\
\vdots \\
\alpha_{k_{2}}(x,y)
\end{array} \right) \Longrightarrow
[\eta_{23}]=\mathcal{A}_{2}|_{y=0}\left(\varphi_{23}(x,0)\right) = \left(\begin{array}{c}
(-1)^{k_{2}-1}\alpha_{k_{2}}(x,0) \\
0 \\
\vdots \\
0
\end{array} \right)
\label{alpha23def}
\end{equation}
Similarly writing \(\varphi_{31}\) as a row vector with entries \(\beta_{i}\)  each of which transforms as an \(n\) under \(SU(n)\) we have
\begin{equation}
[\eta_{31}]=-\left(\varphi_{31}(0,y)\right)\mathcal{A}_{1}|_{x=0}=\left(\begin{array}{cccc}
0 & \cdots & 0 & (-1)^{k_{1}}\beta_{1}(0,y)
\end{array}
\right) \label{beta13def}
\end{equation}
Finally, we can express the \(k_{1}\times k_{2}\) block \(\varphi_{12}(x,y)\) as a matrix with entries \(\rho_{ij}(x,y)\).  Then combining the ingredients \eqref{alpha23def}, \eqref{beta13def}, and the general result \eqref{resideufromfin}, we arrive at the Yukawa
\begin{equation}
W_{Y}=\mathrm{Res}_{(0,0)}\left[\frac{(\beta_{1}\cdot \alpha_{k_{2}})\rho_{k_{1}1}}{(x)(y)}\right].
\end{equation}
\section{The Monodromy Group}
\label{weylorbits}
There is a useful and interesting alternative way to formulate the results of section \ref{Higgsing}.  To motivate this, observe that the reconstructible backgrounds, and their modestly more complicated block diagonal cousins have a simple interpretation in terms of intersecting branes.  This means that these backgrounds, like those of section \ref{ReviewB} are in fact characterized completely by the eigenvalues of the spectral polynomial.  In this section we will push this point of view to its logical conclusion.  This involves developing the study of the spectral equation, in particular the branch structure of the eigenvalues as controlled by the monodromy group.  We will develop the relation of the monodromy group to the general results of section \ref{formalism} for the localized matter and their superpotential couplings.  This approach also makes contact with the way that T-branes have been previously encountered in the literature.  Armed with our general techniques, we will be able to \emph{derive} the rules for computing in backgrounds with monodromy that have previously been postulated in \cite{Hayashi:2009ge,BHSV,EPOINT,Marsano:2009gv}. Further, we will explain when these rules will break down and in this way clarify how T-branes provide a significant generalization of the notion of monodromic branes.  Finally, in section \ref{threebranes} we briefly consider how a three-brane probe explores a T-brane.  This provides an alternative perspective on how T-branes extend the notion of monodromic branes.
\subsection{Matter Counting}
The basic idea we pursue is to try to force a comparison between our T-brane backgrounds with monodromy, and the basic abelian intersecting brane backgrounds studied in section \(\ref{ReviewB}\).  We begin our analysis with an example given by a \(U(3)\) gauge theory deformed by the holomorphic background of section \ref{localizedcharged}
\begin{equation}
\Phi = \left( \begin{array}{cc|c}
0 & 1 & 0 \\
x & 0 & 0 \\
\hline
0 & 0 & 0
\end{array}
\right). \label{nnnnphi}
\end{equation}
Away from \(x=0\) this Higgs field is diagonalizable by conjugation by the matrix
\begin{equation}
g= \left( \begin{array}{cc|c}
\sqrt{x} & 1 & 0 \\
-\sqrt{x} & 1 & 0 \\
\hline
0 & 0 & 1
\end{array}
\right).
\end{equation}
At \(x=0\) itself this Higgs field becomes nilpotent and is therefore not diagonalizable.  In spite of this fact we throw caution to the wind and declare that the Higgs field in \emph{branched gauge} is given by its diagonal form
\begin{equation}
g\Phi g^{-1} = \left( \begin{array}{cc|c}
\sqrt{x} & 0 & 0 \\
0 & -\sqrt{x} & 0 \\
\hline
0 & 0 & 0
\end{array}
\right).
\end{equation}
As it stands, the meaning of this expression is unclear.  The ``gauge transformation'' \(g\) required to put \(\Phi\) in branched gauge is both multivalued and singular.  Nevertheless one can expect that if we can make sense of the spectrum in branched gauge then since the Higgs field is now diagonalized we should be able to phrase our analysis in a language closely parallel to the abelian case.

To begin, notice that the singularities in the gauge transformation \(g\) are confined to the \(2\times 2\) non-trivial block of the background field.  This tells us two things.  First, branched gauge is completely unsuitable for studying the fields which descend from the adjoint of the \(SU(2)\) where \(\Phi\) is non-vanishing.  This is just as well since in section \ref{M101} we learned that the adjoint of \(SU(2)\) does not give rise to any localized fields.  Second and more importantly, since no singularities occur outside this block we can expect that for the study of localized charged fields, passing to branched gauge is simply a peculiar change of basis.  Acting on the charged doublet of section \ref{localizedcharged} we have
\begin{equation}
\varphi \rightarrow g\left( \begin{array}{cc|c}
0 & 0 & \varphi_{+} \\
0 & 0 & \varphi_{-} \\
\hline
0 & 0 & 0
\end{array}
\right)g^{-1}= \left( \begin{array}{cc|c}
0 & 0 & \sqrt{x}\varphi_{+}+\varphi_{-} \\
0 & 0 & -\sqrt{x}\varphi_{+}+\varphi_{-} \\
\hline
0 & 0 & 0
\end{array}
\right). \label{branchedh}
\end{equation}
The available doublet gauge parameters \(\chi\) in branched gauge are similarly obtained by conjugation by \(g\) and thus have identical branch structure to the above.  It follows that we can reach a gauge where \(\varphi_{+}=0\) and \(\varphi_{-}\) depends only on \(y\) and we see that there is a single localized matter field confined to the matter curve \(x=0\).  Of course this fact is merely a tautology obtained by conjugating our previous answer by the matrix \(g\).

How is this change of basis useful?  The answer is that we can obtain a simple group theoretic explanation of the spectrum.  In branched gauge the background Higgs field is not single-valued.  Rather as we circle the origin in the complex \(x\) plane the two eigenvalues \(\pm \sqrt{x}\) interchange.  If we introduce the matrix:
\begin{equation}
\mathcal{W}= \left( \begin{array}{cc|c}
0 & 1 & 0 \\
1 & 0 & 0 \\
\hline
0 & 0 & 1
\end{array}
\right)
\end{equation}
then the branch structure of the eigenvalues is encoded in the following equation:
\begin{equation}
\Phi(e^{2\pi i}x,y)=\mathcal{W}\Phi(x,y)\mathcal{W}^{-1}. \label{wtwist}
\end{equation}
The above has an intuitive interpretation.  The matrix \(\mathcal{W}\) is an element of the Weyl group of \(U(3)\).  Equation (\ref{wtwist}) states that as one circles the branch locus the Higgs field is conjugated by \(\mathcal{W}\).  This means that the invariant data in the Higgs field is not the eigenvalues themselves but rather the Weyl invariant functions of the eigenvalues, i.e the symmetric functions \(\sigma_{i}\).  This is exactly what one should expect in a gauge theory.  Even after diagonalizing the Higgs field there is a residual gauge symmetry given by the permutation of the eigenvalues, and these permutations are carried out by the Weyl group.  The novelty here is that our Higgs field varies over the brane worldvolume and thus as we go around the vanishing locus of the symmetric functions the Weyl group can, and does act.

Branched gauge therefore provides a perspective on T-brane backgrounds where the concept of monodromy really shows its use.  Thus far the monodromy group has simply been a crude invariant of the background Higgs field.  We have seen that a non-zero monodromy group implies that the Higgs field cannot be globally diagonalized, but we have not yet seen how the fact that the monodromy is a \emph{group} really matters for anything.  Now, however, we see that in branched gauge the monodromy group, in this case \(\mathbb{Z}_{2}\), acts on the data of the problem via its embedding in the Weyl group.  The charged perturbations, being perturbations of \(\Phi\) must similarly obey the twisting condition \((\ref{wtwist})\).  Hence for the doublet:
\begin{equation}
\varphi(e^{2\pi i}x,y)=\left(\begin{array}{cc} 0 & 1 \\ 1 & 0 \end{array}\right)\varphi(x,y). \label{nhtwist}
\end{equation}
This explains the pattern of branch cuts found in equation \((\ref{branchedh})\).  Since \(\mathcal{W}\) squares to the identity, a rotation of \(4\pi\) around \(x=0\) leaves the doublet invariant.  It follows that each entry of \(\varphi\) admits an expansion in \(\sqrt{x}\).  Compatibility with equation \((\ref{nhtwist})\) then shows that any allowed mode takes the form
\begin{equation}
\varphi=\left(\begin{array}{c}\sqrt{x}\varphi_{+}+\varphi_{-} \\ -\sqrt{x}\varphi_{+}+\varphi_{-}\end{array}\right)
\end{equation}
with \(\varphi_{\pm}\) single valued exactly as in \((\ref{branchedh})\). In fact, even the terminology ``doublet'' that we have been using to refer to this mode can be explained by this analysis: the mode in question is a doublet under the permutation action of the \(\mathbb{Z}_{2}\) monodromy group.

One can now see how the monodromy group provides a useful organizing principle for calculating charged matter fields.  We start with the roots of the \(U(3)\) group being Higgsed and we restrict our attention to those roots charged under the unbroken gauge group.  Thus we are interested in perturbations of the background \(\Phi\) in the \(R_{13}\) and \(R_{23}\) directions as well as the transposed degrees of freedom.  Since these modes have definite charge under commutation with \(\Phi\) in branched gauge if we were to proceed as in section \(\ref{ReviewMat}\) we would declare that the perturbation in the direction \(R_{13}\) is localized on the ``curve'' \(\sqrt{x}=0\) while the perturbation in \(R_{23}\) resides at \(-\sqrt{x}=0\).  This is almost correct except that now we must take into account the action of the Weyl group.  As we have seen these two modes form a \(\mathbb{Z}_{2}\) Weyl doublet and the effect of the twisting condition \((\ref{nhtwist})\) is to identify the two roots into a single, globally well-defined, charged degree of freedom, localized at \(x=0\).

The line of reasoning given in the previous paragraph is in fact the way that monodromy has been previously studied in the literature. For example, in \cite{Hayashi:2009ge} it was postulated that the localized matter content of a monodromic background was specified by a diagonal Higgs field with branch cuts. Here we have clarified the sense in which this procedure actually works. It is simply to taking an honestly non-diagonalizable Higgs field and forcing a comparison with abelian configurations by going to the singular branched gauge.  However, this analysis also exposes the fact that the monodromy group in general does not completely characterize a given T-brane.  If we relax the assumption that \(\Phi\) is reconstructible then a given spectral polynomial can have  multiple realizations as physically distinct Higgs fields.  As a simple example we can study \(2\times2\) backgrounds with spectral polynomial
\begin{equation}
P_{\Phi}(z)=z^{2}-x^{3}.
\end{equation}
There are two essentially different T-brane configurations which give rise to this spectral data
\begin{equation}
\Phi_{1}=\left(
\begin{array}{cc}
0 & x \\
x^{2} & 0
\end{array}
\right), \hspace{.5in}\Phi_{2}=\left(
\begin{array}{cc}
0 & 1 \\
x^{3} & 0
\end{array}
\right).
\end{equation}
Away from \(x=0\) these Higgs fields are physically identical.  However at the special locus \(x=0\) they are fundamentally different.  For \(\Phi_{1}\), \(x=0\) is a locus of symmetry enhancement, and consistent with this one finds localized matter in this background.  Meanwhile for \(\Phi_{2}\) things are much the same as the \(\mathbb{Z}_{2}\) background of section \ref{M101}, \(x=0\) supports a flux tube, but no trapped matter.  Thus the notion of a T-brane greatly refines the notion of a monodromic brane.  In general, to completely deduce the physics one must use the techniques of section \ref{formalism} as opposed to the monodromy group alone.

Though the spectral equation is not always enough to characterize the localized matter, it is nevertheless true that for the intersecting recombined brane backgrounds of section \ref{Higgsing} everything can be captured in terms of the eigenvalues and the action of the monodromy group.  We can immediately generalize the analysis of the example \eqref{nnnnphi} to this broader setting.  We again consider a block diagonal background of the form \eqref{generalhiggs}
\begin{equation}
\Phi=\left(
\begin{array}{c|c|c|c|c}
\Psi_{1} & 0 & \cdots & 0 & 0 \\
\hline
0 & \Psi_{2} & \cdots & 0 & 0 \\
\hline
\vdots & \vdots & \ddots & \vdots & \vdots \\
\hline
0 & 0 & \cdots & \Psi_{j} & 0 \\
\hline
0 & 0 & \cdots & 0 & 0
\end{array}
\right)
\end{equation}
with each \(\Psi_{i}\) a generic \(k_{i}\times k_{i}\) Higgs field.  In the previous section we found that each block \(B_{i}\) defined in \eqref{bigbdef} gives rise to exactly one localized mode and now we would like to recover this analysis by making use of the monodromy group.  In branched gauge the background above is diagonal with eigenvalues which exhibit a very general and intricate branch structure.  However, from the form of the background, it is clear that the monodromy group is factorized according to the block diagonal structure of \(\Phi\)
\begin{equation}
G_{\mathrm{mono}}=G_{1}\times G_{2} \times \cdots \times G_{j}.
\end{equation}
Each \(G_{i}\) is the subgroup of \(S_{k_{i}}\) which acts via Weyl permutations on the \(k_{i}\times k_{i}\) block \(\Psi_{i}\).  The fact that each block \(\Psi_{i}\) is itself a generic Higgs field means that the associated monodromy group \(G_{i}\) is a \emph{transitive} subgroup of \(G_{i}\).  Indeed if it were non-transitive then the spectral polynomial of \(\Psi_{i}\) would factorize and thus the spectral surface for \(\Psi_{i}\) would be singular.

Now consider the perturbation \(B_{i}\). It is a \(k_{i}\times n\) matrix and hence in the case of a diagonal background of section \(\ref{ReviewB}\) it would have given rise to exactly \(k_{i}\times n\) distinct localized fields.  In branched gauge the Weyl group acts to identify these degrees of freedom.  The matrix \(B_{i}\) is in the permutation representation of \(G_{i}\) with \(G_{i}\) acting on the rows.  To determine the number of modes in the general background we need only to count the number of linear combinations of modes in branched gauge which are Weyl invariant, or equivalently the number of distinct orbits of the matrix entries of \(B_{i}\).  Since we know that each group acts transitively we then deduce that there is exactly one \(\overline{n}\) vector orbit and hence exactly one \(\overline{n}\) localized mode for each \(B_{i}\).

Thus the non-abelian charged matter in intersecting recombined brane backgrounds can be completely characterized in terms of the action via the monodromy group.  Conversely given any finite group \(G_{mono}\) we can engineer a background with this monodromy group by solving the inverse Galois problem\footnote{Unlike the inverse Galois problem over $\mathbb{Q}$, the inverse Galois problem over the field \(\mathbb{C}\left(x,y\right)\) is completely solved \cite{GALOISBOOK}.} for \(G_{mono}\) and writing a block diagonal Higgs field as above.  This method is quite useful when one generalizes to consider breaking patterns for non-unitary gauge groups.  For an example of wide phenomenological interest consider a seven-brane gauge theory with gauge group \(E_{8}\) broken to \(SU(5)_{\mathrm{GUT}}\) by a Higgs field valued in the \(SU(5)_{\perp}\) factor of $SU(5)_{\mathrm{GUT}} \times SU(5)_{\bot} \subset E_{8}$. Even if the Higgs background is block reconstructible as an \(SU(5)_{\perp}\) field, there is no simple brane recombination picture for an \(E_{8}\) brane.  One approach to study such configurations is then to fall back on the general monodromy techniques discussed here and widely applied for example in \cite{BHSV,EPOINT,Marsano:2009gv,Dudas:2009hu,King:2010mq,Leontaris:2010zd}. However our analysis also shows that this approach only describes a limited class of Higgs backgrounds and that to a large extent the phenomenological possibilities of T-branes are unexplored.  There is physically no reason whatsoever to restrict one's attention to Higgs fields which are expressed as block diagonal reconstructible pieces.  These are merely the simplest possibility.  Once one exits this paradigm, the monodromy group and the spectral equation, fail to completely capture the physics, and one must make use of the techniques of section \ref{formalism}.  Exploring the applications of this additional freedom for model building is an interesting question for further research.
\subsection{Yukawas From Monodromy}
The notion of monodromy can also be used to give a heuristic derivation of the general result \eqref{shabadoo} for the Yukawa couplings of block reconstructible backgrounds.  In branched gauge, the matter fields are defined by orbits under the action of the monodromy group, and a natural guess for the Yukawa coupling is to to the sum over all trilinear invariants made from such orbits \cite{BHSV}. To motivate this, let us take seriously the branched gauge picture and attempt to ``guess'' the Yukawa coupling.  In branched gauge a block diagonal background of the form
\begin{equation}
\Phi=\left(
\begin{array}{c|c|c}
\Psi_{1} & 0 & 0 \\
\hline
0 & \Psi_{2} & 0 \\
\hline
0 & 0 & 0
\end{array}
\right),
\end{equation}
 becomes diagonal
\begin{equation}
\Psi_{1} \longrightarrow \left(\begin{array}{cccc}
\lambda_{1} & 0 & \cdots & 0 \\
0 & \lambda_{2} & \cdots & 0 \\
\vdots & \vdots & \ddots & \vdots \\
0 & 0 & \cdots & \lambda_{k_{1}}
\end{array}
\right),
\hspace{.5in}\Psi_{2} \longrightarrow \left(\begin{array}{cccc}
\delta_{1} & 0 & \cdots & 0 \\
0 & \delta_{2} & \cdots & 0 \\
\vdots & \vdots & \ddots & \vdots \\
0 & 0 & \cdots & \delta_{k_{2}}
\end{array}
\right).
\end{equation}
The eigenvalues \(\lambda_{i}\) and \(\delta_{j}\) are branched and the spectrum is quotiented by the associated action of the monodromy group \(G_{mono}=G_{1}\times G_{2}\).

Consider the perturbations \(\varphi_{31}\) and \(\varphi_{23}\) written in branched gauge as row and column vectors respectively
\begin{equation}
\varphi_{31}=\left(\begin{array}{cccc}\beta_{1} & \beta_{2} & \cdots & \beta_{k_{1}}\end{array}\right), \hspace{.5in} \varphi_{23}^{T}=\left(\begin{array}{cccc}\alpha_{1} & \alpha_{2} & \cdots & \alpha_{k_{2}}\end{array}\right).
\end{equation}
Each element \(\beta_{i}\) transforms in the \(n\) of the unbroken \(SU(n)\).  The action of \(G_{1}\) permutes the entries \(\beta_{i}\) resulting in a single orbit and hence a single localized degree of freedom in the $n$ of $SU(n)$.  Similarly, each \(\alpha_{i}\) transforms as an \(\overline{n}\) under \(SU(n)\) and the column vector \(\varphi_{23}\) is permuted by \(G_{2}\).  Finally we write the \(k_{1}\times k_{2}\) matrix \(\varphi_{12}\) as a matrix with entries \(\rho_{ij}\) with \(1 \leq i \leq k_{1}\) and \(1\leq j \leq k_{2}\).

Now we want to write down the superpotential coupling \(W_{Y}\).  Taking branched gauge seriously means that we should write the answer first by ignoring the monodromy action and proceeding as in the intersecting brane solutions of section \ref{ReviewB} and then passing to Weyl invariant quantities. For a single set of modes \(\alpha_{i}, \beta_{j}\), if we ignore the presence of monodromy, the Yukawa would be computed by the residue
\begin{equation}
\mathrm{Res}_{(0,0)}\left[\frac{(\alpha_{i}\cdot \beta_{j})\rho_{ji}}{\delta_{i}\lambda_{j}}\right]
\end{equation}
To take into account the monodromy action, we now follow the prescription given in \cite{BHSV} and
sum over the orbit of the given \(\alpha_{i}\) and \(\beta_{j}\).  This motivates the guess for the Yukawa
\begin{equation}
W_{Y}=\sum_{ij}\mathrm{Res}_{(0,0)}\left[\frac{(\alpha_{i}\cdot \beta_{j})\rho_{ji}}{\delta_{i}\lambda_{j}}\right].
\end{equation}
Of course since the eigenvalues are branched, we do not quite know what to make of the above residue.  We can patch things up by going to a common denominator which is globally well defined.  The obvious choice is
\begin{equation}
\det(\Psi_{1})\det(\Psi_{2})=\left(\prod_{i}\lambda_{i}\right)\left(\prod_{j}\delta_{j}\right).
\end{equation}
Then the Yukawa takes the form
\begin{equation}
W_{Y}=\mathrm{Res}_{(0,0)}\left[\frac{\sum_{i,j}(\alpha_{i}\cdot \beta_{j})\rho_{ji}(\det(\Psi_{1})/\lambda_{j})(\det(\Psi_{2})/\delta_{i})}{\det(\Psi_{1})\det(\Psi_{2})}\right] \label{yukmonod}
\end{equation}
But now we simply observe that in branched gauge, the adjugate matrices \(\mathcal{A}_{i}\) have the simple diagonal form
\begin{eqnarray}
\mathcal{A}_{1} & = & \left(\begin{array}{cccc}
\det(\Psi_{1})/\lambda_{1} & 0 & \cdots & 0 \\
0 & \det(\Psi_{1})/\lambda_{2} & \cdots & 0 \\
\vdots & \vdots & \ddots & \vdots \\
0 & 0 & \cdots & \det(\Psi_{1})/\lambda_{k_{1}}
\end{array}
\right), \\
\mathcal{A}_{2} & = & \left(\begin{array}{cccc}
\det(\Psi_{2})/\delta_{1} & 0 & \cdots & 0 \\
0 & \det(\Psi_{2})/\delta_{2} & \cdots & 0 \\
\vdots & \vdots & \ddots & \vdots \\
0 & 0 & \cdots & \det(\Psi_{2})/\delta_{k_{2}}
\end{array}
\right).
\end{eqnarray}
Thus equation \eqref{yukmonod} is can be written compactly in matrix notation as
\begin{equation}
W_{Y}=\mathrm{Res}_{(0,0)}\left[\frac{\mathrm{Tr}\left(\varphi_{31}\mathcal{A}_{1}\varphi_{12}\mathcal{A}_{2}\varphi_{23}\right)}{\det(\Psi_{1})\det(\Psi_{2})}\right]
\end{equation}
This is \emph{exactly} the answer \eqref{shabadoo} that we derived rigorously in subsection \ref{YukRec}! Since the answer is written as a trace it is insensitive to the distinction between the well-behaved holomorphic gauge, and the singular branched gauge.  One can freely compute in whichever picture one finds more convenient.  The method of derivation given here gives an interesting alternative perspective on the appearances of the adjugate matrices in the Yukawa.  In the formalism of section \ref{formalism} these factors are needed to pass from the \(8D\) field \(\varphi\) to the \(6D\) field \(\eta\). Meanwhile, in branched gauge these adjugate factors enforce the sum over Weyl group orbits and therefore render the result monodromy invariant.
\subsection{Probing Monodromy with Three-Branes}
\label{threebranes}
We conclude our discussion of monodromy with a brief discussion of D3-branes probing a T-brane background, studied for example in \cite{Funparticles,FCFT}. We will focus on the long wavelength limit of the resulting physics. In the four-dimensional probe theory this corresponds to the deep infrared regime, in which all details of the compact geometry of the seven-brane four cycle \(S\) and the internal space are, in a technical sense, irrelevant deformations of the four-dimensional probe theory. Thus, the probe D3-brane provides a way to track ultra-local details of seven-brane monodromy and T-branes. We shall focus on those cases where the D3-brane probe realizes an interacting $\mathcal{N}=1$ or $\mathcal{N}=2$ superconformal field theory (SCFT) in the infrared (IR). As it is the case of primary interest, we consider D3-brane probes of a Yukawa point, so that the Higgs field $\Phi$ is nilpotent at $x=y=0$ where the D3-brane sits.

In many cases of interest the coupling constants of this theory will be an order one parameter and no Lagrangian description of the D3-brane will be available. We can nevertheless, analyze some aspects of this theory, and in particular their deformations by various operators. With terminology as in \cite{Green:2010da} for example, we will refer to superpotential and K\"ahler potential deformations to denote chiral and non-chiral deformations of these possibly non-Lagrangian theories.

Let us first review the main features of the $\mathcal{N}=2$ and $\mathcal{N}=1$ probe theories. In the presence of a probe D3-brane, a stack of parallel seven-branes with gauge group $G$ and $\Phi=0$ will preserve eight real supercharges. In the three-brane theory the coordinates \((x,y,z)\) of the internal space become propagating quantum fields \((X,Y,Z)\)  whose expectation values govern the position of the probe.  Meanwhile the seven-brane gauge group \(G\) becomes a flavor symmetry of the three-brane theory.  The three fields \((X,Y,Z)\) are singlets under this flavor symmetry algebra.  They are accompanied in the probe theory by an operator $O$ which transforms in the adjoint representation of $G$, and can be viewed as setting the flux data of the D3-brane when treated as an instanton of the seven-brane gauge theory.

First consider the $\mathcal{N} = 2$ theories. When $G = SO(8)$ this provides a a realization of the $\mathcal{N} = 2$ $SU(2)$ gauge theory with four flavors \cite{Banks:1996nj}, and for $G=E_{6,7,8}$ the Minahan-Nemeschansky theories \cite{MNI,MNII}. The $\mathcal{N}=2$ moduli space of the probe theory is parameterized in terms of expectation values of the operators \((X,Y,Z)\) and \(O\). This moduli space has two branches which intersect at the point $\langle Z \rangle =\langle O \rangle = 0$. On the Coulomb branch, $\left\langle O\right\rangle=0$, and $\left\langle Z\right\rangle $ is non-zero. On the Higgs branch, $\left\langle Z\right\rangle =0$, and the moduli space is parameterized in terms of $\left\langle O\right\rangle $. Since the D3 can freely sit anywhere in the seven-brane worldvolume, the fields $X$ and $Y$ are decoupled hypermultiplets and have canonical scaling dimension. Meanwhile, the operator $O$ has dimension two, and $Z$ has dimension specified by $G$, so that for example if $G=E_{6,7,8}$ then the dimension of $Z$ is $3,4,6$.

The $\mathcal{N}=1$ probe theories are realized by activating a non-trivial vev for $\Phi$. The F-term coupling of the D3-brane probe to the seven-brane can be determined by passing to holomorphic gauge \cite{Funparticles}. In this gauge, $\Phi$ is a holomorphic function of the quantum fields $X$ and $Y$ valued in the complexified Lie algebra $\mathfrak{g}_{\mathbb{C}}$ of the seven-brane gauge group. The resulting superpotential deformation $\delta W_{\mathrm{D}3}$ is \cite{Funparticles}:
\begin{equation}
\delta W_{\mathrm{D}3}= \ \mathrm{Tr}(\Phi(X,Y)\cdot O) \label{defo}%
\end{equation}
This type of deformation breaks the original flavor symmetry group $G$ to the commutant subalgebra of $\Phi(X,Y)$ in $G$. Assuming that this $\mathcal{N}=1$ deformation realizes another SCFT, one can determine the infrared scaling dimensions for the chiral operators of the theory \cite{FCFT}. In the UV $\mathcal{N}=2$ theory, the operators $O$ transform in the adjoint of $G$, and so the entire adjoint has scaling dimension two. In the IR, the symmetry $G$ is broken, and different elements of the original adjoint multiplet of $O$'s will now have different scaling dimensions \cite{FCFT}.

One significant feature of the superpotential deformation \eqref{defo} is that it shows that the physical operators \(O\) of the three-brane theory couple directly to the seven-brane Higgs field \(\Phi\).  Different T-brane configurations are specified by distinct choices of \(\Phi\) and according to \eqref{defo} the IR physics of the three-brane probe will detect the subtle differences between various T-branes.   This also reinforces a basic point of the previous sections, that the monodromy group, while a useful tool, is not the fundamental feature of a T-brane background.  In branched gauge \(\Phi(X,Y)\) is a multivalued function of the quantum fields \((X,Y)\).  From the probe viewpoint this is physically unnatural. There is no sense in which the operators of the probe theory are \textquotedblleft quotiented by a monodromy group.\textquotedblright  \  Rather the operators \(O\) source directly the globally well-defined, but non-diagonalizable Higgs field.

The D3-brane also detects the elliptic fibration data of an F-theory compactification. Moving the D3-branes away from the locations of the T-branes corresponds to moving to a generic point of the geometry. The low energy theory of the D3-brane is then given by an $\mathcal{N} = 1$ $U(1)$ gauge theory. The value of the holomorphic coupling constant $\tau_{\mathrm{D}3}$ physically corresponds to the IIB holomorphic coupling:
\begin{equation}
\tau_{\mathrm{D}3} = \tau_{IIB}.
\end{equation}
In general, $\tau_{\mathrm{D}3}$ will be a non-trivial function of its position in the compactification. As explained for example in \cite{Intriligator:1994sm} and used in \cite{FCFT}, electric-magnetic duality of this theory implies that we can identify $\tau_{\mathrm{D}3}$ with the complex structure modulus of an elliptic curve. Moreover, it can only depend on $\Phi$ through holomorphic gauge invariant singlets. These singlets correspond to the Casimirs found in the spectral equation for $\Phi$. Translating this spectral equation back to information about the elliptic fibration, we see that the notion of the ``discriminant locus'' as dictated by the spectral equation survives even for T-brane configurations. Note, however, that this provides only incomplete information about the theory of the D3-brane, and thus more generally, the F-theory compactification.

Our aim in the remainder of this section will be to use the theory of a D3-brane probing a Yukawa point as a way to elucidate further details of seven-brane monodromy and T-branes. To this end, we first clarify the sense in which gauge transformations of the seven-brane gauge theory extend to the probe theory. Next, we study how the probe theory detects fluctuations $\varphi$ around the background value of $\Phi(X,Y)$.  Finally, we introduce a physical notion of a coarse-grained T-brane background based on which contributions to $\Phi(X,Y)$ correspond to relevant and marginal deformations of the probe D3-brane theory.

\subsubsection{Holomorphic Gauge and the Chiral Ring}

We now discuss how gauge transformations of the seven-brane descend to the probe theory. Clearly, a remnant of the seven-brane gauge group descends to the physical D3-brane probe theory because the coupling of line (\ref{defo}) is invariant under flavor rotations of the form
\begin{align}
\Phi(X,Y) &  \rightarrow g\cdot\Phi(X,Y)\cdot g^{-1}\\
O &  \rightarrow g\cdot O\cdot g^{-1}%
\end{align}
for $g$ a constant element of the compact realization of $G$. We have seen throughout this paper that at the level of holomorphic data, it is often helpful to pass to complexified gauge transformations valued in $G_{\mathbb{C}}$, and in particular to work in holomorphic gauge. At the level of the superpotential deformations of equation (\ref{defo}), we see that this term is indeed invariant under these complexified flavor rotations.

Even though the superpotential is invariant, such complexified transformations will induce non-chiral D-term deformations of the probe theory. This is in accord with the fact that although holomorphic gauge accurately captures the F-term data, D-term data will in general be sensitive to the distinction between holomorphic and non-holomorphic field redefinitions. Note, however, that since finite D-term deformations do not correspond to relevant deformations of a CFT, it is natural to expect that the IR\ behavior of the D3-brane theory will be insensitive to these distinctions. Thus if we study the IR behavior of the theory we can freely complexify the flavor symmetry.

The idea of the previous paragraph can be extended to include the far broader class of seven-brane gauge transformations of $\Phi(X,Y)$ and $O$ of the form%
\begin{align}
\Phi(X,Y) &  \rightarrow g(X,Y)\cdot\Phi(X,Y)\cdot g(X,Y)^{-1}\equiv\Phi
^{(g)}(X,Y)\\
O &  \rightarrow g(X,Y)\cdot O\cdot g(X,Y)^{-1}\equiv O^{(g)}\label{Og}%
\end{align}
where $g(X,Y)=\exp\chi(X,Y)$, with $\chi(X,Y)$ a holomorphic function of $X$ and $Y$ valued in $\mathfrak{g}_{\mathbb{C}}$. Performing a power series expansion in $X$ and $Y$, we see that $\Phi^{(g)}(X,Y)$ will also be a holomorphic function of $X$ and $Y$. Since the D3-brane can be viewed as a point-like instanton of the seven-brane gauge theory, it is natural to expect the $O$'s parameterizing the Higgs branch to also transform.  In the worldvolume theory of the seven-brane the above transformations are clearly symmetries of the action.  However, from the perspective of the three-brane, this type of gauge transformation leads to a highly non-trivial field redefinition. For example,
in the $\mathcal{N}=2$ and $\mathcal{N}=1$ probe theories considered in \cite{FCFT}, the weight of an operator under the adjoint representation determines its scaling dimension \cite{FCFT}. Note, however, that under the gauge transformation of line (\ref{Og}), the operators $O^{(g)}$ will be linear combinations of operators with different scaling dimensions. Nevertheless, because chiral ring relations are, by definition, covariant
under this more general class of complexified gauge transformations, it follows that these gauge transformations also descend to the chiral sector of the D3-brane probe theory.

Thus \(\Phi\) deformations of the probe theory which differ by a complexified gauge transformation induce the same IR dynamics of the D3-brane probe.  This allows us to bring to bear the full power of holomorphic gauge studied in the previous sections of this paper to constrain and classify the possible SCFTs of D3-branes probing T-branes.

\subsubsection{Coupling To Matter}

Our discussion in the previous section focussed on the chiral couplings of the D3-brane probe to the background field $\Phi(X,Y)$. Let us now turn to fluctuations $\varphi$ around this background. In holomorphic gauge, these fluctuations couple to the operators $O$ via \cite{Funparticles}%
\begin{equation}
W_{3-7}=\mathrm{Tr}(\varphi\cdot O).\label{37coup}%
\end{equation}
The matter field fluctuations \(\varphi\) are characterized in terms of the quotient space \eqref{quo}, and hence an individual element \(\varphi\) has meaning only up to infinitesimal gauge transformation i.e. a shift of the form \(\mathrm{ad}_{\Phi}(\chi)\).  Thus the \(3-7\) superpotential is only well-defined provided that
\begin{equation}
\varphi \hspace{.25in}\mathrm{and} \hspace{.25in}\varphi+\mathrm{ad}_{\Phi}(\chi) \label{matterdifir}
\end{equation}
result in the same superpotential coupling for any holomorphic \(\chi\).  In the IR this is a consequence of the fact that the complexified seven-brane gauge group extends to the three-brane theory.  The two matter fields (\ref{matterdifir}) differ by an infinitesimal complexified gauge transformation, and thus we expect that up to a field redefinition on the \(O\)'s these result in the same IR CFT.

To see this invariance directly in the probe theory we consider the superpotential for the gauge transformed \(\varphi\)
\begin{equation}
W_{3-7}=\mathrm{Tr}\left((\varphi+\mathrm{ad}_{\Phi}(\chi))\cdot O\right)
\end{equation}
Performing a gauge transformation as in line (\ref{Og}), the above superpotential can be recast as
\begin{equation}
W_{3-7}=\mathrm{Tr}(\varphi \cdot O^{(g)}),\label{new37}
\end{equation}
for appropriate $g(X,Y)$.  However, it is easy to see that the deformations $\varphi \cdot O$ and $ \varphi \cdot O^{(g)}$ induce a flow to the \textit{same} theory in the\ IR. Indeed, performing a power series expansion in $X$ and $Y$, we have
\begin{equation}
O^{(g)}=O+... \label{oexp}
\end{equation}
where the \textquotedblleft$...$\textquotedblright\ signify terms linear in $O$ and of order one or higher in $X$ or $Y$.   Now in the \(4D\) probe theory the matter field \(\varphi\) describes a propagating quantum field and hence has dimension at least one.  Since \(O\) has dimension two, it follows that the additional terms in the superpotential induced by ``\(...\)'' of equation \eqref{oexp} result in changes of the theory by irrelevant operators.  Since we are concerned only with the IR dynamics these additional pieces can be ignored.

This shows explicitly that the IR dynamics defined by the coupling \(W_{3-7}\) is well-defined.  Note also that by similar reasoning,  an expansion of the mode $\varphi$ in terms of the coordinates $X$ and $Y$ is also irrelevant: The only candidate relevant or marginal coupling in the IR is given by the constant contribution to $\varphi$.

\subsubsection{Coarse-Grained T-Branes}

The previous subsection indicates an interesting feature of the way that a D3 brane couples to a given T-brane configuration. Keeping track of the IR behavior of the probe theory, this motivates a definition of a \emph{coarse-grained} T-brane where we keep track of only those terms in the Higgs field which the D3-brane CFT detects. In \cite{FCFT} the effects of different $\Phi$ deformations were studied, where it was found that most of these terms drop out. Indeed, since the position coordinates \((x,y)\) of the seven-brane worldvolume are now quantum fields in the probe theory, it follows that in a power series expansion in \(X,Y\) most terms in the background Higgs field \(\Phi(X,Y)\) are irrelevant deformations of the probe theory. More precisely, the flow to the IR is dominated by the operators of the deformation $\mathrm{Tr}(\Phi(X,Y)\cdot O)$ with the lowest scaling dimension and determining the allowed coarse-grained \(\Phi\)'s means determining the anomalous dimensions of the components of \(O\) in the IR.   After we have determined which components of \(O\) have the smallest dimension, we then expand \(\Phi\) in \(X\) and \(Y\) and retain only those operators which are marginal in the IR.

Let us consider in more detail the coarse-grained form of $\Phi(X,Y)$ which is probed by a D3-brane. Our discussion follows that given in \cite{FCFT}. For simplicity, we focus on the case where $\Phi$ takes values in an $sl(m,\mathbb{C})$ subalgebra of the complexified Lie algebra $\mathfrak{g}_{\mathbb{C}}$. We consider the case where $\Phi(0,0)$ is nilpotent, so that the D3-brane probes a $G$-type Yukawa point. Up to a complexified flavor rotation, we can decompose $\Phi$ into the direct sum of nilpotent Jordan blocks:%
\begin{equation}
\Phi(0,0)=\underset{n}{\oplus}N^{(n)}%
\end{equation}
where $N^{(n)}$ denotes an $n\times n$ nilpotent Jordan block. Distinct choices of Jordan decomposition give rise to different anomalous dimensions for \(O\) and hence to different class of coarse-grained T-branes detected by the probe \cite{FCFT}.

For each nilpotent block of length \(n\), there is a canonical $sl(2,\mathbb{C})$ subalgebra of $\mathfrak{g}_{\mathbb{C}}$, with Cartan generator
\begin{equation}
 J_{3}^{(n)}= \mathrm{diag}(j_{n}, j_{n}-1,..., 1-j_{n},-j_{n}), \hspace{.5in} j_{n} \equiv \frac{n-1}{2}.
\end{equation}
Introducing the diagonal generator
\begin{equation}
J_{3}=\underset{n}{\oplus}J_{3}^{(n)}\text{,}
\end{equation}
we can organize all of the operators $O$ according to their $J_{3}$ charge. Given an operator $O_{s}$ of $J_{3}$ charge $+s$, its IR\ scaling dimension is \cite{FCFT}:\footnote{As in \cite{FCFT} we assume that there are no emergent $U(1)$ symmetries in the IR of the $\mathcal{N}=1$ deformed theory.}
\begin{equation}
\Delta_{IR}(O_{s})=3- \frac{3}{2}(s+1)\times t
\end{equation}
where $t>0$ is a parameter which is fixed by a-maximization. Now we consider the three-brane superpotential deformation
\begin{equation}
\delta W_{\mathrm{D}3}=\mathrm{Tr}(\Phi(X,Y)\cdot O)
\end{equation}
Although in the UV, \(X\) and \(Y\) are decoupled hypermultiplets, in the IR they can have in principle distinct scaling dimensions.  Allowing for this possibility it follows that if we perform a power series expansion of \(\Phi(X,Y)\) in \(X,Y\) the contributions to $\delta W_{\mathrm{D}3}$ with the lowest scaling dimension are those in which $O_{s}$ has the highest (and possibly second highest) value of $s$.

We now consider some examples of coarse-grained $\Phi(X,Y)$ backgrounds. Since we know that the complexified gauge symmetry of the seven-brane extends to the IR of the three-brane limit we can further restrict our attention using this symmetry.  For a $2\times2$ Higgs field with a single nilpotent Jordan block, we have%
\begin{equation}
\Phi(X,Y)=\left(
\begin{array}
[c]{cc}%
0 & 1\\
X & 0
\end{array}
\right)
\end{equation}
up to coordinate redefinitions. Similarly, for $\Phi$ a $3\times3$
matrix with constant contribution a single large nilpotent Jordan
block, the generic form is%
\begin{equation}
\Phi(X,Y)=\left(
\begin{array}
[c]{ccc}%
0 & 1 & 0\\
0 & 0 & 1\\
X & Y & 0
\end{array}
\right).
\end{equation}
 More generally, the generic form of $\Phi$ given by an
$n\times n$ matrix with constant contribution a single large nilpotent Jordan block is of the form:%
\begin{equation}
\Phi(X,Y)=\left(
\begin{array}
{ccccc}
0 & 1 &0   & \cdots & 0 \\
0 & 0 &1& ... & 0 \\
\vdots & \vdots   & \vdots  &\ddots& \vdots \\
0 & 0 & 0 & \cdots & 1\\
X & Y & 0 & \cdots & 0
\end{array}
\right)
\end{equation}
up to coordinate redefinitions. In all cases, we observe that over the field of meromorphic functions in $\mathbb{C}(X,Y)$, the Galois group of the characteristic polynomial for $\Phi(X,Y)$ is $S_{n}$, the symmetric group on $n$ letters.  Comparing to our previous notion of a reconstructible background, we see that for a single large Jordan block, a coarse-grained Higgs field is reconstructible of a very special form.  It has \(S_{n}\) monodromy group and all spectral coefficients save two vanish.

Though this provides a characterization in the case of a single nilpotent Jordan block, the case of multiple Jordan blocks is richer. Rather than provide a full characterization, let us discuss the case where $\Phi$ contains two $2\times2$ nilpotent Jordan blocks. Using our previous notion of coarse-graining, we have that the IR\ behavior for $\Phi(X,Y)$ is dictated by the entries%
\begin{equation}
\Phi(X,Y)=\left(
\begin{array}
[c]{cccc}%
0 & 1 & 0 & 0\\
X & 0 & \gamma X+\delta Y & 0\\
0 & 0 & 0 & 1\\
\alpha X+\beta Y & 0 & Y & 0
\end{array}
\right).  \text{.}%
\end{equation}
In this case, the characteristic polynomial of $\Phi$ is:%
\begin{equation}
P_{\Phi}(z)=z^{4}-(X+Y)z^{2}-f(X,Y)=0
\end{equation}
where $f(X,Y)$ is a polynomial quadratic in $X$ and $Y$. The corresponding Galois group is then $Dih_{4}\simeq \mathbb{Z}_{2} \ltimes \mathbb{Z}_{4}$, the symmetry group of the square. This illustrates that for an appropriate Jordan block structure, the notion of a coarse-grained monodromy group can differ from $S_{n}$.

\section{Further Examples and Novelties}
\label{examples}
Our techniques can be extended in a number of ways to produce a plethora of interesting examples.  Our aim in this final section of the paper is not to be exhaustive but to indicate some of the directions for further exploration. Following the general paradigm outlined in section \ref{formalism} in all of our examples we focus on the localized spectra and their couplings.  It is for these that the restriction to trivial brane worldvolumes is really justified.  Thus throughout we will not discuss bulk modes, whose existence or lack there of can only be determined once a compact worldvolume is specified.

In \ref{ResC} we compute some basic examples of superpotentials involving the Higgsing of phenomenologically interesting groups such as $E_6$, $E_{7}$ and $E_8$.  Finally in \ref{beasts} we present some examples illustrating that when singular, the spectral equation provides incomplete information about the localized matter content of a T-brane configuration. We show that there can be matter present even when there is no indication from the spectral equation. Conversely, we also show that even if the spectral equation suggests the presence of a localized mode, none may be present.  We then combine some these themes, showing that depending on the representation type of the matter field, the spectral equation may or may not correctly indicate the presence of a localized mode. This latter point is quite significant for model building in F-theory GUTs because various papers have claimed constraints on the spectra of such models using information derived from the spectral equations.  We conclude the paper with an exciting example of \emph{pointlike} localized matter.  

\subsection{Examples of Superpotentials}
\label{ResC}
\subsubsection{An $E_{6}$ Yukawa \label{E6}}

For phenomenological purposes, an important ingredient in an $SU(5)$ F-theory GUT is the coupling $\mathbf{5}\times\mathbf{10}\times\mathbf{10}$. In terms of \(SU(5)\) group theory the \(\mathbf{5}\) is the fundamental while the \(\mathbf{10}\) is the antisymmetric tensor.  The gauge invariant coupling is then given by the totally antisymmetric contraction with an epsilon tensor
\begin{equation}
\epsilon_{ijklm}\mathbf{5}^{i}\mathbf{10}^{jk}\mathbf{10}^{lm}. \label{51010}
\end{equation}
From the perspective of the gauge theory, this coupling is generated by an appropriate breaking pattern of $E_{6}$ to $SU(5)$ and is localized at a point in the geometry.  The Higgs field in this example preserves an unbroken \(SU(5)\) gauge symmetry and hence takes values in the $sl(2, \mathbb{C})\times u(1,\mathbb{C})$ subalgebra of $sl(5,\mathbb{C})\times sl(2,\mathbb{C})\times
u(1,\mathbb{C})\subset\mathfrak{e}_{6}$%
\begin{equation}
\Phi=\left(
\begin{array}
[c]{cc}%
0 & 1\\
x & 0
\end{array}
\right)  \oplus\left(  y/3\right)  .
\end{equation}
For physical applications, the \(\mathbf{5}\) field describes the up-type Higgs field, while the \(\mathbf{10}\) contains various quarks and leptons.  After GUT breaking \(SU(5)\rightarrow SU(3)\times SU(2)\times U(1)\) the \(\mathbf{10}\) can be seen to contain the up-type quarks.  When the standard model Higgs field develops a vev and breaks electroweak symmetry the Yukawa coupling $\mathbf{5}\times\mathbf{10}\times\mathbf{10}$  is then responsible for the mass of the top quark.  Thus this Yukawa coupling is a key feature of an $SU(5)$ GUT.

To compute the couplings in this example, our first task is to determine the matter curves. To this end, we first determine the charge of each irreducible representation under the adjoint action of $\Phi$. The adjoint representation of $\mathfrak{e}_{6}$ decomposes into irreducible representations of $sl(5,\mathbb{C})\times sl(2,\mathbb{C})\times u(1,\mathbb{C})$
as%
\begin{align}
\mathfrak{e}_{6} &  \supset sl(5,\mathbb{C})\times sl(2,\mathbb{C})\times u(1,\mathbb{C}), \label{e6decomp}\\
78 &  \rightarrow(\mathbf{1},\mathbf{1})_{0} \oplus (\mathbf{1},\mathbf{3})_{0}\oplus(\mathbf{24},\mathbf{1})_{0}\oplus(\mathbf{10},\mathbf{2})_{-3}\oplus(\mathbf{5},\mathbf{1})_{+6}\oplus(\overline{\mathbf{10}},\mathbf{2})_{3}\oplus(\overline{\mathbf{5}},\mathbf{1})_{-6}
\end{align}
Where in the above the subscript refers to the \(u(1,\mathbb{C})\) charge.  Denote by $\varphi_{\mathbf{5}}$ and $\varphi_{\mathbf{10}}$ the matter field fluctuations transforming in the indicated representations of the unbroken gauge group \(sl(5,\mathbb{C})\).  For now, we will suppress the GUT group indices of the modes and focus on their transformation properties under the subgroup \(sl(2,\mathbb{C})\times u(1,\mathbb{C}) \) where the background \(\Phi\) is non-trivial.

For the modes in the \(\mathbf{5}\) things are simple.  These modes are singlets under the non-diagonal \(sl(2,\mathbb{C})\) and hence are not sensitive to the \(\mathbb{Z}_{2}\) monodromy of the background.  Under the adjoint action of \(\Phi\) they transform simply by multiplication by \(2y\).  Thus in the language of section \ref{formalism} they solve the torsion equation \eqref{etadef} with matter curve \(y=0\) and
\begin{equation}
\eta_{\mathbf{5}}=\frac{1}{2}\varphi_{\mathbf{5}}.
\end{equation}
Meanwhile for the modes in the \(\mathbf{10}\) things are more interesting.  These modes are charged under the non-trivial \(sl(2,\mathbb{C})\) piece of \(\Phi\).  If we write the field \(\varphi_\mathbf{10}\) as a doublet
\begin{equation}
\varphi_{\mathbf{10}}= \left(\begin{array}{c} \varphi_{\mathbf{10}_{+}} \\ \varphi_{\mathbf{10}_{-}} \end{array}\right)
\end{equation}
then under gauge transformation with a doublet parameter \(\chi\) we find that
\begin{equation}
\delta \varphi_\mathbf{10}= \left(\begin{array}{cc} -y & 1 \\ x & -y \end{array}\right)\chi.
\end{equation}
Using this gauge freedom we may freely set to zero the upper entry \(\mathbf{10}_{+}\) of \(\varphi_{\mathbf{10}}\).  To find the localized modes we then study the torsion condition \eqref{etadef} for a matter curve defined by \(f=0\)
\begin{equation}
f \varphi_{\mathbf{10}}= \left(\begin{array}{cc} -y & 1 \\ x & -y \end{array}\right) \eta_{\mathbf{10}}. \label{10loc}
\end{equation}
To solve this we proceed as in section \ref{Higgsing} and our study of brane recombination.  The matrix appearing on the right-hand side of \eqref{10loc} is invertible away from its determinant locus, and hence this defines the matter curve
\begin{equation}
f=y^{2}-x.
\end{equation}
Then \eqref{10loc} is solved by acting on both sides with the adjugate matrix yielding
\begin{equation}
\eta_{\mathbf{10}}=\left(\begin{array}{cc} -y & -1 \\ -x & -y \end{array}\right)\varphi_{\mathbf{10}}=-\left(\begin{array}{c} \varphi_{\mathbf{10}_{-}} \\ y \varphi_{\mathbf{10}_{-}}\end{array}\right).
\end{equation}
The residue class of this doublet
\begin{equation}
 [\eta_{\mathbf{10}}] \in \mathfrak{e}_{6}\otimes \mathcal{O}/\langle y^{2}-x \rangle
\end{equation}
is then the \(6D\) gauge invariant localized field which describes the \(\mathbf{10}\)'s in this geometry.

Having determined the profile of the holomorphic zero modes, we now compute the Yukawa.  The general results of section \ref{formalism} indicate that the coupling is computed by the following residue
\begin{equation}
W_{\mathbf{5}\times \mathbf{10}\times \mathbf{10}}=\mathrm{Res}_{(0,0)}\left[\frac{\mathrm{Tr}\left([\eta_{\mathbf{5}},\eta_{\mathbf{10}}] \varphi_{\mathbf{10}}\right)}{(y)(y^{2}-x)}\right]. \label{e6resf}
\end{equation}
To evaluate the above we need one final piece of \(E_{6}\) group theory.  The decomposition \eqref{e6decomp} specifies a decomposition of \(\mathfrak{e}_{6}\) generators.  Let \(t_{\mathbf{10}ij}^{ M}\) and \(t_{\mathbf{5}k}\) denote the generators of \(\mathfrak{e}_{6}\) transforming in the \(\mathbf{10}\) and \(\mathbf{5}\) of \(sl(5,\mathbb{C})\) respectively.  We write \(i,j,k\) for \(sl(5,\mathbb{C})\) indices and \(M,N\) for \(sl(2,\mathbb{C})\) indices.  Then the result we need is that the trace in the adjoint of \(\mathfrak{e}_{6}\) is given as
\begin{equation}
\mathrm{Tr}\left([t_{\mathbf{5}i},t_{\mathbf{10}jk}^{M}] t_{\mathbf{10}lm}^{N}\right)\propto \epsilon_{ijklm}\epsilon^{M N}.
\end{equation}
Thus the \(\mathfrak{e}_{6}\) trace provides the necessary \(sl(5,\mathbb{C})\) epsilon tensor to form the coupling \eqref{51010}.  To evaluate the residue then we need only contract the \(sl(2,\mathbb{C})\) with the two index tensor \(\epsilon^{M N}\).  Plugging into \eqref{e6resf}, restoring the \(sl(5,\mathbb{C})\) indices, and simplifying the result residue yields the final answer
\begin{equation}
W_{\mathbf{5}\times \mathbf{10}\times \mathbf{10}}=\mathrm{Res}_{(0,0)}\left[\frac{\epsilon_{ijklm}\varphi_\mathbf{5}^{i}\varphi_{\mathbf{10}_{-}}^{jk}\varphi_{\mathbf{10}_{-}}^{lm}}{(x)(y)}\right]. \label{e6final}
\end{equation}
Notice that as compared to our examples in the previous sections of the paper, this result is novel in that the coupling involves one field \(\varphi_{\mathbf{10}_{-}}\) participating twice in the trilinear Yukawa.  Indeed, the local geometry of this T-brane configuration has only two intersecting matter curves.  One of these curves supports the \(\mathbf{5}\) and the other supports the \(\mathbf{10}\).  As indicated by the denominator factor in the residue these two curves meet transversally and hence the rank of the associated coupling in the space of \(\mathbf{10}\)'s is exactly equal to one.  In the real world to leading order the top quark is massive and the other generations of up-type quarks are massless.  Thus the rank one \(\mathfrak{e}_{6}\) Yukawa computed \eqref{e6final} in is a reasonable starting point for modeling this feature of our universe.

For comparison it is interesting to observe that this configuration of matter curves and their coupling is somewhat different from the one obtained from Higgsing \(E_{6}\) to \(SU(5)\) by a diagonal Higgs field valued in the Cartan. Indeed, in that case we can essentially repeat the same analysis with background Higgs field%
\begin{equation}
\Phi=\left(
\begin{array}
[c]{cc}%
+x & 0\\
0 & -x
\end{array}
\right)  \oplus\left( y/3\right)  .
\end{equation}
The corresponding $\mathbf{5}$ curve is again $y=0$, but there are now two distinct $\mathbf{10}$ curves,
given by $x+y=0$ and $x-y=0$, which we denote by $\mathbf{10}$ and $\mathbf{10}^{\prime}$.  Now the three matter curves all meet at the origin and the Yukawa coupling involves modes from all three curves
\begin{equation}
W_{\mathbf{5}\times\mathbf{10}\times\mathbf{10}}  =\mathrm{Res}_{(0,0)}\left[ \frac{\epsilon_{ijklm}\varphi_{\mathbf{5}}^{i} \varphi_{\mathbf{10'}}^{jk}\varphi_{\mathbf{10}}^{lm}  }{(x)(y)}\right].
\end{equation}
As noted in \cite{BHVII}, if one views the above coupling as a matrix in the space of \(\mathbf{10}\) zero modes then this leads to a rank two Yukawa matrix, and hence indicates more than one generation of heavy up-type quarks.

\subsubsection{An $E_{7}$ Yukawa}
\label{E7}

Analogously to our previous example one can also study the Yukawa couplings in \(SO(10)\) GUTs.  In these models, the matter of a complete generation of standard model fields, together with a right-handed neutrino, is contained in a single Weyl spinor \(\mathbf{16}\) of \(SO(10)\).  Meanwhile the standard model Higgs field transforms as a vector \(\mathbf{10}\).  The most interesting Yukawa \(\mathbf{16}\times \mathbf{16}\times \mathbf{10}\) is again the one responsible for quark masses, and in this case is generated group theoretically by contraction with a \(\Gamma\) matrix of the \(SO(10)\) Clifford algebra  
\begin{equation}
\left(C\Gamma_{i}\right)_{\alpha \beta}\mathbf{16}^{\alpha}\mathbf{16}^{\beta}\mathbf{10}^{i}. \label{161610def}
\end{equation}
Where in the above \(\alpha, \beta\) are spinor indices, \(i\) is a vector index, and \(C\) denotes the standard charge conjugation matrix.

In a seven-brane model this interaction is generated by breaking an \(E_{7}\) gauge group to \(SO(10)\).  The background Higgs field \(\Phi\) is then valued in an $sl(2,\mathbb{C})\times u(1,\mathbb{C})$ subalgebra of $ so(10,\mathbb{C})\times sl(2,\mathbb{C})\times u(1,\mathbb{C}) \subset \mathfrak{e}_{7}$ and is essentially identical to the \(\mathfrak{e}_{6}\) background studied in the previous example
\begin{equation}
\Phi=\left(
\begin{array}
[c]{cc}%
0 & 1\\
x & 0
\end{array}
\right)  \oplus\left(  y/3\right)  .
\end{equation}
The adjoint representation of \(\mathfrak{e}_{7}\) decomposes as
\begin{align}
\mathfrak{e}_{7} &  \supset so(10,\mathbb{C})\times sl(2,\mathbb{C})\times u(1,\mathbb{C}), \label{e7decomp}\\
\mathbf{133} &  \rightarrow(\mathbf{1},\mathbf{1})_{0} \oplus(\mathbf{1},\mathbf{3})_{0}\oplus(\mathbf{45},\mathbf{1})_{0}\oplus(\mathbf{16},\mathbf{2})_{-3}\oplus(\mathbf{10},\mathbf{1})_{+6} \oplus (\overline{\mathbf{16}},\mathbf{2})_{3}\oplus(\mathbf{10},\mathbf{1})_{-6}
\end{align}
Thus up to changing the GUT group from \(SU(5)\) to \(SO(10)\) this T-brane configuration is identical to the \(\mathfrak{e}_{6}\) Higgsing studied in the previous section.  The standard model Higgs field \(\varphi_{\mathbf{10}}\) is localized on the curve \(y=0\), while a doublet \(\varphi_{\mathbf{16}_{\pm}}\) of spinor fields is localized on the curve \(y^{2}=x\).  Proceeding as in the example \ref{E6}, we then find that the one component of the spinor doublet, $\varphi_{\mathbf{16}_{+}}\) is gauge equivalent to zero, so that this background supports only two matter curves and exactly two \(6D\) fields.  

Now we can easily evaluate the Yukawa.  According to our general results of section \ref{formalism} the coupling is computed by the residue
\begin{equation}
W_{\mathbf{16}\times \mathbf{16}\times \mathbf{10}}=\mathrm{Res}_{(0,0)}\left[\frac{\mathrm{Tr}\left([\eta_{\mathbf{10}},\eta_{\mathbf{16}}] \varphi_{\mathbf{16}}\right)}{(y)(y^{2}-x)}\right].
\end{equation}
To evaluate this, we need to know the analogous result to \eqref{e6traces} for a trace of \(\mathfrak{e}_{7}\) generators.  Let \(t_{\mathbf{16},\alpha}^{M}\), and \(t_{\mathbf{10},i}\) denote \(\mathfrak{e}_{7}\) generators transforming under the \(\mathbf{16}\) and \(\mathbf{10}\) of $so(10,\mathbb{C})$ respectively.  As in \eqref{161610def}, we use \(\alpha, \beta\) for spinor indices of $so(10,\mathbb{C})$, \(i\) for a vector index of $so(10,\mathbb{C})$, and \(M, N\) for \(sl(2,\mathbb{C})\) indices.  Then a trace in the adjoint of $\mathfrak{e}_{7}$ produces the following invariant tensors
\begin{equation}
\mathrm{Tr}\left([t_{\mathbf{10},i},t_{\mathbf{16},\alpha}^{M}] t_{\mathbf{16},\beta}^{ N}\right)\propto \left(C\Gamma_{i}\right)_{\alpha \beta}\epsilon^{MN}. \label{e6traces}
\end{equation}
Plugging into the residue and simplifying then yields the result
\begin{equation}
W_{\mathbf{16}\times \mathbf{16}\times \mathbf{10}}=\mathrm{Res}_{(0,0)}\left[\frac{\left(C\Gamma_{i}\right)_{\alpha \beta}\varphi_{\mathbf{16}_{-}}^{\alpha}\varphi_{\mathbf{16}_{-}}^{\beta}\varphi_{\mathbf{10}}^{i}}{(x)(y)}\right]. \label{e7final}
\end{equation}
Again this coupling involves a single field \(\varphi_{\mathbf{16}_{-}}\) participating twice in the Yukawa coupling.  Since the matter curves meet transversally the final result \eqref{e7final} yields a rank one Yukawa and hence gives mass to exactly one generation of standard model matter \(\mathbf{16}\)'s.

\subsubsection{An $E_{8}$ Yukawa \label{E8}}

The previous two examples all involve Yukawas which arise from a rank two enhancement of the unbroken gauge group.  A more dramatic possibility is to have a Yukawa coupling where the gauge group enhances by more than rank two.  Phenomenologically relevant cases of this idea have been studied in detail in \cite{EPOINT, BHSV}.  Thus for our final example we consider the case of an \(SU(5)\) GUT model which is restored at a point all the way to \(E_{8}\).  We focus on an example considered in both \cite{EPOINT}, and  \cite{FCFT} where the seven-brane Higgs field $\Phi$ takes values in the $sl(5,\mathbb{C})_{\bot}$ factor of $sl(5,\mathbb{C})_{\mathrm{GUT}}\times sl(5,\mathbb{C})_{\bot}\subset\mathfrak{e}_{8}$
\begin{equation}
\Phi=\left(
\begin{array}
[c]{ccccc}%
\lambda_{1} & 1 & 0 & 0 & 0\\
x & \lambda_{1} & 0 & 0 & 0\\
0 & 0 & -2\lambda_{1}-\lambda_{2} & 1 & 0\\
0 & 0 & y & -2\lambda_{1}-\lambda_{2} & 0\\
0 & 0 & 0 & 0 & 2(\lambda_{1}+\lambda_{2})
\end{array}
\right).  \label{PHIBACK}%
\end{equation}
Where in the above the two parameters \(\lambda_{i}\) are taken to be linear functions of the coordinates \((x,y)\)
\begin{equation}
\lambda_{i}=\alpha_{i}x+\beta_{i}y, \hspace{.5in} \alpha_{i}, \beta_{i} \in \mathbb{C}.
\end{equation} 
The constants \(\alpha_{i}\) and \(\beta_{j}\) are local moduli of the configuration.  They determine the geometry of the matter curves near the origin \((x,y)=(0,0)\).  We will assume that our configuration is at a generic point in \(\alpha_{i}, \beta_{j}\) space so that in particular none of these moduli vanish.

Physically, this type of background field configuration describes a Dirac neutrino scenario with $\mathbb{Z}_{2}\times \mathbb{Z}_{2}$ monodromy of the type considered in \cite{EPOINT}. One of the interesting features of this type of higher rank structure is that there are now many matter curves all meeting at the origin.  Of interest to us  are three \(\mathbf{5}\) curves and one \(\mathbf{10}\) curve.  To make contact with phenomenology we will identify the \(\mathbf{10}\) curve as supporting a matter field \(\varphi_{\mathbf{10}_{M}}\) and one of the \(\mathbf{5}\) curves as supporting the \(\varphi_{\overline{\mathbf{5}}_{M}}\) matter field.  The remaining two \(\mathbf{5}\) curves will then support the two Higgs fields of the MSSM GUT, \(\varphi_{\overline{\mathbf{5}}_{H}}\) and \(\varphi_{\mathbf{5}_{H}}\).  As we will see both of the required interaction terms 
\begin{equation}
\epsilon_{ijklm} \mathbf{5}^{i}_{H}\times\mathbf{10}_{M}^{jk}\times\mathbf{10}_{M}^{lm} \hspace{.25in}\mathrm{and}\hspace{.25in} \overline{\mathbf{5}}_{H,i}\times \overline{\mathbf{5}}_{M,j}\times\mathbf{10}_{M}^{ij}
\end{equation}
will be generated at this single \(E_{8}\) point in the geometry.

\begin{figure}
\begin{center}
\framebox{
\includegraphics[totalheight=0.2\textheight, width=.5\textwidth]{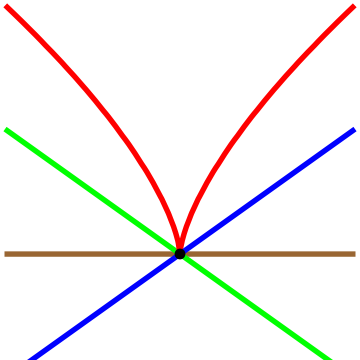}
}
\caption{The local geometry of the $E_{8}$ point with $\mathbb{Z}_{2}\times \mathbb{Z}_{2}$ monodromy specified by the background \eqref{PHIBACK}.  The \(\overline{\mathbf{5}}_{H}\) curve, depicted in red, has a cusp singularity at the origin.  The remaining matter curves are smooth and meet transversally.}
\label{fig:e8point}
\end{center}
\end{figure}

To see how this comes about in more detail, we decompose the adjoint representation of $\mathfrak{e}_{8}$ into irreducible representations of $sl(5,\mathbb{C})_{\mathrm{GUT}}\times sl(5,\mathbb{C})_{\bot}$ as:%
\begin{align}
\mathfrak{e}_{8} &  \supset sl(5,\mathbb{C})_{\mathrm{GUT}}\times sl(5,\mathbb{C})_{\bot},\\
\mathbf{248} &  \rightarrow(\mathbf{24},\mathbf{1})\oplus(\mathbf{1},\mathbf{24})\oplus(\overline{\mathbf{5}},\mathbf{10})\oplus(\mathbf{10},\mathbf{5})+\oplus(\mathbf{5},\overline{\mathbf{10}})\oplus(\overline{\mathbf{10}},\overline{\mathbf{5}}).
\end{align}
Hence, the $\overline{\mathbf{5}}$'s of the GUT\ group correspond to $\mathbf{10}$'s under
$sl(5,\mathbb{C})_{\bot}$, and the $\mathbf{10}$'s of the\ GUT\ group correspond to $\mathbf{5}$'s of
$sl(5,\mathbb{C})_{\bot}$. To determine the resulting matter spectrum, it is helpful to organize these modes according to a weight space decomposition. We introduce basis vectors $e_{1},...,e_{5}$ which span the fundamental of $sl(5,\mathbb{C})_{\bot}$. The basis vectors for the $10$ of $sl(5,\mathbb{C})_{\bot}$ are then $e_{i}\wedge e_{j}$ for $i\neq j$. \ Labeling the components of the $\mathbf{10}$ as $e_{i}\wedge e_{j}$ for $i\neq j$, we have that $\Phi$ acts on the two representations as
\begin{align}
e_{i} &  \rightarrow\Phi\cdot e_{i}\\
e_{i}\wedge e_{j} &  \rightarrow(\Phi\cdot e_{i})\wedge e_{j}+e_{i}\wedge
(\Phi\cdot e_{j}).
\end{align}
To analyze the matter content around the background specified by equation (\ref{PHIBACK}), we need to analyze the action of $\Phi$ on the $\mathbf{5}$ and $\mathbf{10}$ of $sl(5,\mathbb{C})_{\bot}$. Rather than present the full action on each representation, we focus on those pieces of particular phenomenological relevance. Using the identification of matter states performed in \cite{EPOINT}, the subspaces in the $\mathbf{5}$ and $\mathbf{10}$ of~$sl(5,\mathbb{C})_{\bot}$ spanned by the matter fields are
\begin{equation}
\mathbf{5}_{H}:\left(
\begin{array}
[c]{c}%
e_{1}^{\ast}\wedge e_{2}^{\ast}%
\end{array}
\right)  \text{, }\overline{\mathbf{5}}_{M}:\left(
\begin{array}
[c]{c}%
e_{3}\wedge e_{5}\\
e_{4}\wedge e_{5}%
\end{array}
\right)  \text{, }\overline{\mathbf{5}}_{H}:\left(
\begin{array}
[c]{c}%
e_{1}\wedge e_{3}\\
e_{1}\wedge e_{4}\\
e_{2}\wedge e_{3} \\
e_{2}\wedge e_{4}
\end{array}
\right)  \text{, }\mathbf{10}_{M}:\left(
\begin{array}
[c]{c}%
e_{1}\\
e_{2}%
\end{array}
\right)  . \label{smallrdef}
\end{equation}
For each representation \(\mathbf{R}\) appearing in \eqref{smallrdef} we represent the associated action of $\Phi$ as a matrix \(\Phi_{\mathbf{R}}\).  We have
\begin{align}
\Phi_{\mathbf{5}_{H}}   &  =2 \lambda_{1},\label{oneHiggs}\\
\Phi_  {\overline{\mathbf{5}}_{M}}   &  =\left(
\begin{array}
[c]{cc}%
\lambda_{2} & 1\\
y & \lambda_{2}
\end{array}
\right),  \\
\Phi_{  \overline{\mathbf{5}}_{H}}   &  =\left(
\begin{array}
[c]{cccc}%
-(\lambda_{1}+\lambda_{2}) & 1 & 1 & 0\\
y & -(\lambda_{1}+\lambda_{2}) & 0 & 1\\
x & 0 & -(\lambda_{1}+\lambda_{2}) & 1\\
0 & x & y & -(\lambda_{1}+\lambda_{2})
\end{array}
\right),  \\
\Phi_{\mathbf{ 10}_{M}}   &  =\left(
\begin{array}
[c]{cc}%
\lambda_{1} & 1\\
x & \lambda_{1}
\end{array}
\right)  .\label{10M}%
\end{align}
As in our previous study of explicit examples it is helpful to introduce the adjugate matrices \(\mathcal{A}_{\mathbf{R}}\) to \(\Phi_{\mathbf{R}}\). They are defined by the condition that
\begin{equation}
\mathcal{A}_{\mathbf{R}}\Phi_{\mathbf{R}}=\Phi_{\mathbf{R}}\mathcal{A}_{\mathbf{R}}=\mathrm{det} \left(\Phi_{\mathbf{R}}\right) \mathbbm{1}.
\end{equation}
Then the localized \(6D\) fields for each representation is given by
\begin{equation}
\eta_{\mathbf{R}}=\mathcal{A}_{\mathbf{R}}\varphi_{\mathbf{R}}.
\end{equation}
And the matter curves \(f_{\mathbf{R}}=0\) are defined by the determinant loci of the \(\Phi_{\mathbf{R}}\)
\begin{eqnarray}
f_{\mathbf{5}_{H}} &  \equiv &\lambda_{1},\label{5hmatt}\\
f_{\overline{\mathbf{5}}_{M}}   & \equiv & \lambda_{2}^{2}-y,\\
f_{\overline{\mathbf{5}}_{H}}   &   \equiv &(\lambda_{1}+\lambda_{2})^{4}-2(\lambda_{1}+\lambda_{2})^{2}(x+y)+ (x-y)^{2} \\
f_{\mathbf{10}_{M}}   &   \equiv &  \lambda_{1}^{2}-x.\label{10mmatt}%
\end{eqnarray}
Notice in particular that the local geometry of these curves, illustrated in Figure \ref{fig:e8point}, is quite intricate.  The \(\overline{\mathbf{5}}_{H}\) curve has a singularity at the origin.  Nevertheless the gauge theory is still well-behaved and all physical quantities of interest can be computed using our results from section \ref{formalism}.

Let us now turn to the evaluation of the Yukawas. First consider the $\mathbf{5}_{H}\times\mathbf{10}_{M}\times\mathbf{10}_{M}$ coupling. In this case, we note under the action of the internal Higgs field \(\Phi\) the $\mathbf{10}_{M}$ fills out a doublet with components \(\varphi_{\mathbf{10}_{M,\pm}}\), while the $\mathbf{5}_{H}$ corresponds to a singlet. Performing the analogous computation to that presented in the $E_{6}$ example of section \ref{E6}, we now have
\begin{equation}
W_{\mathbf{5}_{H}\times\mathbf{10}_{M}\times\mathbf{10}_{M}}  =  \mathrm{Res}_{(0,0)}\left[  \frac{\mathrm{Tr}\left([\eta_{\mathbf{5}_{H}},\eta_{\mathbf{10}_{M}}]\varphi_{\mathbf{10}_{M}}\right)}{(f_{\mathbf{5}_{H}})(f_{\mathbf{10}_{M}})}\right].
\end{equation}
The \(E_{8}\) trace evaluates identically as the \(E_{6}\) trace and produces the required epsilon tensor for the contraction.  Evaluating and simplifying we find
\begin{equation}
W_{\mathbf{5}_{H}\times\mathbf{10}_{M}\times\mathbf{10}_{M}}  = \mathrm{Res}_{(0,0)}\left[\frac{\epsilon_{ijklm}\varphi_\mathbf{5}^{i}\varphi_{\mathbf{10}_{M,-}}^{jk}\varphi_{\mathbf{10}_{M,-}}^{lm}}{(x)(y)}\right]. \label{e8ftt}
\end{equation}
As in the example \ref{E6} this yields a rank one coupling.

Finally, we consider the evaluation of the $\overline{\mathbf{5}}_{H}\times\overline{\mathbf{5}}_{M}\times\mathbf{10}_{M}$ coupling. This case is somewhat more involved because the $\varphi_{\overline{\mathbf{5}}_{H}}$ states now fill out a four-component vector. We denote the components of this vector as $\varphi_{\overline{\mathbf{5}}_{H, N}}$ with \(N=1,\cdots 4\). Further, denote by $\varphi_{\overline{\mathbf{5}}_{M,\pm}}$ and $\varphi_{\mathbf{10}_{M, \pm}}$ the other modes participating in the Yukawa. Applying our general result we have
\begin{equation}
W_{\overline{\mathbf{5}}_{H}\times \overline{\mathbf{5}}_{M}\times \mathbf{10}_{M}}=\left[\frac{\mathrm{Tr}\left([\eta_{\overline{\mathbf{5}}_{M}}, \eta_{\mathbf{10}_{M}}]\varphi_{\overline{\mathbf{5}}_{H}}\right)}{(f_{\overline{\mathbf{5}}_{M}})(f_{\mathbf{10}_{M}})}\right]. \label{5510com}
\end{equation}
As usual, we need to know how to evaluate a trace of matrices in the adjoint of \(\mathfrak{e}_{8}\) in terms of invariant tensors for \(sl(5,\mathbb{C})_{\mathrm{GUT}}\times sl(5,\mathbb{C})_{\perp}\).  Fortunately due to the symmetry between the internal indices of \(sl(5,\mathbb{C})_{\perp}\) and the GUT indices \(sl (5,\mathbb{C})_{\mathrm{GUT}}\) our previous work already tells us the answer.  Indeed if we think about this coupling from the point of view of \(sl(5,\mathbb{C})\) then it is again \(\mathbf{5}\times\mathbf{10}\times \mathbf{10}\) and hence the internal \(sl(5,\mathbb{C})\) indices are contracted using a totally antisymmetric five index tensor as in \eqref{e8ftt}.  Thus if we continue to use the wedge product notation introduced above then the trace is 
\begin{equation}
\mathrm{Tr}\left([\eta_{\overline{\mathbf{5}}_{M}}, \eta_{\mathbf{10}_{M}}]\varphi_{\overline{\mathbf{5}}_{H}}\right)=\eta_{\overline{\mathbf{5}}_{M}i}\wedge  \eta_{\mathbf{10}_{M}}^{i j} \wedge \varphi_{\overline{\mathbf{5}}_{H}j}.
\end{equation}
All that remains is to explicitly make use of the relevant adjugate matrices substitute into \eqref{5510com}.  In simplifying the residue, it is helpful to note that the two matter curves appearing in the denominator meet transversally, and hence the only non-zero contributions to the residue can come from terms which do not vanish at the origin.  The result, after a small bit of algebra is quite simple
\begin{equation}
W_{\overline{\mathbf{5}}_{H}\times \overline{\mathbf{5}}_{M}\times \mathbf{10}_{M}}=\mathrm{Res}_{(0,0)}\left[\frac{\varphi_{\overline{\mathbf{5}}_{H,4}i}\varphi_{\overline{\mathbf{5}}_{M,-}j}\varphi_{\mathbf{10}_{M,-}}^{ij}}{(x)(y)}\right].
\end{equation}
Again this yields a rank one Yukawa matrix in generation space.  

\subsection{Bestiary}
\label{beasts}

In this section we turn to a collection of examples illustrating some of the novel
phenomena associated with T-branes, and in particular, some of the ways in which a singular spectral equation can miss, or incorrectly predict, the presence of localized matter fields. We also present an example which falsifies some of the assumptions used to claim various constraints on the spectra of F-theory GUTs.

\subsubsection{Nilpotent Matter}
This simplest example where the spectral equation misses a localized mode is to take

\begin{equation}
\Phi=\left(
\begin{array}{cc}
0 & x \\
0 & 0
\end{array}
\right). \label{nilpmat}
\end{equation}
This matrix has a non-trivial spatial variation which is completely invisible to the spectral equation, \(P_{\Phi}(z)=z^{2}\).  Away from \(x=0\) this background breaks \(SU(2)\).  At \(x=0\) the local symmetry group is enhanced and one finds localized matter.  To compute this explicitly we proceed as in section \ref{formalism}.  A localized perturbation at \(x=0\) satisfies the torsion condition
\begin{equation}
x \varphi = \mathrm{ad}_{\Phi}(\eta)
\end{equation}
This equation admits two solutions
\begin{equation}
\varphi_{1}=\left(
\begin{array}{cc}
1 & 0\\
0 & -1
\end{array}
\right),\hspace{.5in} \eta_{1}=\left(
\begin{array}{cc}
0 & 0 \\
1 & 0
\end{array}
\right),
\end{equation}
\begin{equation}
\varphi_{2}=\left(
\begin{array}{cc}
0 & 1\\
0 & 0
\end{array}
\right),\hspace{.5in} \eta_{2}=\left(
\begin{array}{cc}
-\frac{1}{2} & 0 \\
0 & \frac{1}{2}
\end{array}
\right).
\end{equation}
These modes are invisible to the spectral equation.  Developing a detailed theory of this phenomenon of nilpotent T-branes which support localized matter as well as their physical interpretation is an interesting open question.

\subsubsection{Missing Charged Matter}
The previous example illustrates that the spectral equation can miss localized matter.  In that example the background \eqref{nilpmat} completely breaks the symmetry and the missing matter is a neutral localized field.  An even more drastic possibility is that the spectral equation misses a localized \emph{charged} matter field.  An example of this sort is realized by
\begin{equation}
\Phi= \left(
\begin{array}{cc|c}
0 & x & 0 \\
0 & 0 & 0 \\
\hline
0 & 0 & 0
\end{array}
\right).
\end{equation}
This background preserves an unbroken \(U(1)\) gauge symmetry.  If we study the charged doublet perturbation
\begin{equation}
 \left(
\begin{array}{cc|c}
0 & 0 & \varphi_{+} \\
0 & 0 & \varphi_{-} \\
\hline
0 & 0 & 0
\end{array}
\right),
\end{equation}
then we will find localized charged matter invisible to the spectral equation.  The torsion condition for the doublet is
\begin{equation}
x \left(\begin{array}{c} \varphi_{+} \\ \varphi_{-}\end{array}\right) = \left(\begin{array}{cc} 0 & x \\ 0 & 0 \end{array}\right)\left(\begin{array}{c} \eta_{+} \\ \eta_{-}\end{array}\right)
\end{equation}
This is solved by
\begin{equation}
\varphi =\left(\begin{array}{c} 1 \\ 0 \end{array}\right), \hspace{.5in} \eta =\left(\begin{array}{c} 0 \\ 1 \end{array}\right).
\end{equation}
Thus the spectral equation can miss localized charged matter.
\subsubsection{Phantom Curves}
\label{phantom}
The previous example illustrates that the spectral equation can fail to detect a localized charged matter field.  Equally bad, is the fact that the spectral equation can sometimes indicate a matter curve when in fact no localized mode exists.  To demonstrate this let us consider two possible T-branes which describe a Higgsing from \(SU(4)\rightarrow U(1)\)
\begin{equation}
\Phi_{1}=\left(
\begin{array}{ccc|c}
0 & 1 & 0 & 0 \\
0 & 0 & 1 & 0 \\
0 & x & 0 & 0 \\
\hline
0 & 0 & 0 & 0
\end{array}
\right), \hspace{.5in}\Phi_{2}=\left(
\begin{array}{ccc|c}
0 & 1 & 0 & 0 \\
x & 0 & x & 0 \\
0 & 0 & 0 & 0 \\
\hline
0 & 0 & 0 & 0
\end{array}
\right).
\end{equation}
These two Higgs fields preserve the same unbroken gauge symmetry, and have identical spectral equation \(P_{\Phi_{i}}=z^{2}(z^{2}-x)\).  One can see that they are distinct T-brane configurations by noting that at \(x=0\) they have different Jordan decompositions.

Based purely on the spectral equation, one might be tempted to conclude that at \(x=0\), where the two factors of \(P_{\Phi}(z)\) collide one should find localized matter.  To investigate this hypothesis we need only study the torsion equation for localized charged matter in the triplet
\begin{equation}
\left(
\begin{array}{ccc|c}
0 & 0 & 0 & \varphi_{1} \\
0 & 0 & 0 & \varphi_{2} \\
0 & 0 & 0 & \varphi_{3} \\
\hline
0 & 0 & 0 & 0
\end{array}
\right).
\end{equation}
Consider first the background \(\Phi_{1}\).  Using our gauge freedom we can freely set \(\varphi_{1}=\varphi_{2}=0\).  The localization equation then reads
\begin{equation}
x\left(\begin{array}{c}
0 \\
0 \\
\varphi_{3}
\end{array}
\right)=\left(\begin{array}{ccc}
0 & 1 & 0  \\
0 & 0 & 1 \\
0 & x & 0  \\
\end{array}\right)\left(\begin{array}{c}
\eta_{1} \\
\eta_{2} \\
\eta_{3}
\end{array}
\right)
\end{equation}
Which has no solutions.

On the other hand for the background \(\Phi_{2}\) we can reach a gauge where \(\varphi_{1}=0\), and the localization equation reads
\begin{equation}
x\left(\begin{array}{c}
0 \\
\varphi_{2} \\
\varphi_{3}
\end{array}
\right)=\left(\begin{array}{ccc}
0 & 1 & 0  \\
x & 0 & x \\
0 & 0 & 0  \\
\end{array}\right)\left(\begin{array}{c}
\eta_{1} \\
\eta_{2} \\
\eta_{3}
\end{array}
\right)
\end{equation}
This is solved by
\begin{equation}
\varphi=\left(\begin{array}{c}
0 \\
1 \\
0
\end{array}
\right), \hspace{.5in}\eta=\left(\begin{array}{c}
1 \\
0 \\
0
\end{array}
\right).
\end{equation}
Thus for the background \(\Phi_{2}\) the spectral equation indicates correctly that there is matter, while for the background \(\Phi_{1}\) no matter exists at \(x=0\). The basic principle behind these examples is that we have exited the paradigm of reconstructible Higgs fields and the monodromy group is not transitive. This means that at the branch locus \(x=0\) there are multiple Jordan structures of the T-brane which are consistent with the spectral polynomial. The spectral polynomial accurately predicts the localized matter when the Jordan structure is chosen so that the monodromy group is a transitive subgroup of the non-vanishing Jordan block.

\subsubsection{No Correlation Between Representations}

For applications to phenomenology one might also be interested in how the failure of the spectral equation to detect localized matter is correlated across different matter representations. For example, in \cite{Marsano:2009gv} (see also \cite{Dudas:2009hu}) it was found that if one assumes that the spectral equation accurately captures all localized matter, then the homology classes of the matter curves supporting the \(\mathbf{5}\) modes and the \(\mathbf{10}\) modes are correlated.  Activating a hypercharge flux through the Higgs curves to achieve doublet triplet splitting as in \cite{BHVII} we would then find incomplete GUT multiplets descending from the \(\mathbf{10}\) as well.\footnote{As has been noted in previous works (see e.g. \cite{EPOINT}), this result assumes that the entire system can be described globally over a compact $S$ in terms of a single $E_8$ gauge theory. Moreover, one must also assume that there are no ``accidental factorizations'' in the discriminant locus. Neither condition needs to hold in a general model. Here we show that even in a local patch of $S$, the assumptions of \cite{Marsano:2009gv} need not hold.}

At the very least, the previous examples illustrate that a singular spectral equation provides only partial information about the localized matter content of a T-brane configuration. Nevertheless, one might still speculate that whenever the spectral equation fails to detect a localized \(\mathbf{10}\) curve, it also fails to detect a localized \(\mathbf{5}\) curve, so that in any case the matter curves among different representations are still correlated.  To address this latter possibility, we now present an example in which the spectral equation accurately captures the localized \(\mathbf{5}\) matter, but falsely predicts the existence of a localized $\mathbf{10}$ mode. 

Consider a breaking pattern of \(E_{8}\rightarrow SU(5)_{\mathrm{GUT}}\) specified by an \(SU(5)_{\perp}\) Higgs field
\begin{equation}
\Phi=\left(
\begin{array}{ccccc}
0 & 1 & 0 & 0 & 0 \\
0 & 0 & 1 & 0 & 0 \\
0 & 0 & 0 & 1 & 0 \\
0 & 0 & 0 & 0 & 1 \\
0 & 0 & x & 0 & 0
\end{array}
\right).
\end{equation}
Like the example of section \ref{phantom} this Higgs field is not reconstructible.  It has a non-transitive \(\mathbb{Z}_{3}\) monodromy group, but a Jordan structure that is larger than \(3\times 3\).  As explained in section \ref{E8} the \(\mathbf{10}\)'s of the \(SU(5)_{GUT}\) transform as \(\mathbf{5}\)'s under \(SU(5)_{\perp}\) while the \(\overline{\mathbf{5}}\)'s of \(SU(5)_{GUT}\) transform as \(\mathbf{10}\)'s under \(SU(5)_{\perp}\).

Now the spectral equation of \(\Phi\) is:
\begin{equation}
P_{\Phi}(z)=z^{2}(z^{3}-x).
\end{equation}
Based purely on considerations of the spectral equation, one might then be tempted to conclude that there are localized \(\mathbf{10}\)'s when \(x=0\) and the two components of the spectral equation collide.

This is not so.  Like the example of section \ref{phantom}, there is no localized matter for \(\Phi\) acting in the fundamental representation and hence for this representation \(x=0\) is a phantom matter curve.  Meanwhile we can also consider the spectral equation for \(\Phi\) acting in the antisymmetric tensor \(\mathbf{10}\) of \(SU(5)_{\perp}\)
\begin{equation}
P_{\Phi \wedge \Phi }(z)= z(z^{3}+x)(z^{3}-x)^{2}. \label{phiwedgephi}
\end{equation}
This is the relevant spectral equation for charged matter in the \(\overline{\mathbf{5}}\) of \(SU(5)_{GUT}\).
Reasoning based on \eqref{phiwedgephi} one might guess that there is matter localized on the curve \(x=0\).

This guess turns out to be correct.  Suppressing the \(SU(5)_{GUT}\) indices we can write a perturbation \(\varphi\) which transforms in the \(\mathbf{10}\) of \(SU(5)_{\perp}\) as a \(5\times 5\) antisymmetric matrix.  Under a gauge transformation with parameter \(\chi\) the change in the perturbation is
\begin{equation}
\delta \varphi = \Phi \chi +\chi \Phi^{T} .
\end{equation}
To look for localized matter we again study the torsion condition
\begin{equation}
x \varphi = \Phi \eta + \eta \Phi^{T}.
\end{equation}
This admits the solution
\begin{equation}
\varphi= \left(\begin{array}{ccccc}
0 & 0 & 0 & 0 & 0 \\
0 & 0 & 0 & 0 & 1 \\
0 & 0 & 0 & 0 & 0 \\
0 & 0 & 0 & 0 & 0 \\
0 & -1 & 0 & 0 & 0
\end{array}\right), \hspace{.5in} \eta =\left(\begin{array}{ccccc}
0 & 0 & 0 & -1 & 0 \\
0 & 0 & 1 & 0 & 0 \\
0 & -1 & 0 & 0 & 0 \\
1 & 0 & 0 & 0 & 0 \\
0 & 0 & 0 & 0 & 0
\end{array}\right).
\end{equation}
Thus there is localized matter in the \(\mathbf{\bar{5}}\) of \(SU(5)_{GUT}\).  This shows
that the property of being a ``phantom matter curve'' can depend on which representation the Higgs field acts on.

Though this example is clearly not realistic for model building applications, it already shows that from the spectral equation alone, one cannot deduce the homology class of all matter curves, and thus, it is not possible to constrain the spectra of F-theory GUTs, at least using the methods advocated in \cite{Marsano:2009gv}.  To determine whether there are constraints on the matter spectra, it would seem necessary to extend the discussion presented here to a compact $S$. Following \cite{DWIII} one could in principle consider a meromorphic Higgs field, and study the matter content for compact $S$. At some level, there must be some correlation between the matter fields, simply based on various anomaly cancellation considerations in four dimensions, and possibly inflow from higher dimensions. What is not clear is that such constraints must take the form of a relation between the homology classes of matter curves. 

\subsubsection{Pointlike Matter}
\label{points}
The previous examples of this subsection are all intrinsically \(6D\) phenomena.  They involve T-brane configurations which depend only on one coordinate \(x\).  If we study backgrounds which are intrinsically \(4D\) then we find a novel kind of \emph{pointlike} localized matter.  The simplest background which demonstrates this phenomenon is to take
\begin{equation}
\Phi=\left(
\begin{array}{cc}
0 & x \\
y & 0
\end{array}
\right).
\end{equation}
If we proceed naively to study perturbations
\begin{equation}
\varphi=\left(
\begin{array}{cc}
\varphi_{0} & \varphi_{+} \\
\varphi_{-} & -\varphi_{0}
\end{array}
\right).
\end{equation}
then we find that under gauge transformation by a holomorphic \(\chi\) we have
\begin{equation}
\delta \varphi =\left[\left(
\begin{array}{cc}
0 & x \\
y & 0
\end{array}
\right),\left(
\begin{array}{cc}
\chi_{0} & \chi_{+} \\
\chi_{-} & -\chi_{0}
\end{array}
\right) \right]=\left(
\begin{array}{cc}
x\chi_{-}-y\chi_{+} & -2 x\chi_{0} \\
2 y\chi_{0} & y\chi_{+}-x\chi_{-}
\end{array}
\right).
\end{equation}
Thus the gauge invariant data in the perturbation mode \(\varphi_{0}\) is naturally valued in \(\mathcal{O}/\langle x,y\rangle\).  In other words this is a matter mode concentrated at a \emph{point}.  This is quite interesting and deserves to pursued in greater detail. Its existence indicates to us that even working in a small patch, there may still be many strange beasts yet to be discovered.

\section{Conclusions}\label{CONCLUDE}

In this paper we have initiated a study of T-branes, which are bound states of branes characterized by the condition that on some loci the matrices of their normal deformations are upper triangular. We have developed a general formalism for studying the massless matter localized on curves and their associated superpotential couplings. We have also presented a number of examples which explicitly demonstrate that in general such configurations are intrinsically non-abelian and hence are \textit{not} completely captured by the eigenvalues the Higgs field.  These examples themselves deserve further study both to clarify their physical interpretation, and perhaps to make contact with other studies of non-abelian brane physics such as the Myers effect \cite{Myers:1999ps}.

At a practical level, we have seen why the distinction between the eigenvalues of $\Phi$ and the Higgs field itself is so important.  Indeed, using just the data derived from the spectral equation for $\Phi$, some papers have claimed various constraints on the massless spectra of F-theory GUTs. In this paper we have seen that there can be matter curves undetected by the spectral equation, and also no matter curve where the spectral equation would have otherwise indicated one is present. The spectral equation is by itself an incomplete characterization of a theory of seven-branes, and must be supplemented by additional data. It would be quite interesting to consider more realistic models which exploit this additional flexibility in specifying a T-brane configuration.  At the very least, in light of what has been found here, various model building efforts which have relied on singular spectral equations may need to be revisited.

Extending the discussion given in \cite{FGUTSNC}, in this work we have seen that the localized matter content and superpotential interactions of T-branes backgrounds can be characterized in terms of purely holomorphic \textit{algebraic} data.  It would be interesting to consider further deformations to the superpotential, which can be phrased in terms of a holomorphic non-commutative deformation of the geometry \cite{FGUTSNC}. Given our algebraic characterization of matter fields and Yukawa couplings, the extension to this non-commutative case should be straightforward, and likely applies to other non-commutative backgrounds such as those considered recently in \cite{HeckVerlinde}.

Moving beyond particle physics, an important feature of F-theory is that it can be characterized either in terms of open string variables associated with the local gauge theory of seven-branes, or in terms of closed string variables by specifying a compactification on a possibly singular Calabi-Yau fourfold. In this paper we have focussed on the open string description of T-branes. Developing an appropriate closed string description would be quite interesting. For example, this would appear to be a necessary step in coupling such T-brane configurations to gravity.  As we have explained in this paper, a given T-brane fails to be captured by the spectral equation precisely when the spectral surface and hence the Calabi-Yau is singular.  It is thus tempting to speculate that the additional data of a T-brane is encoded in the non-abelian structure of a resolution of singularities.  This would dovetail nicely with previous mathematical studies of nilpotent Higgs fields \cite{Donagi:2003hh}.

Though in this paper we have focussed on T-brane configurations associated with seven-brane gauge theory, the underlying concept and analysis is far more general. The Hitchin-like equations controlling this system should apply quite broadly to branes wrapping complex surfaces probing an ambient normal direction.   Abstractly the dynamics of this gauge theory are described by a topologically twisted version of $\mathcal{N} = 4$ gauge theory in four real dimensions considered in \cite{VafaWitten}.  Localized matter on curves is then a kind of topological surface defect, and our discussion is, at its core, a theory of these defects.  We expect that the idea of T-branes could be applied to many other situations encountered in string theory, and can be extended to different dimensionalities of branes with various amounts of supersymmetry. We hope that this paper will serve as an appetizer for future exploration and application of T-branes in string theory.

\section*{Acknowledgments}
The authors thank F. Denef, Y. Tachikawa, B. Wecht, M. Wijnholt and E. Witten for helpful discussions. In
addition we thank the Eighth Simons Workshop in Mathematics and Physics, where a portion of this work was completed, for hospitality and providing a stimulating environment. JJH thanks the Harvard high energy theory group for hospitality during part of this work. CV thanks the CTP at MIT for hospitality during his sabbatical leave. The work of CV is supported by NSF grant PHY-0244821. The work of JJH is supported by NSF grant PHY-0969448.
\appendix
\section{Classification of Reconstructible Higgs Fields}
\label{classgen}
The goal of this section is to prove the theorem quoted in section \ref{generic} on the uniqueness of reconstructible Higgs fields with a fixed spectral polynomial.  We continue to use the notation \(\mathcal{O}\) for the ring of holomorphic power series in two variables \(x\) and \(y\), and we denote by \(m\subset \mathcal{O}\) the maximal ideal of functions which vanish at the origin.  All matrices and functions are holomorphic unless otherwise stated.
\\
\\
\textbf{Theorem}:  Let \(\Phi\) be an \(n\times n\) matrix with spectral equation:
\begin{equation}
P_{\Phi}(z)=z^{n}+\sigma_{2}z^{n-2}-\sigma_{3}z^{n-3}+\cdots+(-1)^{n}\sigma_{n}.
\end{equation}
Assume as in section \ref{generic} that \(\Phi\) is reconstructible in the sense that $P_{\Phi}(z) = 0$ is a non-singular surface in $\mathbb{C}^{2}$. Further, assume that all of the eigenvalues of \(\Phi\) vanish at the origin.  Then up to conjugation (i.e. holomorphic gauge transformation) we have:
\begin{equation}
\Phi=\left(
\begin{array}{cccccc}
0 & 1 & 0 & \cdots & 0 & 0 \\
0 & 0 & 1& \cdots & 0 & 0 \\
\vdots & \vdots & \vdots & \ddots & \vdots & \vdots \\
0 & 0 & 0 & \cdots & 0 & 1 \\
(-1)^{n-1}\sigma_{n} & (-1)^{n-2}\sigma_{n-1} & (-1)^{n-3}\sigma_{n-2} & \cdots & -\sigma_{2} & 0
\end{array}
\right).
\end{equation}
\\
\\
\textbf{Proof}:  We show by induction that for each \(k\geq0\) we can put \(\Phi\) in the desired form up to terms of order \(m^{k}\).  For the case \(k=0\) we need to show that \(\Phi\) evaluated at the origin can be conjugated to the form asserted in the theorem.  By standard linear algebra the constant matrix \(\Phi |_{(0,0)}\) can be put in Jordan normal form.  By assumption all the eigenvalues vanish at the origin so \(\Phi |_{(0,0)}\) is a pure Jordan block \(J\), its non-vanishing entries are some number of ones on the superdiagonal.  The expansion of \(\Phi\) near the origin is then
\begin{equation}
\Phi=J+\phi_{1}. \label{taylorphi}
\end{equation}
Where in equation (\ref{taylorphi}) the matrix \(\phi_{1}\) vanishes at the origin so that \(\phi_{1}\in m\).  Take the determinant of equation \((\ref{taylorphi})\).  Since the Higgs field is reconstructible we know that \(\det(\Phi)\notin m^{2}\).  Thus since each entry of \(\phi_{1}\) is in \(m\) it must be that the Jordan block \(J\) has maximal length
\begin{equation}
J=\left(
\begin{array}{cccccc}
0 & 1 & 0 & \cdots & 0 & 0 \\
0 & 0 & 1 & \cdots & 0 & 0 \\
\vdots & \vdots & \vdots & \ddots & \vdots & \vdots \\
0 & 0 & 0 & \cdots & 0 & 1 \\
0 & 0 & 0 & \cdots & 0 & 0
\end{array}
\right).
\end{equation}
Thus for \(k=0\) we are done.

Now we proceed to the inductive step.  Assume that we have reached a gauge where the series expansion of \(\Phi\) at the origin takes the form:
\begin{equation}
\Phi=J+C+\phi_{k}.
\end{equation}
In the above the matrix \(C\in m\) has non-vanishing entries only along the bottom row, while \(\phi_{k} \in m^{k}\) is the order \(k\) discrepancy from the desired form.  To proceed we need the following lemma:
\\
\\
\textbf{Lemma:} Let \(\phi\) be any \(n\times n\) matrix.  Then there exists a matrix \(\chi\) such that \([J,\chi]-\phi\) is zero except in the last row.
\\
\\
\textbf{Proof:}  Direct computation.
\\
\\
Now we are essentially done.  Using the claim we choose \(\chi\) such that
\begin{equation}
[J,\chi]-\phi_{k}
\end{equation}
vanishes outside the bottom row.  Notice that since \(\phi_{k}\in m^{k}\) and and all the non-vanishing entries of \(J\) are not in \(m\) we may take \(\chi \in m^{k}\).  Now perform a gauge transformation by \(e^{\chi}\).  We have:
\begin{eqnarray}
\Phi  \longrightarrow e^{\chi}\Phi e^{-\chi} & = & J+C+\phi_{k} - [J,\chi] +\cdots \\
& = & J + C' +\phi_{k+1}
\end{eqnarray}
In the above the matrix \(C'\) denotes a new matrix in \(m\) with non-vanishing entries only in the last row, while the discrepancy \(\phi_{k+1}\) is now in \(m^{k+1}\).  This completes the inductive step and proves the theorem.

\section{Non--Degeneracy of the $6D$ Superpotential}
\label{pairing}

Applying the formulae of section \ref{TheMatterCurveAction} to a variety of explicit examples, we always find the quadratic part of the $6D$ superpotential to have the form
of the $2D$ chiral Dirac theory coupled to a connection on some vector bundle, that is
\begin{equation}\label{action6DDirac}
 W_{6D\: \mathrm{quad.}}=\int_\Sigma \Omega^{ij}\; \phi_i\:\overline{\partial}_{V_j}\,\phi_j,
\end{equation}
where the $\phi_i$'s are the $6D$ fields which transform as sections of the vector bundles $V_i$.
In equation \eqref{action6DDirac}, $\Omega^{ij}$ is a \emph{non--degenerate} symplectic pairing satisfying the selection rule of equation \eqref{omegaselrule}.

The non--degeneracy of the pairing $\Omega^{ij}$ is required if the $6D$ theory is to define a non-singular field theory. In a sense, this is physically obvious since our $6D$ theory is embedded in a consistent model, namely F-theory. However, it is desirable to have a general mathematical proof of this crucial fact as a non--trivial check of the entire circle of ideas. In this Appendix we prove in full generality the non--degeneracy of $\Omega^{ij}$ in the vicinity of a `good' point of  the matter curve, that is a smooth point in  $\Sigma$ which is away from the intersection points with other matter curves and point-like defects.

Since the argument is a bit technical, let us first explain the underlying idea in plain English. Let $\varphi_i$, $i=1,2$ be adjoint valued holomorphic $(2,0)$--forms corresponding to $6D$ modes localized on the same smooth curve $\Sigma$ of (local) equation $f=0$, which satisfy $f\,\varphi_i =[\Phi,\eta_i]$ for certain holomorphic sections
$\eta_i$ of $\mathrm{ad}(P)\otimes\mathcal{O}_S(\Sigma)$. We write $\chi_i$ for the (adjoint valued) $(1,0)$--form on the matter curve
$\Sigma$ given by the Poincar\'e residue
\begin{equation}
 \chi_i = \text{Poincar\'e Residue of }\  \frac{\varphi_i}{f}.
\end{equation}
Extending the modes \emph{off--shell} by replacing the holomorphic section with smooth sections of the sheaves of $\mathcal{A}^\infty$--modules generated by the
$\eta_i\big|_\Sigma$'s and $\chi_i$'s, the formulae of section
\ref{TheMatterCurveAction} give for the pairing of the two modes
\begin{equation}\label{paring3}
 \int_\Sigma \mathrm{Tr}\Big(\chi_1\:\overline{\partial}_{V_2}\,\eta_2\Big).
\end{equation}
The $6D$ fields $\eta_i$ and $\chi_i$ are not independent; the $\chi_i$'s are linear functions of the corresponding $\eta_i$'s.

The pairing being non--degenerate means that, given a localized zero--mode $\varphi_1$, inducing the $6D$ field $\chi_1$, we may find a localized zero--mode $\varphi_2$, which induces a $6D$ field $\eta_2$ such that the pairing in equation \eqref{paring3} is not zero. In terms of adjoint representation matrices, this amounts to  $\mathrm{Tr}(\chi_1\,\eta_2)\neq 0$. The obvious strategy to show this is to take a matrix $\eta_2$ such that
$\mathrm{Tr}(\chi_1\, \eta_2)\neq 0$, which always exists, and identify the $8D$ mode $\varphi_2$ with $[{\Phi},\eta_2/f]$.
However such a $\varphi_2$ would be a valid $8D$ zero--mode only if $[{\Phi},\eta_2/f]$ has no pole along the curve $f=0$. For generic $\eta_2$ satisfying $\mathrm{Tr}(\chi_1\, \eta_2)\neq 0$, we get indeed a pole. So, what one really has to show is that there is one choice of the matrix $\eta_2$ such that
$\mathrm{Tr}(\chi_1\, \eta_2)\neq 0$ while the pole in $[{\Phi},\eta_2/f]$ cancels.
In order to do that, one filters the sheaf $\mathrm{ad}_\Phi(\mathfrak{g}\otimes \mathcal{O}_S)$
according to the order of zero along the curve $f=0$; in this ways one checks that there are enough holomorphic matrices
of the form $[\Phi,\eta]$ which are divisible by $f^k$ to pair up all the localized zero--modes.
\medskip

Since we aim to prove a local result near a `good' point, we may as well take $S=\mathbb{C}^2$ and the trivial gauge bundle
 $\mathrm{ad}(P)\simeq
\mathfrak{g}\otimes\mathcal{O}$. The $(2,0)$ forms are then identified with the scalars.  We assume the Higgs background to be such that $x=0$ is a matter curve, and invariant by translation in the $y$ direction. So
$\Phi(x)$ is a $N\times N$ traceless matrix depending holomorphically on  $x$.
We write $\mathbb{M}(N,K)$ for the space of $N\times N$ matrices with coefficients in a ring $K$. Then, from equation \eqref{quo},
\begin{equation}
\text{zero modes}=\mathbb{M}(N,\mathcal{O})\Big/ \mathrm{ad}_\Phi\:\mathbb{M}(N,\mathcal{O})\equiv \mathcal{Q}.
\end{equation}
The mode represented by the matrix $\Upsilon\in \mathbb{M}(N,\mathcal{O})$ is localized on the line $x=0$ iff there is a {positive integer} $\ell$ such that the matrix
\begin{equation}
x^\ell\,\Upsilon\in \mathrm{ad}_\Phi\:\mathbb{M}(N,\mathcal{O}).
\end{equation}
We call the elements of the subspace $\mathfrak{F}_\ell\equiv \ker\{ \mathcal{Q}\xrightarrow{\ x^\ell\ } \mathcal{Q}\}$ the localized zero--modes (on $x=0$) \textit{having weight} $\leq\ell$. These are \emph{on shell} (holomorphic) modes. Replacing $\mathcal{O}$ with the ring $R$ of the functions $f(x,y)$ depending holomorphically on $x$ and smoothly on $y$ we get the \emph{off--shell} modes (in the $6D$ sense).

We have $x^j\,\mathfrak{F}_k\subset \mathfrak{F}_{k-j}$; consequently we define the weight $\ell$  \emph{primary} modes as
the elements of the coset
 \begin{equation*}
 \mathfrak{F}_\ell^\mathrm{prim}:= \mathfrak{F}_\ell\Big/ x\, \mathfrak{F}_{\ell-1}.
 \end{equation*}
 As $\mathbb{C}$--spaces, $\mathfrak{F}_\ell=\bigoplus_{r\geq 0}\: x^r\,\mathfrak{F}_{\ell+r}^\mathrm{prim}$.
 The elements of the subspaces $x^r\,\mathfrak{F}^\mathrm{prim}_{\ell+r}$ with $r\geq 1$ are called descendent modes.
 Everything is determined just by the {primary} modes.
Indeed,
 let $\chi$ be a matrix representing a primary weight $\ell$  mode.
A representative of the full set of its descendent is given by the matrix
 \begin{equation}\label{descendents}
 \chi(x,y)_\mathrm{descendents}= \Big(\phi_0(y)+\phi_1(y)\, x+ \phi_2(y)\, x^2+\cdots + \phi_{\ell-1}(y)\, x^{\ell-1}\Big)\chi(x,y)_\mathrm{primary}.
 \end{equation}
 where the $\phi_k(y)$ are (scalar) smooth functions, namely the $6D$ fields.

As a matter of convention, we extend the concept of weight to
 non--localized modes by stating that they have weight $\infty$. Indeed, in these cases, the polynomial in $x$ of degree $\ell-1$ of equation
 \eqref{descendents} is replaced by an infinite power series.
 We also extend the notion of weight to the pure gauge modes by giving them weight \emph{zero}. Indeed, if $\chi$ is pure gauge, $x^0\cdot \chi$ is already zero in the coset $\mathcal{Q}$.

Saying that $\chi$ represents a \emph{primary} weight $\ell$ mode is equivalent to the existence of an element
$\eta\in \mathbb{M}(N,R)$ such that
\begin{equation}\label{defw1}
x^\ell\:\chi =\big[\Phi(x),\eta\big]
\end{equation}
while for \emph{all} $\eta^\prime\in \mathbb{M}(N,R)$
\begin{equation}\label{defw2}
x^{\ell-1}\:\chi \not=\big[\Phi(x),\eta^\prime\big].
\end{equation}
In particular, a primary mode satisfies $\chi\big|_{x=0}\not \equiv 0$,
since otherwise we may write it as $\chi= x\cdot (\chi/x)$, \textit{i.e.}\! as a descendent of the (regular) mode $\chi/x$. Thus the map
$\mathfrak{F}_\ell^\mathrm{prim}\rightarrow \mathbb{M}(N,\mathbb{C})$ given by $\chi\mapsto \chi|_{x=0}$ is injective.
From this observation it follows that we may \emph{choose} the representative matrices of a basis of $\mathfrak{F}_\ell^\mathrm{prim}$ in such a way that the corresponding $\chi$'s are $x$--independent (just replace $\chi$ by its image $\chi\big|_{x=0}$\,). We call this \textit{representative I}. Dually, as discussed in section  \ref{local6d}, for primaries the map $\eta\mapsto \eta\big|_{x=0}$ is also injective, and  we may choose the representatives in such a way that the $\eta$'s are $x$--independent. We call this \emph{representative II}. All other representatives differ by terms vanishing as $x\rightarrow 0$.

Given two {primary} modes with representative matrices $\chi^{(1)}$ and $\chi^{(2)}$
of weights (respectively) $\ell_1$ and $\ell_2$ we introduce an $x$--independent pairing
\begin{equation}
\langle \chi^{(1)}\,|\, \chi^{(2)}\rangle :=\mathrm{Tr}\Big[\eta^{(1)}_{II}\,\chi^{(2)}_{I}\Big].
\end{equation}
where the subscript $I$ or $II$ stand for the two special choices of representatives defined above.
\medskip

\textbf{Lemma.} If $\ell_1>\ell_2$, $\langle \chi^{(1)}\,|\, \chi^{(2)}\rangle=0$.
\vglue 9pt
\noindent Indeed, by the cyclic property of the trace and the definition of $\eta$'s in terms of the $\chi$'s,
\begin{equation}\label{anti}
\mathrm{Tr}\Big[\eta^{(1)}_{II}\: \chi^{(2)}_{I}\Big] = - x^{\ell_1-\ell_2}\:
\mathrm{Tr}\Big[\eta^{(2)}_{I}\: \chi^{(1)}_{II}\Big]=
- x^{\ell_1-\ell_2}\:
\Big(\mathrm{Tr}\Big[\eta^{(2)}_{II}\: \chi^{(1)}_{I}\Big]+O(x)\Big).
\end{equation}
Since the \textsc{lhs} is independent of $x$, it is identically zero.
\medskip

We write  convenient representatives\footnote{\ Being an F-term the $6D$ action pairing 
is independent of the choice of representatives; this is crucial for our argument.} 
of the full set of descendent modes of the two primary $\chi^{(1)}$ and $\chi^{(2)}$,
\begin{align}
\eta^{(1)}_\mathrm{desc}&=
\Big(\phi_0^{(1)}(y)+x\, \phi_1^{(1)}(y)+\cdots +x^{\ell_1-1}\,\phi^{(1)}_{\ell_1-1}(y)\Big)\eta^{(1)}_I\label{liftI}\\
\chi^{(2)}_\mathrm{desc}&=
\Big(\phi_0^{(2)}(y)+x\, \phi_1^{(2)}(y)+\cdots +x^{\ell_2-1}\,\phi^{(2)}_{\ell_2-1}(y)\Big)\chi^{(2)}_{II},\label{liftII}
\end{align}
where the $\phi_k^{(a)}(y)$'s are the independent $6D$ fields.

Consider the $6D$ kinetic pairing of these modes
\begin{equation}
\int dy\, d\bar y \:\oint
 \frac{\mathrm{Tr}\big[\,\eta^{(1)}_\mathrm{desc}\, \partial_{\bar y} \,\chi^{(2)}_\mathrm{desc}\,\big]}{x^{\ell_1}}\: dx
\end{equation}
It has two properties:
\begin{enumerate}
\item it is symmetric under $(1)\leftrightarrow (2)$ since it can written (choosing, say, representatives $II$ for both sets of modes) in the manifestly symmetric form
\begin{equation}
\int dy\, d\bar y \:\oint
 \frac{\mathrm{Tr}\big\{\,\eta^{(1)}_\mathrm{desc}\cdot \mathrm{Ad}(\Phi)\cdot \partial_{\bar y} \,\eta^{(2)}_\mathrm{desc}\,\big\}}{x^{\ell_1}\: x^{\ell_2}}\: dx
\end{equation}
($\mathrm{Ad}(\Phi)$ is antisymmetric and holomorphic).

\item It is proportional to the pairing of the corresponding primaries.
Indeed choosing representatives as in eqns.\eqref{liftI}\eqref{liftII}, it is equal to
\begin{equation}\label{exexp}
\int dy\, d\bar y \;\langle \chi^{(1)}\,|\,\chi^{(2)}\rangle\:\oint
 \frac{\sum_{k=0}^{\ell_1-1}x^k\, \phi_k^{(1)}\, \partial_{\bar y} \,\sum_{j=0}^{\ell_1-1}x^j\, \phi_k^{(2)}}{x^{\ell_1}}\: dx.
\end{equation}
\end{enumerate}
\smallskip

Putting together these two properties and the \textbf{lemma}, we see that, with respect to the $6D$ kinetic pairing, modes descending from primaries of different weights are \emph{orthogonal}.

Now we are ready to show that the $6D$ action pairing is non--degenerate.
With our conventions about the weight of pure gauge modes ($\ell=0$) and non--localized ones ($\ell=\infty$) we have a complete direct sum decomposition
\begin{equation}
\mathbb{M}(N,\mathbb{C})=\bigoplus_{\ell=0}^\infty  X_\ell
\end{equation}
such that
\begin{equation}
\chi_{I}\in X_\ell\quad\Rightarrow\quad \chi_{I}\ \text{is a primitive mode of weight } \ell.
\end{equation}

Likewise, we have a second direct sum decomposition
\begin{equation}
\mathbb{M}(N,\mathbb{C})=
\left(\bigcap_{x\in \mathbb{C}^2} \ker \mathrm{ad}\,\Phi(x)\right)\oplus\left(\bigoplus_{\ell=0}^\mathrm{finite}  H_\ell\right)
\end{equation}
such that
\begin{equation}
\eta_{II}\in H_\ell\quad\Rightarrow\quad x^{-\ell}\,[\Phi(x),\eta_{II}]\ \text{is a primitive mode of weight } \ell.
\end{equation}
We set
\begin{equation}
H_\infty=
\bigcap_{x\in \mathbb{C}^2} \ker \mathrm{ad}\,\Phi(x).
\end{equation}
Then
\begin{equation}\label{comdim}
\# \,\text{primitive modes of weight }\ell=\dim X_\ell=\dim H_\ell,\quad \ell=0,1,2,\dots, \infty.
\end{equation}

Let $\mathrm{Kill}$ be the bilinear form on $\mathbb{M}(N,\mathbb{C})$ defined by the trace. It is a \emph{non--degenerate} pairing. Let
\begin{equation}
\mathrm{Kill}_{\ell_1,\ell_2}\colon H_{\ell_1}\otimes X_{\ell_2}\rightarrow \mathbb{C},
\end{equation}
be the bilinear form induced by $\mathrm{Kill}$ under restriction,
which we rewrite as
\begin{equation}
\mathrm{Kill}^\vee_{\ell_1,\ell_2}\colon X_{\ell_2}\rightarrow H_{\ell_1}^*.
\end{equation}
The \textbf{Lemma} may be rephrased as the statement
\begin{equation}
\mathrm{Kill}^\vee_{\ell_1,\ell_2}\equiv 0\quad \text{if } \ell_1>\ell_2.\label{leema}
\end{equation}
Consider the map
\begin{equation}
\mathrm{Kill}^\vee_{\bigstar,\infty}\colon
 X_\infty\rightarrow \left(\bigoplus_{m=0}^\infty H_m\right)^{\!\bigstar}.
\end{equation}
Since the Killing form is non--degenerate, this map is an isomorphism on its image. By \eqref{leema} the image is contained in $H_\infty^*$ and then, by comparing dimensions via equation \eqref{comdim}, it is $H_\infty^*$.

Consider next the Killing map of the filtration at level $\ell$
\begin{equation}
\mathrm{Kill}^\vee_{\bigstar,\geq \ell}\colon
\left(\bigoplus_{k\geq \ell} X_k\right)\rightarrow \left(\bigoplus_{m=0}^\infty H_m\right)^{\!\bigstar}.
\end{equation}
Again, it is an isomorphism on its image. By \eqref{leema} the image is contained in
\begin{equation}\label{whichsapaceL}
\bigoplus_{m\geq \ell}H_m^\ast.
\end{equation}
Comparing dimension with the help of \eqref{comdim}, we see that the image is equal to the space \eqref{whichsapaceL}.
Thus,
comparing the different $\ell$'s, we conclude that for all $\ell$'s, the map
\begin{equation}
\mathrm{Kill}^\vee_{\ell,\ell}\colon
X_\ell\rightarrow H_\ell^*,
\end{equation}
is an isomorphism.
This is equivalent to the statement that, in the space of primitive modes of
weight $\ell$, the primary pairing
\begin{equation}
\omega_{ab}^{(\ell)}:=\langle\, \chi^{(1)}\,|\, \chi^{(2)}\rangle,
\end{equation}
is non--degenerate and hence, by equation
\eqref{anti} \emph{antisymmetric}. For $\Phi(x)$ independent of $y$, $\omega_{ab}^{(\ell)}$ is just a constant symplectic matrix.

Using equation \eqref{exexp} we get the final formula for $W_{6D}$ (quadratic part)
\begin{equation}\label{W6Dfinal}
\int_\Sigma dy\, d\bar y\;\sum_\ell\sum_{k+j=\ell-1}\omega_{ab}^{(\ell)}\:
\phi^{(a)}_k \:\partial_{\bar z}\, \phi^{(b)}_j,
\end{equation}
which is the formula it was to be shown with the symplectic pairing
\begin{equation}
\Omega =\sum_\ell\omega^{(\ell)} \otimes S_\ell, \qquad\text{where }\  (S_\ell)_{ij}=\delta_{i,\ell-j-1}.
\end{equation}

The above argument shows that the pairing is perfect when the gauge Lie
algebra is $\mathfrak{u}(N)$. The general case is easily reduced to this
one: let the gauge algebra be $\mathfrak{g}=\mathfrak{a}\oplus
\mathfrak{s}_1\oplus \cdots \oplus \mathfrak{s}_k$ with $\mathfrak{a}$
Abelian and $\mathfrak{s}_j$ simple.
In the Abelian part there are no localized zero--modes, while the
localized zero--modes arising from the adjoint of $\mathfrak{s}_j$ pair
between themselves. So it is enough to consider the case $\mathfrak{g}$
simple. Let $R\colon \mathfrak{g}\rightarrow \mathfrak{su}(N)$ be any
faithful representation. The traces of the $N\times N$ matrices reproduce
(up to normalization) the Killing form of $\mathfrak{g}$ which is
non--degenerate. Then defining the spaces of $N\times N$ matrices $X_\ell$
and $H_\ell$ as before, we have the decompositions
\begin{equation}
\mathfrak{g}=\bigoplus_\ell \Big(X_\ell\cap \mathfrak{g}\Big)=
\bigoplus_\ell \Big(H_\ell\cap \mathfrak{g}\Big),
\end{equation}
while the fact that the Killing form of $\mathfrak{su}(N)$ restricted to
$\mathfrak{g}$ is non--degenerate implies that the restricted map
\begin{equation}
\mathrm{Kill}^\vee_{\ell,\ell}\colon
X_\ell\cap \mathfrak{g}\rightarrow \Big(H_\ell\cap\mathfrak{g}\Big)^{\!*},
\end{equation}
is still an isomorphism. Thus the pairing $\Omega$ remains non--degenerate
when restricted to the subspace $\mathfrak{g}$.

\section{The General Residue Formula for the Yukawa}
\label{resformgen}

In this Appendix we prove the residue formula \eqref{resideufromfin}.
In reference \cite{FGUTSNC} the residue formula was proven under two assumptions: \textit{i)} $\mathrm{ad}_\Phi$ is diagonal, and
\textit{ii)} the matter curves meet transversely. Dropping assumption \textit{i)} the argument of \cite{FGUTSNC} would still apply, whereas the generalization to non--transverse crossing requires a bit more work.

Physically one expects that the residue formula for the Yukawa holds whenever it is a mathematically well-defined expression, namely when the curves meet at an isolated point $p$, whatever their local intersection number is. The curves are simply required not to have \emph{components} in common. This generalization is relevant since in the presence of monodromy, typically the matter curves do not meet transversely.
\medskip

Our setting is the following: We have three zero--modes of the $8D$ YM theory on $\mathbb{R}^{3,1} \times \mathbb{C}^2$, $\Upsilon_i\equiv (a_i,\varphi_i)$, $i=1,2,3$, and the first two modes are assumed to be localized, respectively, on the divisor $f_1=0$ and $f_2=0$. These divisors are {not assumed} to be (necessarily) {prime} (they may have several irreducible components and/or multiplicities $> 1$), but we assume that their \emph{set--theoretical} intersection
\begin{equation*}
 \{f_1=0\}\cap \{f_2=0\} \in \mathbb{C}^2
\end{equation*}
consists of just one point which we identify with the origin $0\in\mathbb{C}^2$. We stress that their analytic intersection number at $0$ may be any (positive) integer, since the intersection is \underline{not} assumed to be transverse. We make no assumption on the third mode $\Upsilon_3$.

\smallskip

Following \cite{FGUTSNC}, the Yukawa coupling of the three modes is given by the cohomology invariant
\begin{equation}
 \mathrm{Yuk}= \int_{\mathbb{C}^2} \mathbf{yuk},
\end{equation}
where the the $(2,2)$ form $\mathbf{yuk}$ has the expression (conventions as in \cite{FGUTSNC})
\begin{equation}
 \mathbf{yuk}= \frac{1}{2}\mathrm{Tr}\Bigg(\Upsilon_1\wedge \Upsilon_2\wedge \Upsilon_3\Big|_{(2,2)}+
\Upsilon_2\wedge \Upsilon_1\wedge \Upsilon_3\Big|_{(2,2)}\Bigg).
\end{equation}
Notice that this expression is automatically invariant under permutations of the three zero--modes.

As discussed in the main body of the paper, we may choose representatives so that the two localized modes, $\Upsilon_1$ and $\Upsilon_2$, have support in some tiny neighborhood of the respective curves. Then the products $\Upsilon_1\wedge \Upsilon_2$ and
$\Upsilon_2\wedge \Upsilon_1$ will have support inside some ball $B(0,r)$ of radius $r$ centered at $0$. As illustrated in the paper, we may also take a representative of the third mode $\Upsilon_3$ of pure type $(2,0)$ (that is, we work in the gauge $a_3=0$)
\begin{equation}
 \Upsilon_3= (0,\varphi_3)\qquad \varphi_3\in \Gamma(\mathbb{C}^2, \mathrm{ad}(P)\otimes \Omega^2).
\end{equation}

With the above choice of representatives,
\begin{gather}\label{supp}
 \mathrm{supp}\, \mathbf{yuk}\subset B(0,r)\\
\intertext{where}
\mathbf{yuk}= \frac{1}{2}\, \mathrm{Tr}\Bigg\{ \Bigg(\Upsilon_1\Big|_{(0,1)}\wedge \Upsilon_2\Big|_{(0,1)}+
\Upsilon_2\Big|_{(0,1)}\wedge \Upsilon_1\Big|_{(0,1)}\Bigg)\wedge \varphi_3\Bigg\}.
\end{gather}

 Since $\mathbf{yuk}$ is a $(2,2)$--form, by the $\overline{\partial}$--Poincar\'e theorem there exists in $\mathbb{C}^2$ a smooth $(2,1)$ form
$\boldsymbol{\beta}$ such that
\begin{equation}
 \mathbf{yuk}= \overline{\partial}\,\boldsymbol{\beta}.
\end{equation}
Let $R> r$, and write $S_R$ for the sphere of radius $R$ centered at $0$. By equation \eqref{supp},
\begin{equation}\label{yuksph}
 \mathrm{Yuk}= \int_{\mathbb{C}^2}\mathbf{yuk}=\int_{B(0,R)}\mathbf{yuk}=\int_{S_R}\boldsymbol{\beta}.
\end{equation}

Let $U=\mathbb{C}^2\setminus B(0,r)$. One has $S_R\subset U$. Again by equation \eqref{supp}, one has
\begin{equation}
 \overline{\partial}\,\boldsymbol{\beta}\big|_U=0,
\end{equation}
so $\boldsymbol{\beta}$ represents a class
in $H^{2,1}_{\overline{\partial}}(U)$. Consider the \emph{trace map} $H^{2,1}_{\overline{\partial}}(U)\rightarrow \mathbb{C}$
induced by the sequence of natural maps\footnote{\ Compare reference \cite{Griffiths} page 651.}
\begin{equation}\label{chiso}
 H^{2,1}_{\overline{\partial}}(U) \rightarrow H^3_{dR}(U) \simeq H^3_{dR}(S_R) \simeq \mathbb{C},
\end{equation}
given explicitly by $\alpha\mapsto \int_{S_R}\alpha$.
From equation \eqref{yuksph}, the Yukawa coupling is just the image of $\boldsymbol{\beta}$ under this
trace map.

The next step is to exploit the \v Cech--Dolbeault isomorphism
\begin{equation}\label{chekdr}
 H^{2,1}_{\overline{\partial}}(U) \simeq \check H^1(U, \Omega^2).
\end{equation}
 Following reference \cite{Griffiths} we introduce the following open cover of $U$
\begin{gather}\label{cover}
 U= U_1\cup U_2,\\
 U_i =\{ z\in U,\ |f_i(z)|\geq \epsilon\},
\end{gather}
where $\epsilon$ is chosen small enough (or, alternatively, $R$ big enough) so that \eqref{cover} holds.

Since we have only two open sets in our cover, a \v Cech one--cochain $C^1(U, \Omega^2)$ is just a holomorphic $(2,0)$ form
\begin{equation}
 \mathfrak{h} \in \Gamma( U_1\cap U_2, \Omega^2),
\end{equation}
which is automatically a cocycle $\delta\mathfrak{h}=0$.

The isomorphism \eqref{chekdr} works explicitly as follows. Let $\gamma\equiv\{\gamma_i\in C^\infty(U_i, \Omega^{2,0})\}$ be a $0$--cochain such that\footnote{\ $\gamma$ exists, since $C^\infty$ sheaves are fine, i.e they admit partitions of unity.}
\begin{equation}\label{cechDal}
  \mathfrak{h} = \delta \gamma.
\end{equation}
One has $\delta\,{\overline{\partial}}\,\gamma=0$. Hence the local $(2,1)$--forms ${\overline{\partial}}\,\gamma_i$'s glue in a global ${\overline{\partial}}$--closed $(2,1)$--form
$\boldsymbol{\beta}$ which represents the $H^{2,1}_{\overline{\partial}}(U)$ class corresponding to the $\check H^1(U, \Omega^2)$ class $\mathfrak{h}$.

One has
\begin{equation}
 \mathbf{yuk}\Big|_{\mathbb{C}^2\setminus\{ f_i=0\}} = {\overline{\partial}}\, \beta_i
\end{equation}
where
\begin{align}\label{beta1beta2}
 &\beta_1 = \frac{1}{2}\: \varrho_1\:\mathrm{Tr}\left\{\left(\frac{\eta_1}{f_1}\:a_2-a_2\: \frac{\eta_1}{f_1}\right)\wedge \varphi_3\right\}\\
&\beta_2 = \frac{1}{2}\: \varrho_2\:\mathrm{Tr}\left\{\left(\frac{\eta_2}{f_2}\:a_1-a_1\: \frac{\eta_2}{f_2}\right)\wedge \varphi_3\Big)\right\}
\end{align}
and the $\eta_i$ are defined as in the main body of the paper
(\textit{i.e.}\! $f_i\, \varphi_i = [\Phi, \eta_i]$). In equation \eqref{beta1beta2}, $\varrho_i$ is some smooth function which is zero for $|f_i|< \epsilon/2$ and $1$ for $|f_i|>\epsilon$. The difference $\beta_1-\beta_2$ is the $\overline{\partial}$ of a globally defined form $\sigma$ which is easy to write explicitly.

 One has
\begin{equation}
 \beta_1\Big|_{U_1}= \overline{\partial}\left(\frac{1}{2}\, \varrho_2\wedge \mathrm{Tr}\left\{\frac{\big[\eta_1,\,\eta_2\big]}{f_1\: f_2}\wedge \varphi_3\right\}\right)
\end{equation}
\begin{equation}
 \beta_2\Big|_{U_2}= -\overline{\partial}\left(\frac{1}{2}\, \varrho_1\wedge \mathrm{Tr}\left\{\frac{\big[\eta_1,\,\eta_2\big]}{f_1\: f_2}\wedge \varphi_3\right\}\right)
\end{equation}
and
\begin{equation}
 \Big(\beta_1-\beta_2\Big)\Big|_{U_1\cap U_2}=0
\end{equation}
since $\overline{\partial}\,\varrho_i\big|_{U_i}=0$ (in fact $\varrho_i =1$ in $U_i$).\smallskip

Comparing with the explicit form of the \v Cech--Dolbeault isomorphism illustrated around equation \eqref{cechDal}, we get the identifications
\begin{equation}
 \gamma_1= \frac{1}{2}\, \varrho_2\wedge \mathrm{Tr}\left\{\frac{\big[\eta_1,\,\eta_2\big]}{f_1\: f_2}\wedge \varphi_3\right\}\in C^\infty(U_1, \Omega^{2,0})
\end{equation}
\begin{equation}
\gamma_2= -\frac{1}{2}\, \varrho_1\wedge \mathrm{Tr}\left\{\frac{\big[\eta_1,\,\eta_2\big]}{f_1\: f_2}\wedge \varphi_3\right\}\in C^\infty(U_2, \Omega^{2,0})
\end{equation}
This is the \v Cech $0$--cochain we are looking for. Then the \v Cech $1$--cocycle $\mathfrak{h}$, corresponding to the
Dolbeault class $\boldsymbol{\beta}\in H^{2,1}_{\overline{\partial}}(U)$ is
\begin{equation}\label{whath}
 \mathfrak{h}\equiv \mathrm{Tr}\left\{\frac{\big[\eta_1,\,\eta_2\big]}{f_1\: f_2}\: \varphi_3\right\}\in \Gamma(U_1\cap U_2, \Omega^2).
\end{equation}

 Finally, we are in the position to apply the \textbf{Lemma} on page 651 of \cite{Griffiths},
which in the present notations states
\begin{equation}
 \int_{S_R}\boldsymbol{\beta} = \frac{1}{(2\pi i)^2}\int_\Gamma \mathfrak{h}.
\end{equation}
The expression in the \textsc{rhs} of this equation is the definition of the residue of the meromorphic $(2,0)$--form
$\mathfrak{h}$ (as a meromorphic form, the \textsc{rhs} of equation \eqref{whath} is defined in $\mathbb{C}^2$).
Therefore, using eqns.\eqref{yuksph} and \eqref{whath}, we get the formula we are after
\begin{equation}
 \mathrm{Yuk} = \mathrm{Residue}\: \mathrm{Tr}\left\{\frac{\big[\eta_1,\,\eta_2\big]}{f_1\: f_2}\: \varphi_3\right\}.
\end{equation}

\section{General Worldvolumes, Gauge Bundles, and Matter Curves}
\label{gens}
In this Appendix we briefly indicate how to generalize the definition of localized modes to the case of an arbitrary worldvolume \(S\), gauge bundle \(\mathrm{ad}(P)\), and matter curve \(\Sigma\).  To do this we make heavy use of the sheaf theoretical interpretation of the localized zero-modes developed in equations \((\ref{quo})-(\ref{torsionr})\).  In this general setting the background Higgs field is a holomorphic adjoint valued \((2,0)\) form
\begin{equation}
\Phi\in \Gamma(S,K_S\otimes\mathrm{ad}(P)).
\end{equation}
We consider the following two Dolbeault complexes of $C^\infty$ sheaves of adjoint--valued $(p,q)$ forms
\begin{align}
 &\dots \xrightarrow{\ \overline{\partial}\ } \Omega^{0,k}(\mathrm{ad}(P)) \xrightarrow{\ \overline{\partial}\ }
\Omega^{0,k+1}(\mathrm{ad}(P))\xrightarrow{\ \overline{\partial}\ } \cdots\\
 &\dots \xrightarrow{\ \overline{\partial}\ } \Omega^{2,k}(\mathrm{ad}(P)) \xrightarrow{\ \overline{\partial}\ }
\Omega^{2,k+1}(\mathrm{ad}(P))\xrightarrow{\ \overline{\partial}\ } \cdots
\end{align}
Since $\Phi$ is holomorphic, the natural map $\mathrm{ad}_\Phi\colon \Omega^{0,\ast}(\mathrm{ad}(P))\rightarrow \Omega^{2,\ast}(\mathrm{ad}(P))$ commutes with $\overline{\partial}$, and hence gives rise to a chain map between the two complexes above.  There is a standard construction in homological algebra, the \emph{mapping cone}, that is relevant in this situation.  By definition \cite{MS}, the mapping cone of $\mathrm{ad}_\Phi$ is the complex
\begin{equation}
\cdots\xrightarrow{\ \overline{D}\ } \boldsymbol{M}^k \xrightarrow{\ \overline{D}\ } \boldsymbol{M}^{k+1}\xrightarrow{\ \overline{D}\ }\cdots, \label{mcomplex}
\end{equation}
where
\begin{equation}
\boldsymbol{M}^k = \Omega^{0,k+1}(\mathrm{ad}(P))\oplus \Omega^{2,k}(\mathrm{ad}(P)),
\end{equation}
and
\begin{equation}
\overline{D}=\begin{pmatrix}\overline{\partial} & 0\\ \mathrm{ad}_\Phi & -\overline{\partial}\end{pmatrix}.
\end{equation}
As a consequence of the fact that \(\bar{\partial}^{2}=0\) and that \(\Phi\) is holomorphic, it follows that \(\overline{D}^{2}=0\) and hence it is meaningful to consider the cohomology of the mapping cone \((\ref{mcomplex})\).\footnote{By definition the group \(\Omega^{2,-1}(S,\mathrm{ad}(P))\) is taken to be the zero group.}

Let us consider the zeroth \(\overline{D}\) cohomology group of the complex \(\boldsymbol{M}^{\ast}\).  An element of \(\boldsymbol{M}^{0}\) which is \(\overline{D}\) closed is represented by a pair of adjoint-valued forms $(a,\varphi)$ of type, $(0,1)$ and $(2,0)$, respectively which satisfy the equations
\begin{eqnarray}
\bar{\partial}a &= &0, \label{eomm1}\\
\bar{\partial}\varphi & = & \mathrm{ad}_\Phi (a).
\label{eomm2}
\end{eqnarray}
Equations \eqref{eomm1}-\eqref{eomm2} are exactly the $8D$ F-term zero mode equations and justify our choice of notation \cite{BHVI,FGUTSNC}. Meanwhile the elements of \(\boldsymbol{M}^{0}\) which are \(\overline{D}\) exact are such that there exists a \(\chi \in \boldsymbol{M}^{-1}\) with
\begin{eqnarray}
a & = & \bar{\partial}{\chi} \\
\varphi & = & \mathrm{ad}_{\Phi}(\chi)
\end{eqnarray}
These are exactly the modes that are infinitesimal gauge transformations.  It follows that the zeroth cohomology group of the mapping cone coincides with the space of zero-modes solutions modulo complexified gauge transformations.  Thus we have the basic identification
\begin{equation}
\text{zero-modes} = H^{0}(S,\boldsymbol{M}).
\end{equation}
This is the most general definition of the zero-modes for the $8D$ gauge theory, valid in any circumstance (in particular, for $S$ compact \emph{and} non--compact).

Now, the basic properties of the mapping cone give rise to a long exact sequence
\begin{equation}
 \cdots \longrightarrow H^0(S, \mathrm{ad}(P))\xrightarrow{\mathrm{ad}_{\Phi}} H^0(S, \Omega^{2,0}(\mathrm{ad}(P)) \longrightarrow H^0(S, \boldsymbol{M})\longrightarrow H^1(S, \mathrm{ad}(P))\longrightarrow \cdots
\end{equation}
If we make the assumption\footnote{In phenomenological applications, this assumption is typically made to kill unwanted exotic bulk modes and to allow the possibility of a decoupling limit \cite{BHVI,DWI,BHVII,DWII,Cordova:2009fg, Grimm:2009yu}} $H^1(S, \mathrm{ad}(P))=0$, then the above sequence simplifies and we see that
\begin{equation}\label{coh1}
 \text{zero-modes}\equiv H^0(S,\boldsymbol{M})=H^{0}(S,K_S\otimes \mathrm{ad}(P))\Big/ \mathrm{ad}_\Phi\left( H^{0}(S,\mathrm{ad}(P))\right),
\end{equation}
which has precisely the same form as equation \eqref{quo} for the local geometries.  Thus provided $H^1(S, \mathrm{ad}(P))=0$ the passage from our local analysis to the case of a general brane worldvolume and adjoint bundle is completely trivial.  One simply takes the local expression \eqref{quo} and computes global sections over \(S\) valued in \(\mathrm{ad}(P)\).

Continuing with these assumptions, it is then natural to introduce the sheaf of modes \(\mathcal{Q}\) defined as
\begin{equation}\label{coh2}
\mathrm{ad}(P) \xrightarrow{\ \mathrm{ad}_\Phi\ } K_S\otimes \mathrm{ad}(P)\rightarrow\mathcal{Q}\rightarrow 0.
\end{equation}
The space of the zero-modes may be identified with a subspace $Z(S)$ of $H^{0}(S,\mathcal{Q})$.  The simplest possibility is that $Z(S)\equiv H^{0}(S,\mathcal{Q})$.  This occurs automatically if $H^1\big(S, \mathrm{ad}(P)/\ker \mathrm{ad}_\Phi\big)=0$, for example if $S$ is Stein, as in the local geometries of the previous subsection. Then, the localized modes correspond to the torsion part of the sheaf of modes \(\mathcal{Q}\). To be completely formal, one introduces a sub-sheaf $\mathfrak{Loc}$ by the exact sequence \cite{Friedman}
\begin{equation}
 0\rightarrow \mathfrak{Loc}\rightarrow \mathcal{Q}\rightarrow \mathcal{Q}^{\ast\ast}
\end{equation}
where $\mathcal{Q}^{\ast\ast}$ is the double dual. If \(\mathcal{Q}\) were a vector bundle, then \(\mathcal{Q}\) would simply equal \(\mathcal{Q}^{\ast\ast}\) and the sheaf \(\mathfrak{Loc}\) would vanish.  Thus when \(\mathfrak{Loc}\) is non-trivial, it measures the failure of \(\mathcal{Q}\) to be a vector bundle and hence captures the torsion of \(\mathcal{Q}\).  It follows that in general we have
\begin{gather*}
\text{\emph{localized} zero modes}= H^{0}(S,\mathfrak{Loc})\cap Z(S)\\
\text{potential matter curves}= \text{irreducible components of } \mathrm{supp}\,(\mathfrak{Loc}).
\end{gather*}

We can similarly extend our analysis of the $6D$ superpotential $W_{6D}$ to the setting of a general matter curve \(\Sigma\) and gauge bundle \(\mathrm{ad}(P)\).  In complete generality, the \(6D\) action takes the form of a non-degenerate $2D$ chiral Dirac Lagrangian coupled to suitable connections on holomorphic vector bundles $V_i\rightarrow \Sigma$,
\begin{equation}
W_{6D}= \int_\Sigma \Omega^{ij}\, \eta_i \,\overline{\partial}_{V_j}\eta_j,
\end{equation}
where the $6D$ field $\eta_i$ transforms as a section of $V_i$. It is easy to see that the symplectic pairing $\Omega^{ij}$ satisfies the selection rule
\begin{equation}\label{omegaselrule}
\Omega^{ij}\neq 0 \quad\Rightarrow \quad V_i= K_\Sigma\otimes V_j^*.
\end{equation}
This implies that the integrand in \(W_{6D}\) is gauge invariant and naturally a \((1,1)\) form on \(\Sigma\).  In particular, this means that the \(6D\) superpotential is independent from the K\"ahler metric. Thus provided we correctly identify the bundles \(V_{i}\) we may count the net number of localized zero-modes on $\Sigma$ with standard $2D$ index theorems.

To determine the $V_i$'s, one uses the adjunction formula as in reference \cite{BHVI}. Again we assume $H^1(S,\mathrm{ad}(P))=0$.  From equation \eqref{basic2} we see that from the $8D$ viewpoint $\eta_i$ is a section of some sub-bundle (depending on the specific mode $\eta_i$)
\begin{equation}
\mathcal{F}_i\otimes \mathcal{O}(\Sigma)^{m_i}\subset \mathrm{ad}(P)\otimes \mathcal{O}(\Sigma)^{m_i},
\end{equation}
where the matter curve $\Sigma$ is given locally by $f=0$. Analogously, $\varphi_i$ is a section
of some sub-bundle $K_S\otimes \mathcal{E}_i\subset K_S\otimes \mathrm{ad}(P)$. The bundles $\mathcal{F}_i$,
$\mathcal{E}_i$ correspond to decomposition of the adjoint of $\mathfrak{g}$ into irreducible representations of the unbroken gauge subgroup.  In a tubular neighborhood of $\Sigma$ we write\footnote{\ Here we are cavalier with a subtlety which has already appeared in the literature about branes with triangular Higgs backgrounds \cite{Donagi:2003hh}, namely the fact that, while in the $C^\infty$ sense a tubular neighborhood of $\Sigma$ is diffeomorphic to its normal bundle, this is not true in the complex-analytic sense. The discrepancy is related to obstructions of moduli, and hence it is expected to be relevant for cubic (and higher) terms in $W_{6D}$, which are outside the scope of the present paper.} (locally)
\begin{align}
\varphi_i &=\left(\sum_{s=0}^{m_i-1} f^s\, \chi_i^{(s)}\right)\wedge df\\
\eta_i &=\sum_{s=0}^{m_i-1} f^s\, \xi_i^{(s)}
\end{align}
where the $6D$ fields are sections
\begin{align}
\chi_i^{(s)}\Big|_\Sigma&\in
C^\infty\!\!\left(\Sigma, K_\Sigma\otimes \mathcal{E}_i\otimes\mathcal{O}(\Sigma)^{-s-1}\Big|_\Sigma\right)\\
\xi_i^{(s)}\Big|_\Sigma &\in C^\infty\!\!\left(\Sigma, \mathcal{F}_i\otimes\mathcal{O}(\Sigma)^{m_i-s}\Big|_\Sigma\right),
\end{align}
(in the first line we used the adjunction formula
$K_S\otimes\mathcal{O}(\Sigma)\big|_\Sigma =K_\Sigma$). On--shell, smooth sections over $\Sigma$ get replaced by holomorphic ones.  From equation \eqref{W6Dfinal} of Appendix \ref{pairing} we have (we omit writing the restriction $(\cdot)\big|_\Sigma$ which is everywhere implied)
\begin{equation}
 W_{6D}= \int_\Sigma \sum_{i,j} \varrho(\mathcal{E}_i,\mathcal{F}_j) \sum_{s=0}^{m_i-1} \xi^{m_i-1-s}\;
\overline{\partial}_{V_j^{(s)}}\,\chi_j^{(s)},
\end{equation}
where $\varrho(\mathcal{E}_i,\mathcal{F}_j)$ is a group theory factor which vanishes unless the two vector bundles $\mathcal{E}_i$, $\mathcal{F}_j$ are induced by dual representations of the unbroken gauge group. Therefore
\begin{equation}
 V_j^{(s)}= K_\Sigma\otimes \left.\left(\mathcal{E}_j\otimes \mathcal{O}(\Sigma)^{-s-1}\right)\right|_\Sigma,
\end{equation}
and the selection rule \eqref{omegaselrule} is verified.

\newpage
\addcontentsline{toc}{section}{References}
\bibliographystyle{utphys}
\bibliography{tbranes}

\end{document}